\documentclass[%
reprint,
amsmath,amssymb,
aps,
]{revtex4-2}

\usepackage{graphicx}
\usepackage{dcolumn}
\usepackage{bm}
\usepackage{siunitx}
\usepackage{hyperref}
\usepackage{comment}

\usepackage{color}
\usepackage{graphicx}
\usepackage{mathpazo}
\definecolor{JG}{rgb}{0.17, 0.74, 0.17}
\definecolor{IB}{rgb}{0.12, 0.47, 0.71}
\usepackage{ulem}
\usepackage{overpic}
\usepackage{float}
\usepackage{placeins}

\begin{document}
	
	\preprint{APS/123-QED}
	
	\title{Impact of Charge Conversion on NV-Center Relaxometry}
	
	\author{Isabel Cardoso Barbosa}
	\affiliation{
		Department of Physics and State Research Center OPTIMAS, University of Kaiserslautern-Landau, Erwin-Schroedinger-Str. 46, 67663 Kaiserslautern, Germany\\
	}

	\author{Jonas Gutsche}
	\affiliation{
		Department of Physics and State Research Center OPTIMAS, University of Kaiserslautern-Landau, Erwin-Schroedinger-Str. 46,
        67663 Kaiserslautern, Germany\\
	}
	
	\author{Artur Widera}
	\email{Author to whom correspondence should be addressed: widera@physik.uni-kl.de}
    \affiliation{
		Department of Physics and State Research Center OPTIMAS, University of Kaiserslautern-Landau, Erwin-Schroedinger-Str. 46,
        67663 Kaiserslautern, Germany\\
	}

	\date{\today}

    \begin{abstract}
	Relaxometry schemes employing nitrogen-vacancy (NV) centers in diamonds are essential in biology and physics to detect a reduction of the color centers' characteristic spin relaxation ($T_1$) time caused by, e.g., paramagnetic molecules in proximity. However, while only the negatively-charged NV center is to be probed in these pulsed-laser measurements, an inevitable consequence of the laser excitation is the conversion to the neutrally-charged NV state, interfering with the result for the negatively-charged NV centers' $T_1$ time or even dominating the response signal. 
	In this work, we perform relaxometry measurements on an NV ensemble in nanodiamond combining a \SI{520}{\nano\meter} excitation laser and microwave excitation while simultaneously recording the fluorescence signals of both charge states via independent beam paths. Correlating the fluorescence intensity ratios to the fluorescence spectra at each laser power, we monitor the ratios of both charge states during the $T_1$-time  measurement and systematically disclose the excitation-power-dependent charge conversion. Even at laser intensities below saturation, we observe charge conversion, while at higher intensities, charge conversion outweighs spin relaxation. 
	These results underline the necessity of low excitation power and fluorescence normalization before the relaxation time to accurately determine the $T_1$ time and characterize paramagnetic species close to the sensing diamond.  
	\end{abstract}

	\maketitle
	
	\section{\label{sec:introduction}INTRODUCTION}
	The negatively-charged nitrogen-vacancy (NV) center in diamond constitutes a versatile tool for the detection of magnetic \cite{Acosta.2009, Balasubramanian.2008, Maze.2008, Degen.2008, Taylor.2008, Laraoui.2010, Schirhagl.2014, Thiel.2019, Dix.2022} and electric \cite{Dolde.2011} fields with high sensitivity and spatial resolution.
	Measurement of the NV centers' spin relaxation ($T_1$) time is widely applied in different fields of science to detect magnetic noise \cite{Rollo.2021, Sigaeva.2022}.
	Various so-called relaxometry measurement schemes employ a reduction of the NV centers' $T_1$ time with the host nanodiamond exposed to paramagnetic molecules fluctuating at the NV centers' resonance frequency \cite{Cole.2009, Hall.2009, Steinert.2013}.
	Thus, relaxometry schemes have been used to detect a superparamagnetic nanoparticle \cite{SchmidLorch.2015}, or paramagnetic $\mathrm{Gd}^{3+}$ ions \cite{Steinert.2013, Tetienne.2013, Sushkov.2014, Pelliccione.2014, Gorrini.2019}.
	Further, relaxometry with $\mathrm{NV}^-$ centers has been utilized to trace chemical reactions involving radicals \cite{Barton.2020, PeronaMartinez.2020}.
	Also, the NV centers' $T_1$ time as a measure for the presence of paramagnetic noise gains momentum in biological applications \cite{Schirhagl.2014, Sigaeva.2022}.
	Individual ferritin proteins have been detected \cite{SchaferNolte.2014} and relaxometry has been applied to detect radicals even inside cells \cite{Nie.2021, Sharmin.2021, Sigaeva.2022b, Norouzi.2022}.
	
	\begin{figure}[b]
    \includegraphics[width=86mm]{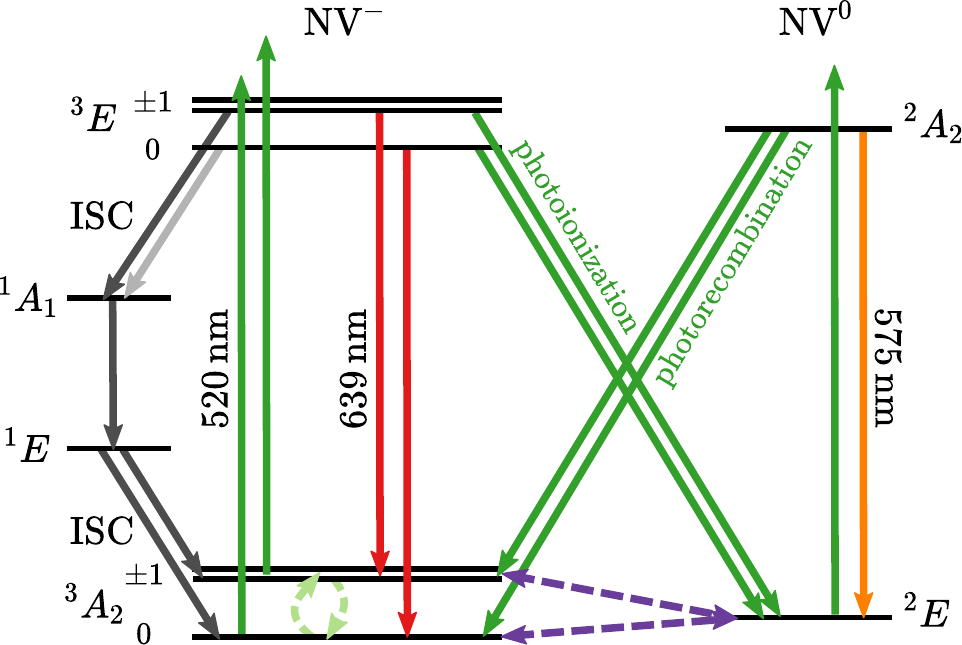}
    \caption{\label{fig:level_scheme} 
     Level scheme of the NV center in diamond. Depicted are levels of the negatively-charged $\mathrm{NV}^-$ and the neutrally-charged $\mathrm{NV}^0$ and transitions between the two charge states. Gray arrows show transitions between $\mathrm{NV}^-$'s triplet and singlet states, mediated via intersystem crossing (ISC). Green arrows denote transitions driven by a green laser, red and orange arrows mark the fluorescence of the NV charge states. Light-green dashed arrows between $m_S$ states are transitions driven by microwave radiation at \SI{2.87}{\giga\hertz} at zero magnetic field. Additionally, the light-green dashed arrows represent the relaxation of the spin-polarized state to a thermally mixed state without illumination ($T_1$). The purple dashed arrows denote charge transfer processes in the dark.
    }
    \end{figure}
    
    Especially in the field of biology, $T_1$ measurement schemes are often conducted only with optical excitation of the $\mathrm{NV}^-$ centers, while the readout of their spin states is realized by detection of the ensemble's fluorescence intensity. 
    This all-optical NV relaxometry avoids microwave pulses for convenience and undesired heating of biological samples \cite{Barton.2020, Sigaeva.2022b}.
	However, recent results indicate that a second process impeding the $\mathrm{NV}^-$ centers' fluorescence signal is present in relaxometry measurements \cite{Choi.2017, Giri.2018, Giri.2019, Gorrini.2019, Gorrini.2021}. 
	The laser pulse that is fundamental for preparation of the $\mathrm{NV}^-$ centers' spin state can additionally ionize the $\mathrm{NV}^-$ center to its neutrally-charged state, $\mathrm{NV}^0$.	
	The physics of this NV-center charge conversion has been studied in \cite{RamanNair.2020}. Here, we show the impact of this charge-state switching on relaxometry.
	Conversion under illumination and back-conversion in the dark influence the $\mathrm{NV}^-$ centers' fluorescence signal, complicating a seemingly simple measurement.
    A quantitative determination of the unwanted contribution of the $\mathrm{NV}^0$ state to the $\mathrm{NV}^-$ relaxometry data is, however, elusive.
    In this work, we compare the results of two relaxometry schemes well-known in literature for the same nanodiamond at varying laser powers. Additionally, we introduce a novel method to extract the ratio of the two NV charge states from the NV centers' fluorescence spectra throughout the entire measurement sequence to give an insight into the vivid NV charge dynamics we observe in our data.
    
	A level scheme of the NV center in diamond is depicted in Fig.~\ref{fig:level_scheme}, including the negatively-charged $\mathrm{NV}^-$ \cite{Doherty.2011, Felton.2008, Levine.2019}, the neutrally-charged $\mathrm{NV}^0$ and transitions from $\mathrm{NV}^-$ to $\mathrm{NV}^0$ under green illumination \cite{Chen.2015, Meirzada.2018}.
	We include transitions from the $\mathrm{NV}^0$'s ground state to $\mathrm{NV}^-$ without green illumination, reflecting the observation of recharging processes in the dark in \cite{Choi.2017, Giri.2018} and in this work.
	Using a \SI{520}{\nano\meter} laser, we non-resonantly excite the $\mathrm{NV}^-$ centers from their triplet ground state $^3A_2$ to the electronically-excited state $^3E$. 
	Because $^3E$'s states $m_S = \pm 1$ are preferentially depopulated via the  $\mathrm{NV}^-$ centers' singlet states $^1A_1$ and $^1E$, illumination with a green laser will spin polarize the $\mathrm{NV}^-$ centers into their ground spin state $m_S = 0$ \cite{Levine.2019}.
	The $T_1$ time describes how long this spin polarization persists until the spin population decays to a thermally mixed state \cite{Levine.2019}.
	It can reach up to \SI{6}{\milli\second} in bulk diamonds at room temperature \cite{Naydenov.2011} and is influenced by paramagnetic centers within the host diamond or on its surface \cite{Jarmola.2012, Romach.2015}.
	In the simplest $T_1$ measurement scheme, spin polarization is achieved by a laser pulse, followed by a second readout-laser pulse after a variable relaxation time $\tau$.
	Besides different durations, the two laser pulses are identical. 
	Therefore, the readout pulse is capable of spin-polarizing and ionizing the NV-center ensemble as well as the initialization pulse.
	Additionally, the spin-polarization pulse provides information about the charge-conversion processes during laser excitation.
	
	To determine the $T_1$ time of $\mathrm{NV}^-$ centers of a specific orientation in the diamond crystal, coherent spin manipulation is introduced in these measurements \cite{Jarmola.2012}.
	Here, a resonant microwave $\pi$ pulse transfers the population of these $\mathrm{NV}^-$ centers from $m_S = 0$ to $m_S = + 1$ or $m_S = - 1$ after the spin-polarization pulse. 
	A second laser pulse is used for the readout of the spin state. 
	Repetition of the sequence with the $\pi$ pulse omitted and subtracting the readout signals from each other yields a spin-polarization signal as a function of $\tau$ that is robust against background fluorescence \cite{Jarmola.2012, Mrozek.2015}.
	
	In the following, we present our experimental system in Section~\ref{sec:exp_sys}.
	Our results are divided into two main parts. 
	We first analyze fluorescence spectra of NV centers in a single nanodiamond to assign concentration ratios to count ratios measured with SPCMs in Section~\ref{sec:fl_spec}.
    This knowledge allows us to quantify the $\mathrm{NV}^0$ contribution during the spin-relaxation dynamics in Section~\ref{sec:Relaxometry}.

	\section{\label{sec:exp_sys}Experimental system}	
	
	\begin{figure}[b]
    \begin{overpic}[width=86mm]{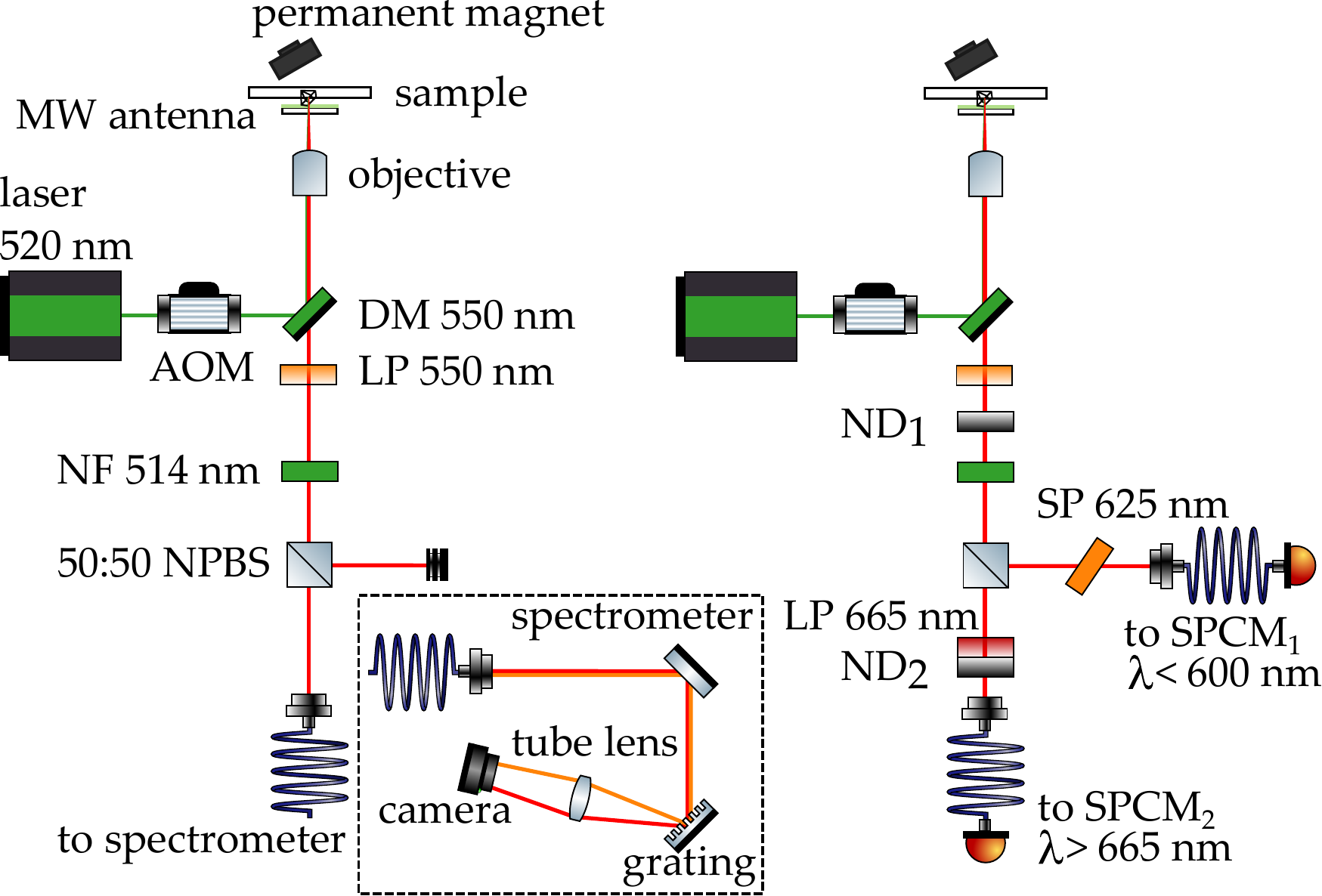}
         \put(0,63){(a)}
         \put(52,63){(b)}
    \end{overpic}
    \caption{\label{fig:exp_setup} 
    Experimental setup for recording NV fluorescence spectra and relaxometry data. In both setups, the excitation is the same, but the detection sections are different for the respective application.
    (a) NV centers in a single crystal nanodiamond are excited by a \SI{520}{\nano\meter}-laser in combination with an acousto-optic modulator (AOM). The light stemming from the sample is filtered by a dichroic mirror (DM), a longpass filter (LP) and a notch filter (NF) with given wavelenghts and passes a non-polarizing beamsplitter (NPBS). The remaining fluorescence is spectrally resolved on a camera chip. This setup is used for the measurement of the NV fluorescence spectra.
    (b) The NV fluorescence is split into two arms of a beamsplitter, additionally filtered with an LP or a shortpass filter (SP) and detected with fiber-coupled SPCMs. The SP is tilted to only transmit fluorescence below \SI{600}{\nano\meter}. To keep the detectors below saturation, neutral-density (ND) filters are used. Luminescence above \SI{665}{\nano\meter} ($\mathrm{NV}^-$ fluorescence) is detected in $\mathrm{SPCM_2}$, while light below \SI{600}{\nano\meter} ($\mathrm{NV}^0$ fluorescence) is detected in $\mathrm{SPCM_1}$. Transitions of the $\mathrm{NV}^-$ centers' spin states $m_S$ are driven with a microwave (MW) antenna.
    This setup is used for the measurement of the charge-state dependent relaxometry.
    }
    \end{figure}
    
	We perform our studies on a single nanodiamond crystal of size \SI{750}{\nano\meter} commercially available from Adamas Nano as water suspension (NDNV/NVN700nm2mg). 
	As specified by the manufacturer, the nanodiamonds' NV concentration is $[\mathrm{NV}] \approx \SI{0.5}{ppm}$, which is about \SI{2e4}{} NV centers per diamond. 
	For sample preparation, the suspension is treated in an ultrasonic bath to prevent the formation of crystal agglomerates. 
	We spin-coat the nanodiamonds to a glass substrate and subsequently remove the solvent by evaporating the residual water on a hot contact plate.
	
	To probe the NV centers in a single nanodiamond, we use a microscope consisting of an optical excitation and detection section and a microwave setup, as shown in Fig.~\ref{fig:exp_setup}.
	A CW-laser source of wavelength $\lambda = \SI{520}{\nano\meter}$ is used to optically excite the NV centers with a maximum laser power of \SI{4.9}{\milli\watt}. 
	The laser power was measured directly in front of the microscope objective (Nikon N20X-PF).
	The laser beam is focused to a spot-size diameter of \SI{700}{\nano\meter} ($1/e^2$ diameter), reaching a maximum intensity of $\sim \SI{2500}{\kilo\watt\per\centi\meter\squared}$.
	Pulses are generated by an AOM with an edge width of about \SI{120}{\nano\second}. Laser light is guided through an objective (NA = \SI{0.5}{}, WD = \SI{2.1}{\milli\meter}) and focused at the position of the nanodiamond. 
	Fluorescent light stemming from the sample is guided back through the objective and filtered by a dichroic mirror with a cut-on wavelength of \SI{550}{\nano\meter}. 
	Next, the fluorescence light is filtered by an additional \SI{550}{\nano\meter}-longpass filter and a \SI{514}{\nano\meter}-notch filter to prevent detection of reflected laser light. 
	The filtered fluorescence light is branched at a 50:50 non-polarizing beamsplitter, giving the possibility to further filter the luminescence and collect it in two separate detectors. 
	In particular, our setup allows for tailoring the transmitted wavelengths to the spectral regions, where either photon emission from the neutral or the negative NV charge state dominates in each beam path individually.
	Thus, we can easily discriminate between the emission of both charge states in our measurements.
	In this work, we make use of different detectors. 
	While for spectral analysis of the NV centers' fluorescence, we use a spectrometer (Fig.~\ref{fig:exp_setup}~(a)), we employ two single-photon counting modules (SPCMs) as detectors for our spin-relaxation measurements (Fig.~\ref{fig:exp_setup}~(b)) in combination with a time-to-digital converter.
	
	Microwave signals are generated, amplified, and brought close to the nanodiamond using a microwave antenna structure written on a glass substrate.
	All experiments are carried out under ambient conditions and in an external magnetic field in the order of \SI{12}{\milli\tesla} caused by a permanent magnet to split the NV centers' ODMR resonances. 
	In our ODMR spectrum, eight resonances appear because of the four existing orientations of NV centers in the single diamond crystal.
    We select one resonance to drive Rabi oscillations, from which we determine a $\pi$-pulse length of \SI{170}{\nano\second}.
    An ODMR spectrum and Rabi oscillations are provided in Fig.~\ref{fig:ODMR}.

	\section{\label{sec:fl_spec}Fluorescence spectra}
	
	\subsection{Setup}
	
	To spectrally resolve the NV centers' fluorescence, we use a spectrometer. 
	Details on the setup can be found in Appendix ~\ref{sec:spectrometer}.
	
	\subsection{Concentration ratio assignment}
    
    \begin{figure}[b]
        \begin{overpic}[width=86mm]{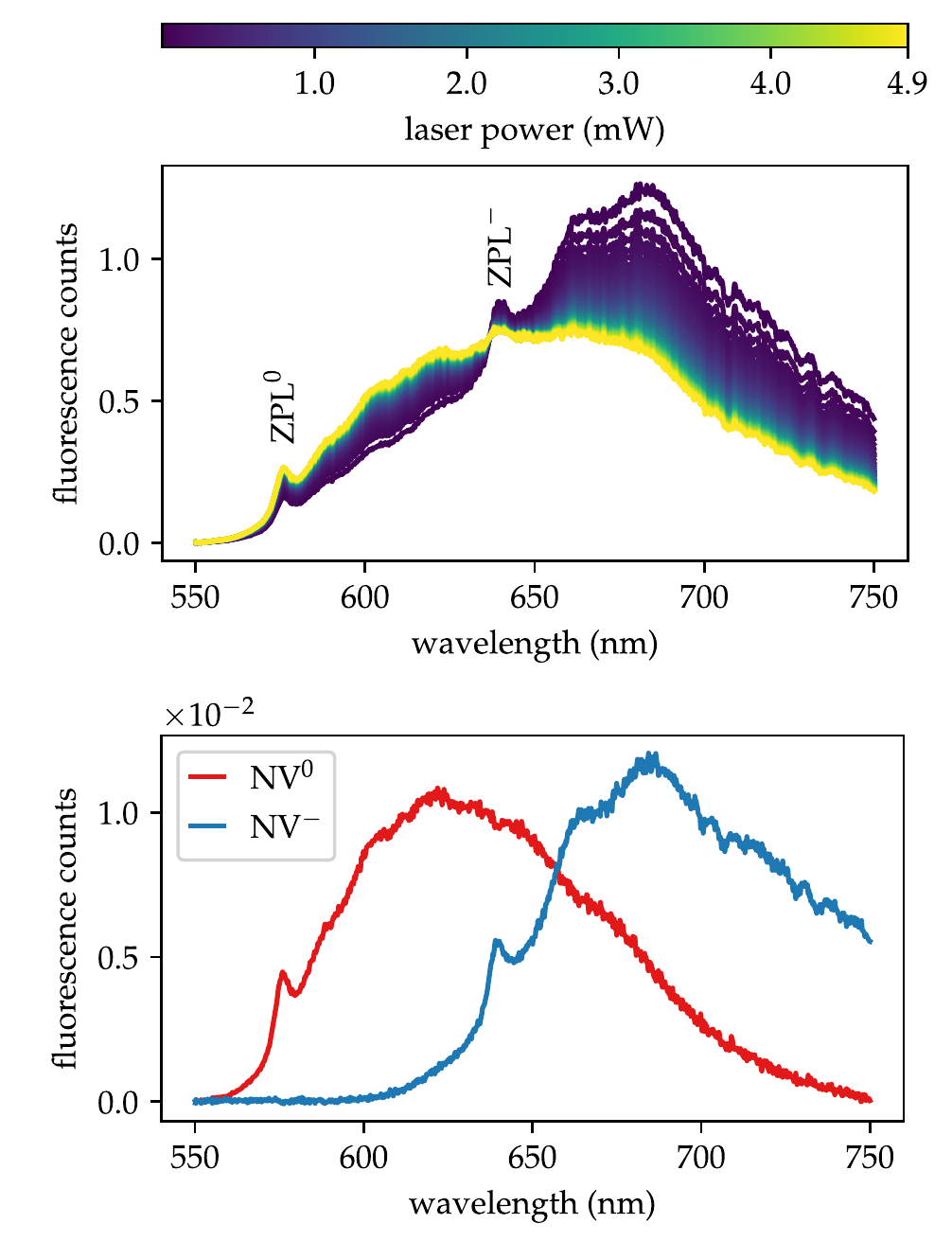}
         \put(6,96.5){(a)}
         \put(6,45){(b)}
        \end{overpic}
    \caption{\label{fig:spectra} 
    NV fluorescence spectra. 
    (a) Spectra recorded at laser powers from \SI{0.05}{\milli\watt} to \SI{4.9}{\milli\watt}. One hundred spectra were recorded at different laser powers in steps of $\sim$ \SI{0.05}{\milli\watt}. 
    The $\mathrm{NV}^0$s' ZPL at $\sim \SI{575}{\nano\meter}$ and the $\mathrm{NV}^-$s' ZPL at $\sim \SI{639}{\nano\meter}$ are evident and marked in the spectrum.
    For better visibility, spectra were normalized to the sum of the NV charge states' ZPL intensities.
    (b) Area-normalized decomposed basis functions for $\mathrm{NV}^0$ and $\mathrm{NV}^-$.
    }
    \end{figure}
    
    Corrected fluorescence spectra of a monocrystalline nanodiamond for excitation laser powers from \SI{0.05}{\milli\watt} to \SI{4.9}{\milli\watt} are depicted in Fig.~\ref{fig:spectra}~(a). 
    Two features, the $\mathrm{NV}^0$s' ZPL at $\sim \SI{575}{\nano\meter}$ \cite{Manson.2018} and the $\mathrm{NV}^-$s' ZPL at $\sim \SI{639}{\nano\meter}$ \cite{Juan.2017} are clearly visible. 
    The overlapping fluorescence spectra of both NV charge states show phonon broadening. 
    Conforming with the observation in \cite{Acosta.2009}\textcolor{IB}, the $\mathrm{NV}^0$s' ZPL intensity increases with higher laser power with respect to the $\mathrm{NV}^-$s' ZPL in our sample. 
    These results indicate a lower $[\mathrm{NV}^-]/[\mathrm{NV}^0]$ ratio at higher laser powers and thus an increasing charge conversion for higher powers.
    In \cite{Gorrini.2021}, similar experiments were performed on shallow NV centers, and the opposite effect was observed. 
    However, due to the different samples used in \cite{Gorrini.2021}, our results do not contradict the findings in this study.
    Moreover, our results perfectly agree with the results previously recorded in \cite{Acosta.2009}.
    
    \begin{figure}[b]
        \begin{overpic}[width=86mm]{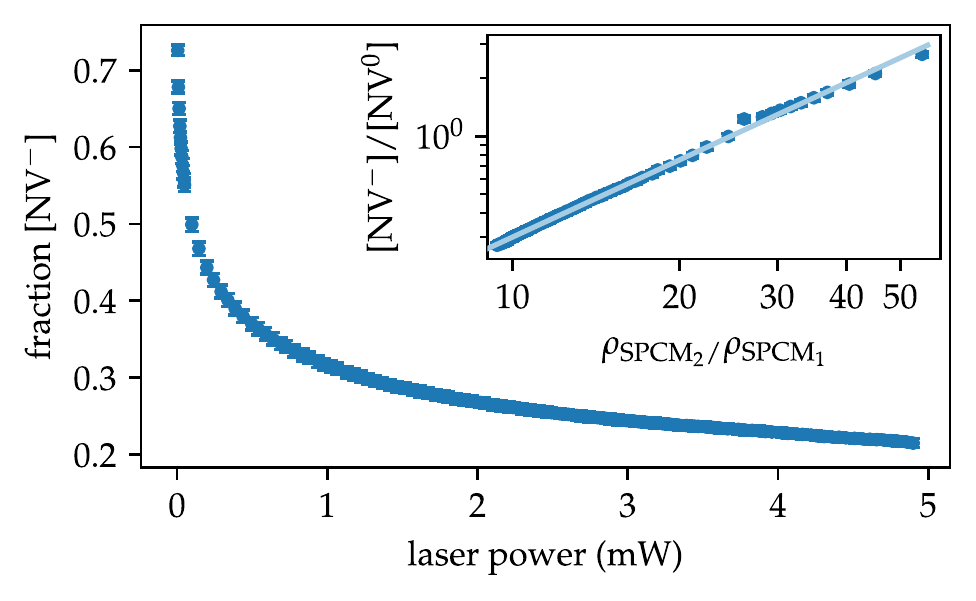}
        \end{overpic}
   \caption{\label{fig:ratio_LP} 
    Fraction of $[\mathrm{NV}^-]$ as a function of the laser power we derived from spectral analysis.
    The inset shows the ratio $[\mathrm{NV}^-]/[\mathrm{NV}^0]$ as a function of the fluorescence count ratio in the two SPCMs applied as detectors. Using the fit curve, we map the fluorescence count ratio to an NV ratio during relaxometry measurements. Fitting the $[\mathrm{NV}^-]/[\mathrm{NV}^0]$ concentration ratio with $f(x) = ax^n$, we obtain $a = \SI{0.0143 \pm 0.0002}{}$ and $n = \SI{1.325 \pm 0.005}{}$. 
    }
    \end{figure}
    
    We obtain area-normalized extracted spectra for $\mathrm{NV}^-$ and for $\mathrm{NV}^0$ from our recorded data as shown in Fig.~\ref{fig:spectra}~(b). 
    We conduct the spectra decomposition analysis of our spectra according to \textit{Alsid et al.} and follow the nomenclature given in reference \cite{Alsid.2019}. The fraction of $[\mathrm{NV}^-]$ of the total NV concentration $[\mathrm{NV_{total}}]$ is defined by
    
    \begin{equation}
        \frac{[\mathrm{NV}^-]}{[\mathrm{NV_{total}}]} =
        \frac{[\mathrm{NV}^-]}{[\mathrm{NV}^-]+[\mathrm{NV}^0]} = 
        \frac{c_-}{c_- + \kappa_{520} c_0}.
    \end{equation}
    
    Thus, the concentration ratio between NV charge states $[\mathrm{NV}^-]/[\mathrm{NV}^0]$ can be described with
    
    \begin{equation}
        \frac{[\mathrm{NV}^-]}{[\mathrm{NV}^0]} = \frac{c_-}{c_0} \frac{1}{\kappa_{520}}.
    \end{equation}
    
    Here, $c_-$ and $c_0$ describe the coefficients of the basis functions of $\mathrm{NV}^-$ and $\mathrm{NV}^0$ used to assemble an area-normalized composed spectrum at arbitrary laser power with the condition $c_- + c_0 = 1$. 
    The correction factor $\kappa_{520}$ translates this fluorescence ratio $c_-/c_0$ to the ratio of NV concentrations $[\mathrm{NV}^-]/[\mathrm{NV}^0]$, taking into account the different lifetimes and the absorption cross sections of the two NV charge states \cite{Alsid.2019}. 
    Note the different subscript in our work for the excitation wavelength of \SI{520}{\nano\meter} compared to $\kappa_{532}$ in \cite{Alsid.2019}. 
    Using ten spectra recorded at laser powers below the saturation intensity and the deviations from the linearity of the charge states' fluorescence intensity with the applied laser power, we find $\kappa_{520} = 1.97 \pm 0.07$. The error denotes the statistical error from a weighted fit we performed on our measurement data. 
    For a detailed description of the determination of $\kappa_{520}$, see Appendix~\ref{sec:520}.
    This value is within the reported value for $\kappa_{532} = 2.5 \pm 0.5$ for an excitation wavelength of \SI{532}{\nano\meter} \cite{Alsid.2019}. 
    We use our value for $\kappa_{520}$ to calculate the fractions of $[\mathrm{NV}^-]$ and $[\mathrm{NV}^0]$ and the concentration ratio $[\mathrm{NV}^-]/[\mathrm{NV}^0]$ as a function of the laser power. 
    
    In Fig.~\ref{fig:ratio_LP}, we show the fraction of $[\mathrm{NV}^-]$ as a function of the laser power.
    Since we neglect any other charge states of the NV center in this analysis, the sum of $[\mathrm{NV}^-]$ and $[\mathrm{NV}^0]$ is assumed constant.
    As shown in Fig.~\ref{fig:ratio_LP}, the fraction of $[\mathrm{NV}^-]$ is high for low laser powers and decreases with higher laser powers. At the lowest laser power of \SI{5}{\micro\watt}, about \SI{73}{\percent} of the total NV concentration is $[\mathrm{NV}^-]$, while at the highest laser power, only about \SI{21}{\percent} $[\mathrm{NV}^-]$ remain. Already at laser powers of \SI{0.10}{\milli\watt} ($\sim \SI{51}{\kilo\watt\per\cm\squared}$), which is below saturation intensity ($\approx \SI{100}{\kilo\watt\per\cm\squared}$) \cite{Wolf.2015}, $[\mathrm{NV}^0]$ outweighs $[\mathrm{NV}^-]$. 
    Therefore, a significant influence due to charge conversion is to be considered in relaxometry measurements. 
    
    To verify this laser-power dependent charge conversion in our nanodiamond samples, we perform this experiment for ten additional nanodiamonds of similar sizes and provide the results in Fig.~\ref{fig:10_spectra} in the Appendix. Overall, we observe similar behavior in all examined nanodiamonds. Additionally, we derive $\kappa_{520} = \SI{2.2 \pm 0.5}{}$ as a mean value for all nanodiamonds. The error denotes the standard deviation.
    
    Together with the recorded fluorescence-count-rate ratio of both SPCMs for each laser power, we assign each count-rate ratio $\rho_{\mathrm{SPCM}_2}/\rho_{\mathrm{SPCM}_1}$ a ratio $[\mathrm{NV}^-]/[\mathrm{NV}^0]$. The results are shown in Fig.~\ref{fig:ratio_LP} in the inset. 
    With an increasing ratio of $\rho_{\mathrm{SPCM}_2}/\rho_{\mathrm{SPCM}_1}$, the ratio $[\mathrm{NV}^-]/[\mathrm{NV}^0]$ increases. 
    We fit a power law (inverse-variance-weighted fit) to the ratio $[\mathrm{NV}^-]/[\mathrm{NV}^0]$ to be able to trace the NV-concentration ratio over a broad range of count-rate ratios during the spin-relaxation measurements.  
    Thereby we are able to quantitatively trace the contribution of $\mathrm{NV}^0$ during the spin-relaxation dynamics of the $\mathrm{NV}^-$ centers in the following.
    
	\section{\label{sec:Relaxometry}spin-relaxation measurements}
	
	\subsection{\label{sec:T1}Setup and measurement sequences}
	
	\begin{figure}[b]
	\begin{overpic}[width=86mm]{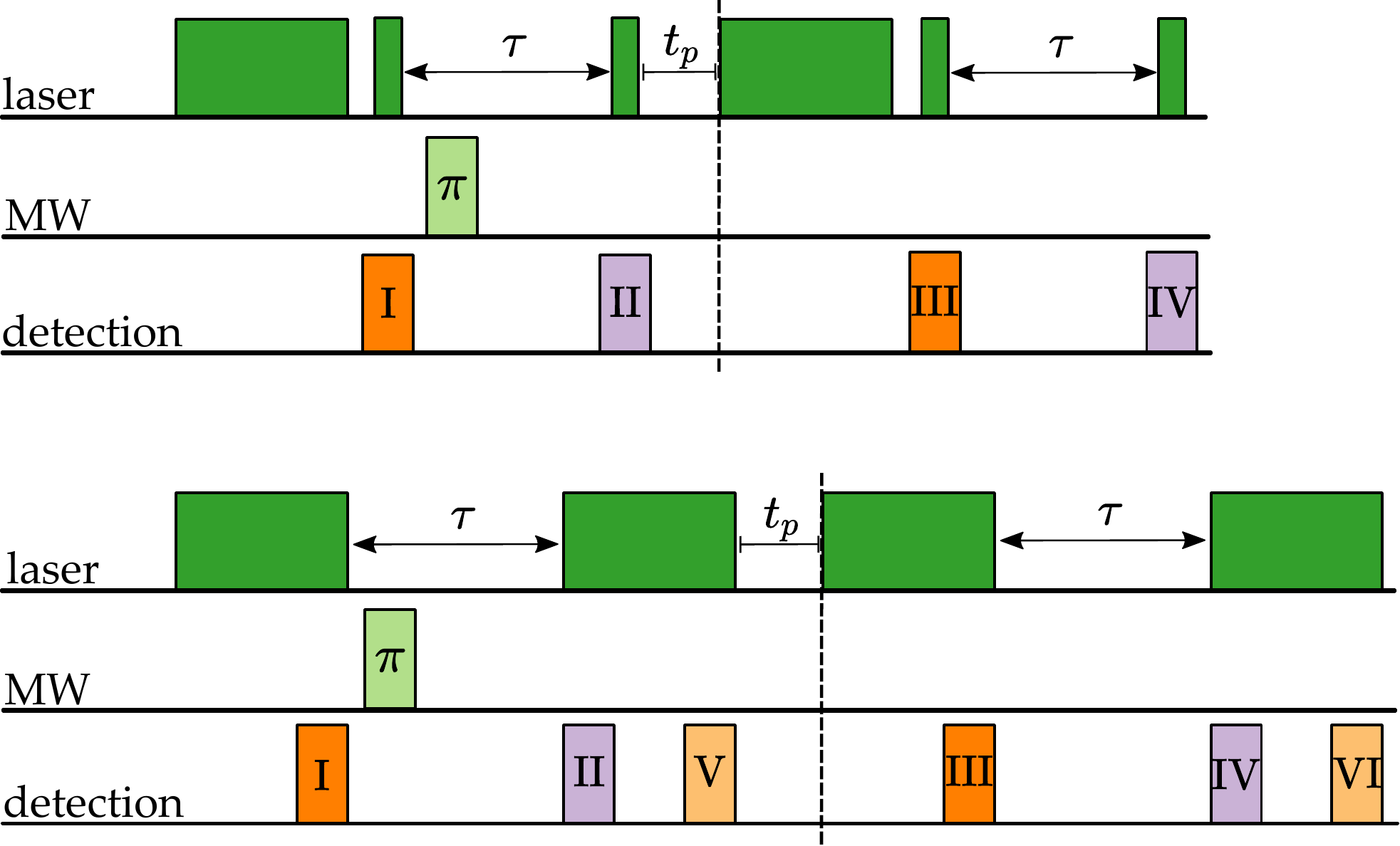}
         \put(0,60.7){(a) Sequence $P_1$}
         \put(0,27){(b) Sequence $P_2$}
    \end{overpic}
    \caption{\label{fig:sequence} 
    Longitudinal spin relaxation time ($T_1$) measurement schemes applied in this work. The beginning of the second half of each sequence is indicated by a dashed line. 
    (a) Sequence $P_1$. The NV ensemble is spin polarized by a laser pulse. Next, the fluorescence is detected by a control pulse (orange). Within the variable relaxation time $\tau$, a resonant $\pi$ pulse is applied (light-green). The spin state is read out by a third laser pulse (purple). The sequence is repeated with the $\pi$ pulse omitted after a pause time $t_p$.
    (b) Sequence $P_2$. As opposed to $P_1$, the readout pulse has the same length as the spin-polarization pulse. The control collection windows III within the initialization pulse and VI within the readout pulse will be compared in this work.
    }
    \end{figure}
	
	To separately detect the fluorescence of $\mathrm{NV}^-$ and $\mathrm{NV}^0$ throughout our measurement, different filters are used in the optical beam path as depicted in Fig.~\ref{fig:exp_setup}~(b). After passing a 50:50 non-polarizing beamsplitter, the sample's transmitted fluorescence light is guided through a \SI{665}{\nano\meter}-longpass filter, and mainly $\mathrm{NV}^-$ fluorescence is detected. For the luminescence reflected by the beamsplitter, we use a tilted \SI{625}{\nano\meter}-shortpass filter to collect $\mathrm{NV}^0$ fluorescence below \SI{600}{\nano\meter}. Neutral-density filters are added in front of the beamsplitter and within its transmitted beam path to keep the SPCMs below saturation.
	
	For determining the longitudinal spin relaxation time $T_1$, we conduct and compare two different and frequently used pulsed-measurement schemes, which we term $P_1$ and $P_2$ in the following. These two pulse sequences are depicted in Fig.~\ref{fig:sequence}.
	
	In the pulsed sequence $P_1$, we choose an initialization pulse of \SI{200}{\micro\second} duration to spin polarize the NV-center ensemble to their spin states $m_S = 0 $. 
	We apply a \SI{5}{\micro\second} normalization pulse \SI{1}{\micro\second} after the initialization pulse to probe the fluorescence intensity before a variable relaxation time $\tau$ in collection window I. 
	To assure a depopulation of the $\mathrm{NV}^-$ centers' singlet states, we choose the time between the two pulses to be longer than the singlet lifetimes of $\tau_\mathrm{meta} \approx\SI{150}{\nano\second}$ at room temperature \cite{Robledo.2011}.
	Approximately \SI{1.5}{\micro\second} into $\tau$, a resonant $\pi$ pulse is applied. 
	After $\tau$, a readout pulse of duration \SI{5}{\micro\second} probes the fluorescence of both NV centers' charge states in collection window II. 
	Subsequently, the sequence is repeated with the $\pi$ pulse omitted, obtaining fluorescence intensities in collection windows III and IV. 
	The spin polarization as a function of $\tau$ for $\mathrm{NV}^-$ is obtained by subtracting the fluorescence counts in II from the counts in collection window IV.
	Details on measurement sequence $P_1$ can be found in \cite{Jarmola.2012, Mrozek.2015}.
	Sequence $P_1$ provides a technique for determination of the $\mathrm{NV}^-$ centers' $T_1$ time robust against background fluorescence \cite{Jarmola.2012, Mrozek.2015} and is believed to be unaffected by charge-state conversion \cite{Giri.2018}.
	
	Analysis of the second half of $P_1$ represents an all-optical $T_1$ measurement scheme as often applied in biology \cite{Schirhagl.2014, Nie.2021, Norouzi.2022}. 
	Further, using only the second half of this sequence, we are able to obtain the fluorescence evolution as a function of $\tau$ for $\mathrm{NV}^-$ and $\mathrm{NV}^0$, including effects caused by the charge-state conversion. 
	Only taking into account the signal without the $\pi$ pulse applied, we obtain the fluorescence evolution by dividing the fluorescence counts in collection window IV by the counts in collection window III.
	Charge conversion during the relaxometry measurement has an effect on the $\mathrm{NV}^-$ fluorescence as well as on the $\mathrm{NV}^0$ fluorescence during the relaxation time $\tau$.
	Therefore, by only evaluating $P_1$'s second half, we gain information about the charge conversion taking place alongside the spin relaxation. 
	However, to obtain the $\mathrm{NV}^-$ centers' $T_1$ time, the full sequence $P_1$ is evaluated.
	
	As opposed to $P_1$, $P_2$ uses a normalization probe \textit{after} the readout of the NV centers' fluorescence \cite{Pelliccione.2014, Barton.2020, Guillebon.2020}. 
	We choose the laser readout pulse to have the same duration as the initialization pulse (\SI{200}{\micro\second}) and carry out the readout collection windows II and IV in the first \SI{5}{\micro\second} and the normalization probes V and VI in the last \SI{5}{\micro\second} of the readout pulse. 
	Scheme $P_2$ assumes the NV centers to have the same fluorescence intensity at the end of the second pulse as at the end of the first pulse. 
	To test this notion, we apply second normalization collection windows, I and III, within the last \SI{5}{\micro\second} of the initialization pulse and compare the results for both normalized data.
	
	Between readout and the upcoming initialization pulse, we insert a pause time $t_p$ between the sequences of \SI{1}{\milli\second}, which is in the order of $T_1$, to minimize build-up effects from spin polarization during the cycle for both sequences. Each cycle is repeated \SI{50000} times, and the whole measurement is swept multiple times.
	The sequences are repeated for different laser powers, ranging from \SI{5}{\micro\watt} to \SI{0.54}{\milli\watt}. 
	
	\subsection{\label{sec:Results}Results and discussion}
  
    In this section, we present and compare the experimental results for the spin-relaxation measurements for both sequences, $P_1$ and $P_2$.
    Using our experimental setup as described in Section~\ref{sec:T1}, we observe laser-power-dependent dynamics in the $\mathrm{NV}^-$ and $\mathrm{NV}^0$ fluorescence throughout our measurement. 
    
    \subsubsection{Sequence $P_1$}
    
\begin{figure}[b]
    \includegraphics[width=86mm]{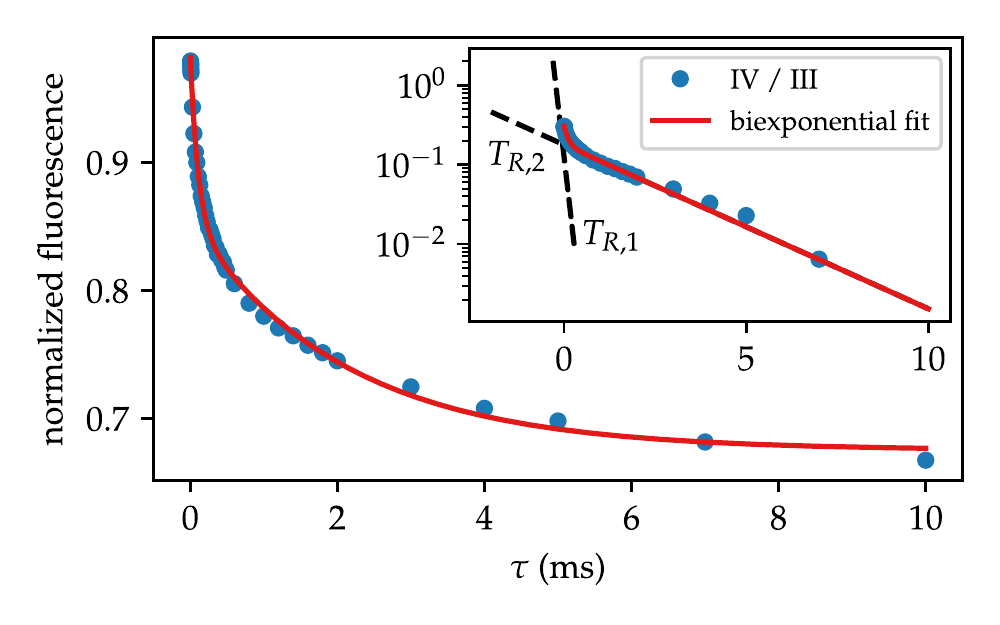}
    \caption{\label{fig:TR} 
    $\mathrm{NV}^0$ fluorescence as a function of $\tau$ as recorded with sequence $P_1$ (second half) by division of the fluorescence counts in IV by the counts in III for \SI{0.54}{\milli\watt} laser power. With a biexponential fit function, we find $T_{R,1} =  \SI{109 \pm 7}{\micro\second}$ and $T_{R,2} =\SI{2.1 \pm 0.1}{\milli\second}$. The inset shows the same measurement data and fit functions with the offset subtracted in a semi-logarithmic plot to illustrate the necessity of a biexponential fit function in our analysis.}
    \end{figure}

    Fig.~\ref{fig:TR} depicts an example for the fluorescence as a function of $\tau$ for the $\mathrm{NV}^0$ fluorescence recorded at a laser power of \SI{0.54}{\milli\watt} with sequence $P_1$. 
    These results show the normalized fluorescence as a function of $\tau$  obtained from the second part of the measurement sequence without a microwave $\pi$ pulse, dividing the fluorescence counts in collection window IV by the counts in collection window III. 
    The normalized fluorescence as a function of $\tau$ decays exponentially. 
    Different from the dynamics of the $\mathrm{NV}^-$ center, we observe similar behavior for the $\mathrm{NV}^0$ fluorescence at all laser powers. We fit a biexponential function of type 
    \begin{equation}
       f^0(\tau) = A\,e^{-\tau/T_{R,1}} + B\,e^{-\tau/T_{R,2}} + d
    \end{equation}
    to our measurement data and obtain two recharge times in the order of $T_{R,1} = \SI{100}{\micro\second}$ and $T_{R,2} = \SI{2.0}{\milli\second}$ for all laser powers. 
    We plot the $\mathrm{NV}^0$ fluorescence as well as $f^0(\tau)$ with the offset $d$ subtracted in a semi-logarithmic plot as an inset in Fig.~\ref{fig:TR}. 
    This presentation shows that a biexponential fit function is required to describe our measurement data.
    We assign these time constants $T_{R}$ to an electron-recapturing process of  $\mathrm{NV}^0$ during the dark time $\tau$, after an ionization from $\mathrm{NV}^-$ to $\mathrm{NV}^0$ has previously taken place in the initializing laser pulse. 
    Remarkably, this process occurs even at the lowest laser power. 
    Presumably, the presence of two components of $T_{R}$ is due to the different environments of NV centers concerning charge transfer sites.
    Vacancies or electronegative surface groups on the diamond surface are known to promote a charge conversion of $\mathrm{NV}^-$ to $\mathrm{NV}^0$ \cite{Rondin.2010, Wilson.2019}. 
    We assume that the NV environment similarly affects the recharging process in the dark.
    Therefore, we attribute one component of $T_R$ to NV centers closer to the nanodiamond surface and the other to NV centers more proximate to the center of the crystal.
    We emphasize that both $T_{R,1}$ and $T_{R,2}$ we report match previously reported values for $T_R$ of \SI{100}{\micro\second} \cite{Choi.2017} and \SI{3 \pm 1}{\milli\second}  \cite{Gorrini.2019} and underline that they simultaneously appear as two components in our sample.
    We find that neither $T_{R,1}$ nor $T_{R,2}$ changes as a function of the laser power. 
    The coefficients of the exponential functions $A$ and $B$ do not change significantly from \SI{0.05}{\milli\watt} to \SI{0.54}{\milli\watt} laser power. However, for the lowest laser power of \SI{5}{\micro\watt} $A$ and $B$ are smaller.
    We attribute this to little $\mathrm{NV}^0$ fluorescence observed at this low laser power due to less charge conversion, resulting in a lower signal-to-noise ratio (SNR) for the $\mathrm{NV}^0$ fluorescence.
    To confirm this observed biexponential decay in the $\mathrm{NV}^0$ fluorescence, we conduct the same sequence $P_1$ on ten additional nanodiamonds of similar sizes under similar conditions. We provide the results in the Appendix in Fig.~\ref{fig:Relaxometry_NV0_10}. In all examined nanodiamonds, the $\mathrm{NV}^0$ fluorescence decays biexponentially as a function of $\tau$, and the two components we find for $T_{R,1}$ and $T_{R,2}$ are in accordance with the values we previously observed. We find $T_{R,1} = \SI{91 \pm 26}{\micro \second}$ and $T_{R,2} = \SI{1.7 \pm 0.5}{\milli \second}$ as mean values for the ten additional nanodiamonds, the error denotes the standard deviation.

    \begin{figure}
        \begin{overpic}[width=86mm]{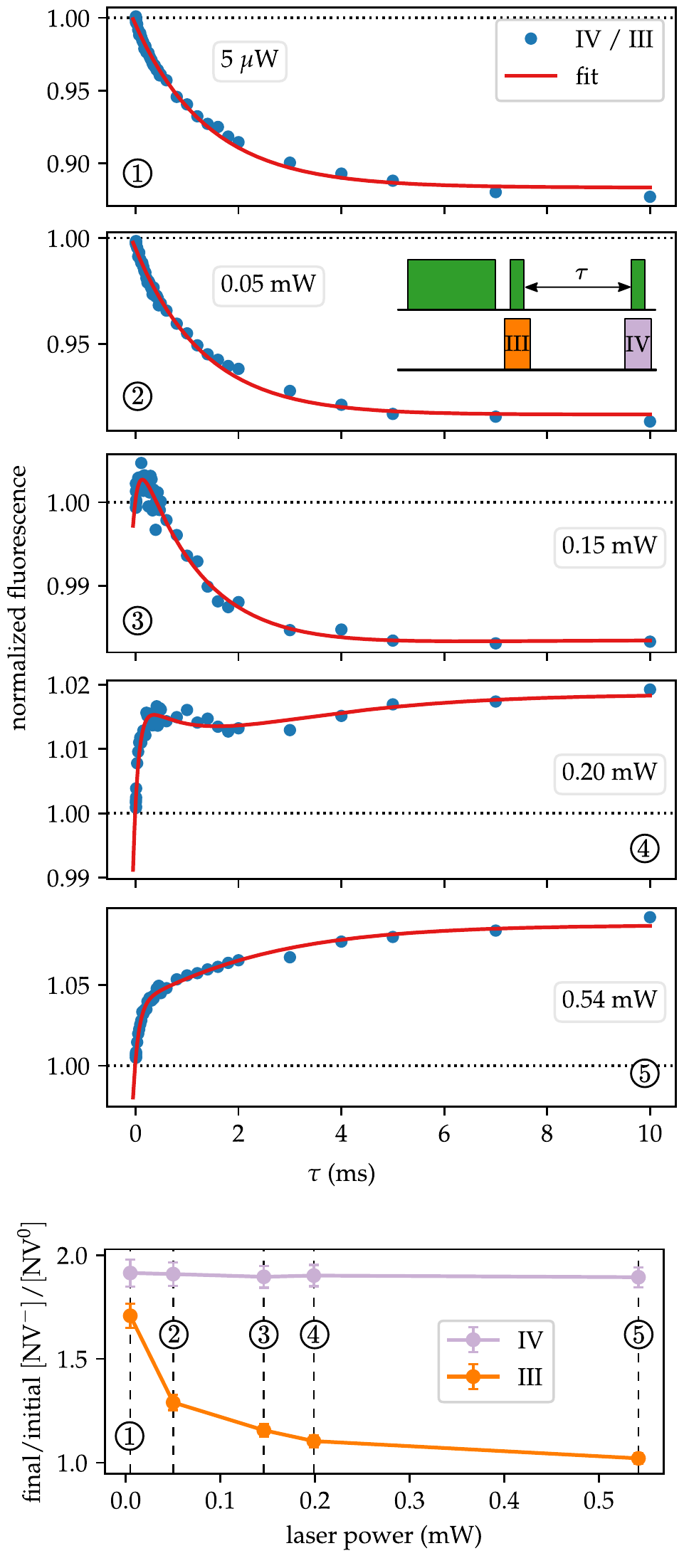}
         \put(0,98.5){(a)}
         \put(0,23.4){(b)}
        \end{overpic}
   \caption{\label{fig:relaxometry}
    (a) $\mathrm{NV}^-$ fluorescence as a function of $\tau$, obtained from the second half of $P_1$ at different laser powers. 
    We observe a transition from an exponential fluorescence decay to an inverted exponential profile with rising laser powers. For \SI{5}{\micro\watt} laser power, we perform a monoexponential fit and obtain $T_1 = \SI{1.42 \pm 0.06}{\milli\second}$. For laser powers above, we fit a sum of three exponentials as explained in the text.
    (b) Changes of the NV-charge-state ratio during the relaxometry measurement from lowest to highest $\tau$ in sequence $P_1$.}
    
    \end{figure}
    
    Further, we present the results for the normalized $\mathrm{NV}^-$ fluorescence as a function of $\tau$ in Fig.~\ref{fig:relaxometry}~(a) for ascending laser powers. 
    We conducted the experiment with sequence $P_1$, and this data refers to the results with the $\pi$ pulse omitted. 
    The laser-power-dependent dynamics of $\mathrm{NV}^-$ and $\mathrm{NV}^0$ result in a drastic change of shape of the normalized fluorescence as a function of $\tau$. 
    While we observe an exponential decay in the lowest laser power, we find an inverted exponential profile of the $\mathrm{NV}^-$ fluorescence at \SI{0.54}{\milli\watt} laser power. 
    In-between laser powers show both an exponential decay and an increase, present in the fluorescence. This phenomenon of inverted exponential components in the recorded normalized fluorescence during a $T_1$ measurement has been reported by \cite{Giri.2018} and attributed to a recharging process of $\mathrm{NV}^0$ to $\mathrm{NV}^-$ during $\tau$. 
    However, a complete flip of the fluorescence alone by a laser power increase has not been reported so far. 
    Remarkably, this behavior indicates that $\mathrm{NV}^0$ to $\mathrm{NV}^-$ charge dynamics outweigh the $\mathrm{NV}^-$ ensemble's spin relaxation at high laser powers in our sample.

    \begin{figure}[b]
    \includegraphics[width=86mm]{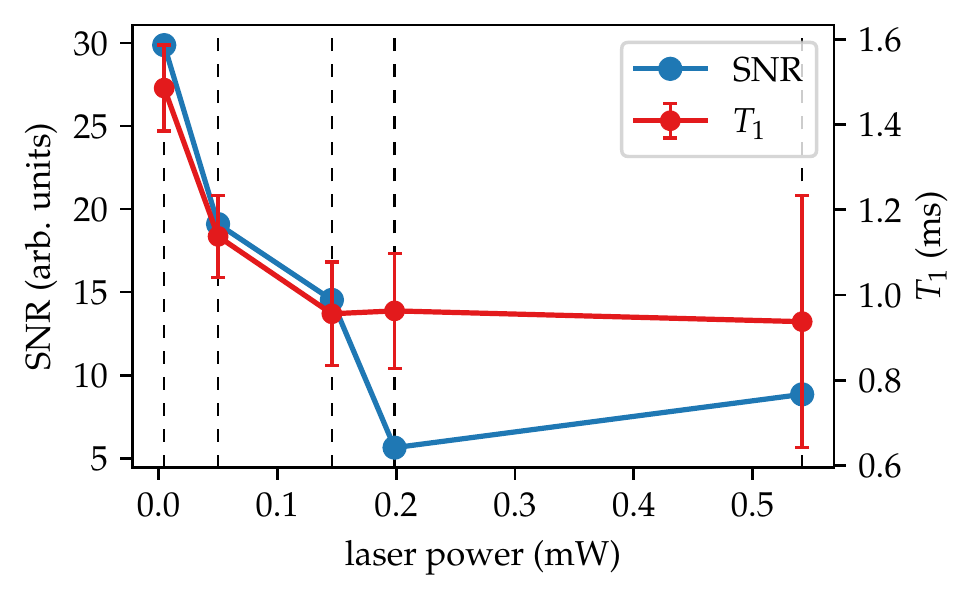}
    \caption{\label{fig:contrast} 
    SNR and $T_1$ time as a function of the laser power, obtained from measurements performed in sequence $P_1$. 
    With higher laser power, the SNR decreases, and so does the $T_1$ time. The standard error of $T_1$ that we obtain from monoexponential fits increases with higher laser power.
    The data is extracted from equal numbers of repetitions of relaxometry measurements for each laser power.
    }
    \end{figure}
   
    To better understand the $\mathrm{NV}^-$ power-dependent behavior, we use the results from the spectral analysis to map the ratios of $[\mathrm{NV}^-]/[\mathrm{NV}^0]$ to our relaxometry measurement data of sequence $P_1$. 
    Thus, we trace $[\mathrm{NV}^-]/[\mathrm{NV}^0]$ as a function of $\tau$ for all laser powers. 
    Fig.~\ref{fig:relaxometry}~(b) shows the ratio $[\mathrm{NV}^-]/[\mathrm{NV}^0]$ at the final $\tau$ divided by the ratio at the initial $\tau$.
    The result for $[\mathrm{NV}^-]/[\mathrm{NV}^0]$ as a function of $\tau$ can be found in Fig.~\ref{fig:Relaxometry_ratios} in the Appendix.
    For all laser powers, even for the lowest, which lies well below saturation intensity, we observe an increase of $[\mathrm{NV}^-]/[\mathrm{NV}^0]$ from shortest to longest $\tau$ in the readout pulse IV by a factor of $\sim 2$, see Fig.~\ref{fig:relaxometry}~(b). 
    We conclude that during $\tau$ a re-conversion from $\mathrm{NV}^0$ to $\mathrm{NV}^-$ takes place in the dark, after ionization of $\mathrm{NV}^-$ had occurred in the initialization pulse. 
    The ratios $[\mathrm{NV}^-]/[\mathrm{NV}^0]$ we find in control pulse III as a function of $\tau$ also show a power-dependent behavior. 
    While the ratio increases from shortest to longest $\tau$ at the lowest laser power, it is constant in the control pulse for the highest power.
    These power-dependent recharge processes in the control pulse we observe appear most likely due to build-up effects during the measurement cycle, as we explain in the following. 
    At low powers, the initializing laser pulse spin polarizes the $\mathrm{NV}^-$ centers but does not ionize to a steady state of $\mathrm{NV}^-$ and $\mathrm{NV}^0$. 
    For short $\tau$, the re-conversion in the dark of $\mathrm{NV}^0$ to $\mathrm{NV}^-$ has not completed, and the following laser pulse continues to ionize the $\mathrm{NV}^-$ centers. 
    However, at the highest power, each initialization pulse efficiently ionizes to a steady state of the two NV charge states, reaching a constant ratio $[\mathrm{NV}^-]/[\mathrm{NV}^0]$.
    These results clearly show that the normalization in the sequence we perform is mandatory to only detect the change in the relative fluorescence during the relaxation time $\tau$ and minimize influences due to charges passed through cycles. 
    
    At the lowest laser power of \SI{5}{\micro\watt}, we observe the highest ratio of $[\mathrm{NV}^-]/[\mathrm{NV}^0]$, see Fig.~\ref{fig:Relaxometry_ratios}, and therefore expect the most negligible influences of charge conversion on the $\mathrm{NV}^-$s' spin relaxation. Thus, we fit a monoexponential function to the relative fluorescence as a function of $\tau$ and obtain $T_1 = \SI{1.42\pm 0.06}{\milli\second}$ for the $\mathrm{NV}^-$ ensemble in the nanodiamond. 
    To further underline the necessity of a normalization of the fluorescence intensity, we fit a monoexponential function to the non-normalized bare $\mathrm{NV}^-$ fluorescence detected in IV at \SI{5}{\micro\watt} laser power.
    We obtain a $T_1$ time of \SI{0.94 \pm 0.05}{\milli\second}, see Fig.~\ref{fig:Relaxometry_ohne_norm}, which is drastically lower than the $T_1$ time retrieved with normalization by the fluorescence counts in III.
    
    For higher laser powers, we fit the normalized data with a function of type
    \begin{equation}
        f^- (\tau) = - A\,e^{-\tau/T_{R,1}} - B\,e^{-\tau/T_{R,2}} + C\,e^{-\tau/T_1} + d
    \label{eq:triexp}
    \end{equation}
    and restrict the time constants to $T_{R,1} = \SI{100}{\micro\second}$, $T_{R,2} = \SI{2.0}{\milli\second}$ and $T_{1} = \SI{1.4}{\milli\second}$. With this, we assume that the decay of $[\mathrm{NV}^0]$ causes an increase of $[\mathrm{NV}^-]$ and, therefore, their fluorescence. Thus, the $\mathrm{NV}^-$ fluorescence is best described by a sum of an exponential decay due to the loss of spin polarization and a biexponential inverted component due to the recharging process of $\mathrm{NV}^0$ to $\mathrm{NV}^-$ in the dark.
    As shown in Fig.~\ref{fig:relaxometry}~(a), our fit function Eq.~(\ref{eq:triexp}) describes the measurement data from \SI{0.05}{\milli\watt} to \SI{0.54}{\milli\watt}  laser power very well.
    We emphasize that the measurement data for \SI{0.05}{\milli\watt} laser power does not visibly appear to show this triexponential behavior. 
    Fitting a monoexponential function to the $\mathrm{NV}^-$ fluorescence at \SI{0.05}{\milli\watt} laser power, however, results in $T_1 = \SI{1.28 \pm 0.06}{\milli\second}$, see Fig.~\ref{fig:Relaxometry_64_mV_monoexp}, which deviates significantly from the value obtained at lower laser power. 
    
    Measurement sequence $P_1$ is a well-established method to accurately measure the $T_1$ time of the $\mathrm{NV}^-$ centers excited by a resonant $\pi$ pulse \cite{Jarmola.2012}. Since the $\pi$ pulse only acts on the negatively-charged NV centers, it is said to be independent of charge conversion processes alongside the spin polarization \cite{Giri.2018}. We compare the results we obtain in the complete measurement sequence $P_1$, subtracting fluorescence intensities in II from the counts in IV, to the result we gave for the $T_1$ time above without the $\pi$ pulse taken into account.
    Remarkably, although in Fig.~\ref{fig:relaxometry}~(a) we observe vivid dynamics ranging from exponential decay to an inverted exponential profile in the $\mathrm{NV}^-$ fluorescence as a function of $\tau$, the complete sequence $P_1$ yields a monoexponential decrease for all laser powers, see Fig.~\ref{fig:Relaxometry_MW} in the Appendix.
    For the lowest laser power, we obtain $T_1 = \SI{1.5 \pm 0.1}{\milli\second}$ for sequence $P_1$ comparing the fluorescence intensity with and without the resonant $\pi$ pulse.
    This value matches the previously determined $T_1$ time when only considering the normalized signal without the $\pi$ pulse for the lowest laser power.
    It does not match the $T_1$ time obtained from the monoexponential fit we performed on the measurement data for \SI{0.05}{\milli\watt} laser power, stressing the effects of NV charge conversion within this measurement and the necessity for consideration of the two components $T_{R,1}$ and $T_{R,2}$ in a triexponential fit function.
    
    However, the measurement sequence $P_1$ is not entirely unaffected by the charge conversion process. 
    Although the resonant $\pi$ pulse does not directly act on the $\mathrm{NV}^0$ center (we observe no difference in the signals with and without the $\pi$ pulse), the fluorescence contrast in the measurement decreases because of $\mathrm{NV}^0$ to $\mathrm{NV}^-$ conversion.
    This lower contrast becomes noticeable in Fig.~\ref{fig:Relaxometry_MW} due to the decaying amplitude of the monoexponential function with increasing laser power. 
    The effect of $\mathrm{NV}^-$ spin depolarization due to charge conversion has been previously investigated in \cite{Chen.2015, Choi.2017}.
    As a measure for the reliability of our measurement result, we use the area under the curves showing spin polarization as a function of $\tau$ for each laser power as a fluorescence contrast in the respective measurement. 
    We divide this value by the Root mean squared error (RMSE) value we obtain from the fit result to account for fluctuations in our measurement data and define this value contrast/RMSE as the SNR.
    In Fig.~\ref{fig:contrast}, we show the SNR as a function of the laser power. 
    In addition, we display the value for $T_1$ we obtain in the same graph. With the SNR decreasing, we observe a decrease in $T_1$, accompanied by a larger standard deviation with higher laser power. 
    We conclude that the $T_1$ time we measured at the lowest laser power is the most reliable one due to the highest SNR.
    In addition, we note that $T_1$ seems to decay as a function of the laser power, although $T_1$ should be independent of the excitation power.
    We attribute this decay of $T_1$ to the lower SNR in the measurements at higher laser power due to increased charge conversion. 
    Additionally, effects of spin polarization may play a role next to charge conversion during illumination.
    
    From the results of sequence $P_1$, we conclude that the normalization in the $T_1$ measurement is essential to reflect the charge-state processes alongside the $\mathrm{NV}^-$ ensemble's spin relaxation.
    
    \subsubsection{Sequence $P_2$}
     \begin{figure}
        \begin{overpic}[width=86mm]{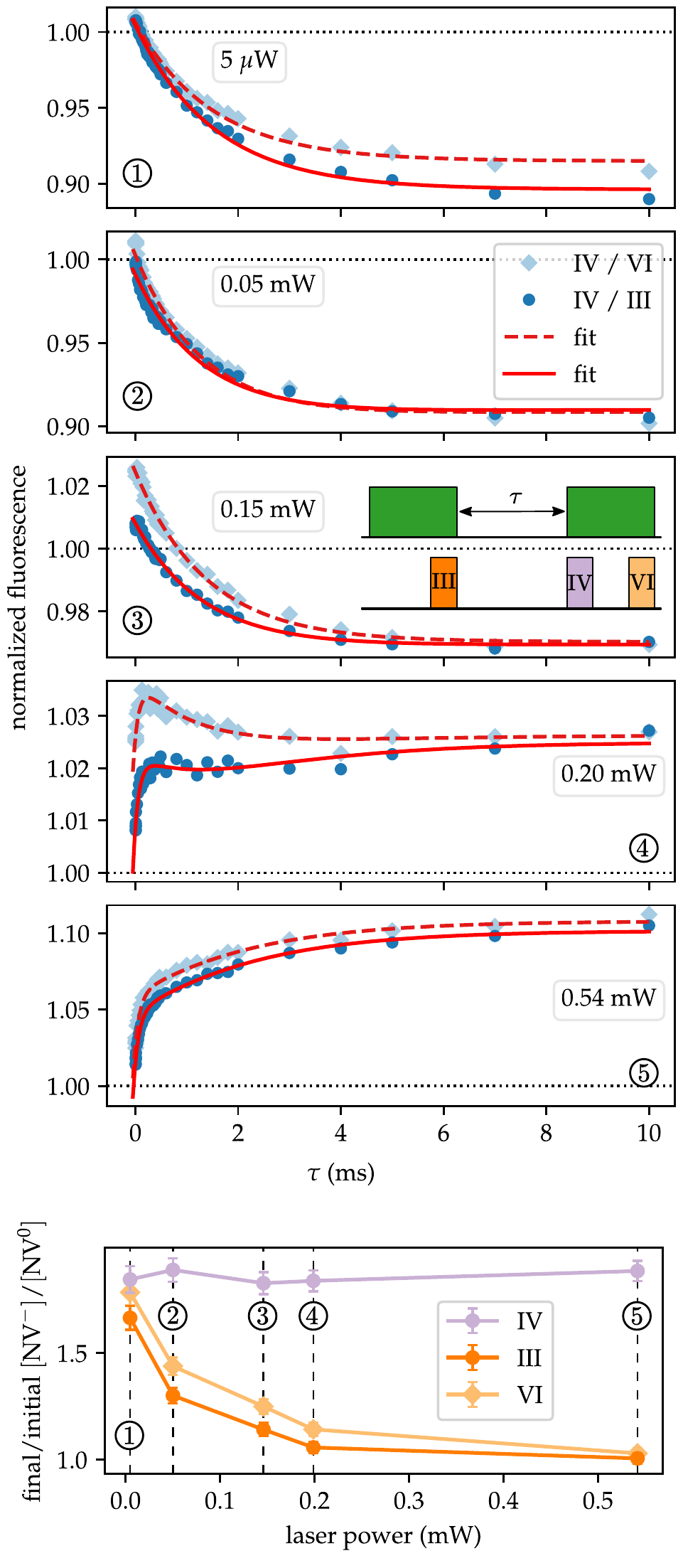}
         \put(0,98.5){(a)}
         \put(0,23.4){(b)}
        \end{overpic}
   \caption{\label{fig:relaxometry_GWR}
    (a) $\mathrm{NV}^-$ fluorescence as obtained from the second half of $P_2$ for both normalization detection windows.
    Fitting monoexponential functions to the fluorescence normalized by the counts in III or VI yields $T_1 = \SI{1.54 \pm 0.06}{\milli\second}$ or $T_1 = \SI{1.50 \pm 0.07}{\milli\second}$ for the lowest laser power, respectively.
    The $\mathrm{NV}^-$ fluorescence at higher laser powers is fit with a sum of three exponential functions.
    (b) Changes of the NV-charge-state ratio during the relaxometry measurement from lowest to highest $\tau$ in sequence $P_2$.} 
    \end{figure}
    Besides $P_1$, sequence $P_2$ is used in literature to determine a single NV center's \cite{Guillebon.2020} or an ensemble's \cite{Pelliccione.2014} $T_1$ time. 
    While the $\pi$ pulse is often omitted in these sequences, we chose to implement it for low laser powers for better comparison to the results obtained in $P_1$. 
    For laser powers starting from \SI{0.15}{\milli\watt}, we repeated the sequence without the $\pi$ pulse and calculated the mean values of the control and readout data taken.
    The results for sequence $P_2$ with a $\pi$ pulse included for \SI{5}{\micro\watt} laser power are shown in Fig.~\ref{fig:Relaxometry_GWR_OD_1}. 
    Using the data for \SI{5}{\micro\watt} laser power and subtracting II from IV, we obtain $T_1 = \SI{1.45 \pm 0.09}{\milli\second}$, which is the same result as in sequence $P_1$. 
    Since both sequences are used in the literature to measure an $\mathrm{NV}^-$ ensemble's $T_1$ time, we expect them to produce the same result for our NV ensemble when neglecting additional effects due to charge conversion.
    At this low laser power, charge conversion is inferior to spin relaxation. 
    Therefore, the $T_1$ times we obtain from both sequences do not differ. 
    However, with higher laser power, charge conversion prevails, and both NV charge states' fluorescence signals are greatly affected by recharge in the dark.
    
    In order to evaluate the result of sequence $P_2$ without the $\pi$ pulse applied, we normalize the fluorescence intensities. To this end, we divide the counts in IV obtained by the counts measured during the two control collection windows III or VI, yielding two normalized fluorescence signals for each NV charge state.
    This way, we obtain two normalized fluorescence signals as a function of $\tau$. If no charge conversion effects were present in this measurement, both signals for the normalized fluorescence should be equal. 
    However, as pointed out, charge conversion is prominent in our sample, not only for high laser powers. 
    We show the $\mathrm{NV}^-$ fluorescence as a function of $\tau$ we obtain from sequence $P_2$ in Fig.~\ref{fig:relaxometry_GWR}~(a). 
    Qualitatively similar to sequence $P_1$, we see a smooth transition from an exponential decay at low laser powers to an inverted exponential profile at high laser powers. 
    Similarly as in $P_1$, we derive $T_1 = \SI{1.54 \pm 0.06}{\milli\second}$ for normalization with III and $T_1 = \SI{1.50 \pm 0.07}{\milli\second}$ for normalization with VI for the lowest laser power. 
    We emphasize that all $T_1$ times we derive from the normalized $\mathrm{NV}^-$ fluorescence in both sequences are equal within their standard errors.
    In addition, the values for $T_{R,1}$ and $T_{R,2}$ we obtain from the $\mathrm{NV}^0$ fluorescence with sequence $P_2$ are the same as in sequence $P_1$. 
    We fit the $\mathrm{NV}^-$ fluorescence for laser powers from \SI{0.05}{\milli\watt} to \SI{0.54}{\milli\watt} in the same manner as for $P_1$ using Eq.~(\ref{eq:triexp}) and restrict $T_1$, $T_{R,1}$ and $T_{R,2}$ to the aforementioned values. 
    This triexponential fit function models our data well, regardless of the normalization we use. 
    
    However, the amplitudes of the respective exponential functions differ depending on the normalization, III or VI, employed. 
    Thus, the shapes of the fluorescence as a function of $\tau$ differ with the collection windows used for normalization, which is especially visible at \SI{0.20}{\milli\watt} laser power. 
    To understand the difference in the measurement results that the positions of the normalization collection window cause, we take the ratios $[\mathrm{NV}^-]/[\mathrm{NV}^0]$ into account. 

    We trace the ratio $[\mathrm{NV}^-]/[\mathrm{NV}^0]$ as a function of $\tau$ for sequence $P_2$ and summarize this data as the ratio at longest $\tau$ divided by the ratio at shortest $\tau$ in Fig.~\ref{fig:relaxometry_GWR}~(b) for each laser power.
    Additionally, in Fig.~\ref{fig:Relaxometry_ratios_GWR} and Fig.~\ref{fig:Relaxometry_ratios_GWR_2}, $[\mathrm{NV}^-]/[\mathrm{NV}^0]$ as a function of $\tau$ for sequence $P_2$ is displayed.
    The ratios as a function of $\tau$ behave similarly to as observed with sequence $P_1$ discussed above.
    
    However, we note that the ratios and their change from shortest to longest $\tau$ we obtain in our measurement for the two control collection windows III and VI are different.
    We find that in Fig.~\ref{fig:relaxometry_GWR}~(b) the changes of $[\mathrm{NV}^-]/[\mathrm{NV}^0]$ as a function of the laser power are higher for VI than for III for low powers and converge to the same value for higher laser powers. 
    We therefore attribute the difference in the normalized fluorescences in Fig.~\ref{fig:relaxometry_GWR}~(a) when normalizing to III or VI to the differences in $[\mathrm{NV}^-]/[\mathrm{NV}^0]$ for III and VI, respectively. 
    
    To explain the behavior described above in more detail, we analyze the ratio $[\mathrm{NV}^-]/[\mathrm{NV}^0]$ as a function of $\tau$ in Fig.~\ref{fig:Relaxometry_ratios_GWR_2}.
    For $\tau \lesssim \SI{1}{\milli\second}$ the ratio $[\mathrm{NV}^-]/[\mathrm{NV}^0]$ is smaller for VI than for III, while for values $\tau \gtrsim \SI{1}{\milli\second}$ the opposite is the case.
    For the same reasons discussed in sequence $P_1$, this effect is prominent in laser powers up to \SI{0.20}{\milli\watt}. In contrast, for the highest laser power, the ratios in the control collection windows are approximately constant with $\tau$ and do not differ significantly. 
    As pointed out in the discussion of $P_1$, the results indicate that the first laser pulse does not ionize into a steady state of $[\mathrm{NV}^-]/[\mathrm{NV}^0]$, and the second laser pulse continues to ionize $\mathrm{NV}^-$ into $\mathrm{NV}^0$.
    Therefore, especially for small values of $\tau$, the ratio $[\mathrm{NV}^-]/[\mathrm{NV}^0]$ is smaller in VI than in III.
    For larger values of $\tau$, recharge dynamics of $\mathrm{NV}^0$ to $\mathrm{NV}^-$ in the dark add to the different ratios of $[\mathrm{NV}^-]/[\mathrm{NV}^0]$ for both control collection windows. 
    We do not exclude additional effects due to continued spin polarization of $\mathrm{NV}^-$ in the second laser pulse, especially for low laser powers.

    Both the results from measurement sequences $P_1$ and $P_2$ and the simultaneous mapping of $[\mathrm{NV}^-]/[\mathrm{NV}^0]$ indicate that a charge conversion from $\mathrm{NV}^-$ to $\mathrm{NV}^0$ during the spin-polarization pulse of a spin-relaxation measurement is inevitable. 
    We emphasize that a normalization collection window is mandatory to correctly display the fluorescence dynamics of $\mathrm{NV}^0$ and $\mathrm{NV}^-$ as a function of $\tau$.
    Comparison of the two control collection windows III and VI shows that the normalized fluorescence signal depends on the positions of the collection window used for normalization because of charge conversion processes that take place alongside the $\mathrm{NV}^-$ ensemble's spin relaxation.
   
	\section{\label{sec:Concl}CONCLUSIONS}
	
	This work examines laser-power-dependent dynamics of NV charge conversion within spin-relaxation measurements of the negatively-charged NV centers in a single nanodiamond.
	We present a new method of tracing the ratio of $[\mathrm{NV}^-]$ to $[\mathrm{NV}^0]$ during our sequence, in which we extract the relative concentrations of $\mathrm{NV}^-$ to $\mathrm{NV}^0$ from their fluorescence spectra and perform a mapping to fluorescence count ratios in two separate detectors.
	From the analysis of low-excitation intensity spectra of several nanodiamonds, we find $\kappa_{520} = \SI{2.2 \pm 0.5}{}$.
	This correction factor $\kappa_{520}$ allows us to translate the fluorescence ratio of $\mathrm{NV}^-$ to $\mathrm{NV}^0$ to a concentration ratio, taking into account different lifetimes and absorption cross sections for the two charge states.
	Combining our results, we conclude that ionization of $\mathrm{NV}^-$ to $\mathrm{NV}^0$ during the optical initialization and readout is inevitable and occurs even at low laser powers. 
	A recharge process in the dark of $\mathrm{NV}^0$ to $\mathrm{NV}^-$ significantly affects the $\mathrm{NV}^-$ ensemble's fluorescence during the spin-relaxation measurement.
	We find the recharging in the dark to be biexponential with two components $T_{R,1}$ and $T_{R,2}$ in all examined nanodiamonds.
	At high laser powers, the effect of charge conversion outweighs spin relaxation, making it impossible to accurately measure a $T_1$ time, even with a scheme involving a $\pi$ pulse for two reasons.
	Firstly, recharging effects of $\mathrm{NV}^0$ to $\mathrm{NV}^-$ in the dark dominate the $\mathrm{NV}^-$ fluorescence signal.
	Secondly, the measurement of $T_1$ is crucially impeded by a diminished fluorescence contrast due to charge conversion.
	To determine the $\mathrm{NV}^-$ centers' $T_1$ time at low laser powers, we find it necessary to conduct a pulsed sequence with a normalization collection window included. 
	We prove the normalization mandatory to accurately reflect the charge-state dynamics as a function of $\tau$ and mitigate additional effects due to charge-state accumulation during the measurement cycle.
	Additionally, comparing two pulsed sequences often used in the literature, we find that the position of the normalization collection windows plays an essential role due to charge conversion during the measurement. 
	We emphasize that including a normalization collection window directly after the spin polarization before the relaxation time $\tau$ is a simple method to accurately display the fluorescence dynamics during the relaxation time.
	This way, comparing the fluorescence counts in the readout collection window to the counts in the control collection window reliably reflects the spin relaxation and the charge dynamics in the relaxometry measurement.
	
	Overall, we emphasize that the results presented in this work impact relaxometry schemes widely used in biology, chemistry, and physics.
	To further extend this work, the effects of different duration of the spin-polarization pulse and the readout pulse can be examined and give insight into the steady-state dynamics of the NV centers.
	Further, the excitation of $\mathrm{NV}^-$ can be conducted at longer wavelengths, changing the charge-state dynamics \cite{Dhomkar.2018} and impacting the spin relaxation results.
	The influence of different NV and nitrogen concentrations in diamonds of different sizes on the charge dynamics can be considered to unravel the mechanisms of charge conversion in the dark.
	In addition, the SNR reduction observed in our measurements at high laser powers deserves systematic studies on several nanodiamonds.
	
	\section{Data availability}
    The data plotted in the figures is available on Zenodo \cite{Zenodo}.
    	
	\begin{acknowledgments}
    We acknowledge support by the nano-structuring center NSC.
    This project was funded by the Deutsche Forschungsgemeinschaft
    (DFG, German Research Foundation)—Project-ID No. 454931666.
    Further, I.~C.~B. thanks the Studienstiftung des deutschen Volkes for financial support.
    We thank Oliver Opaluch and Elke Neu-Ruffing for providing the microwave antenna in our experimental setup.
    Furthermore, we thank Sian Barbosa, Stefan Dix, and Dennis Lönard for fruitful discussions and experimental support.
    \end{acknowledgments}
	
    \newpage
    \appendix
    \renewcommand\thefigure{\thesection.\arabic{figure}}  
    \renewcommand{\thefigure}{A\arabic{figure}}
    \renewcommand{\theHfigure}{A\arabic{figure}}
    \setcounter{figure}{0}    
    \section{\label{sec:ODMR}ODMR and Rabi oscillations}
    \begin{figure}[ht]
        \begin{overpic}[width=86mm]{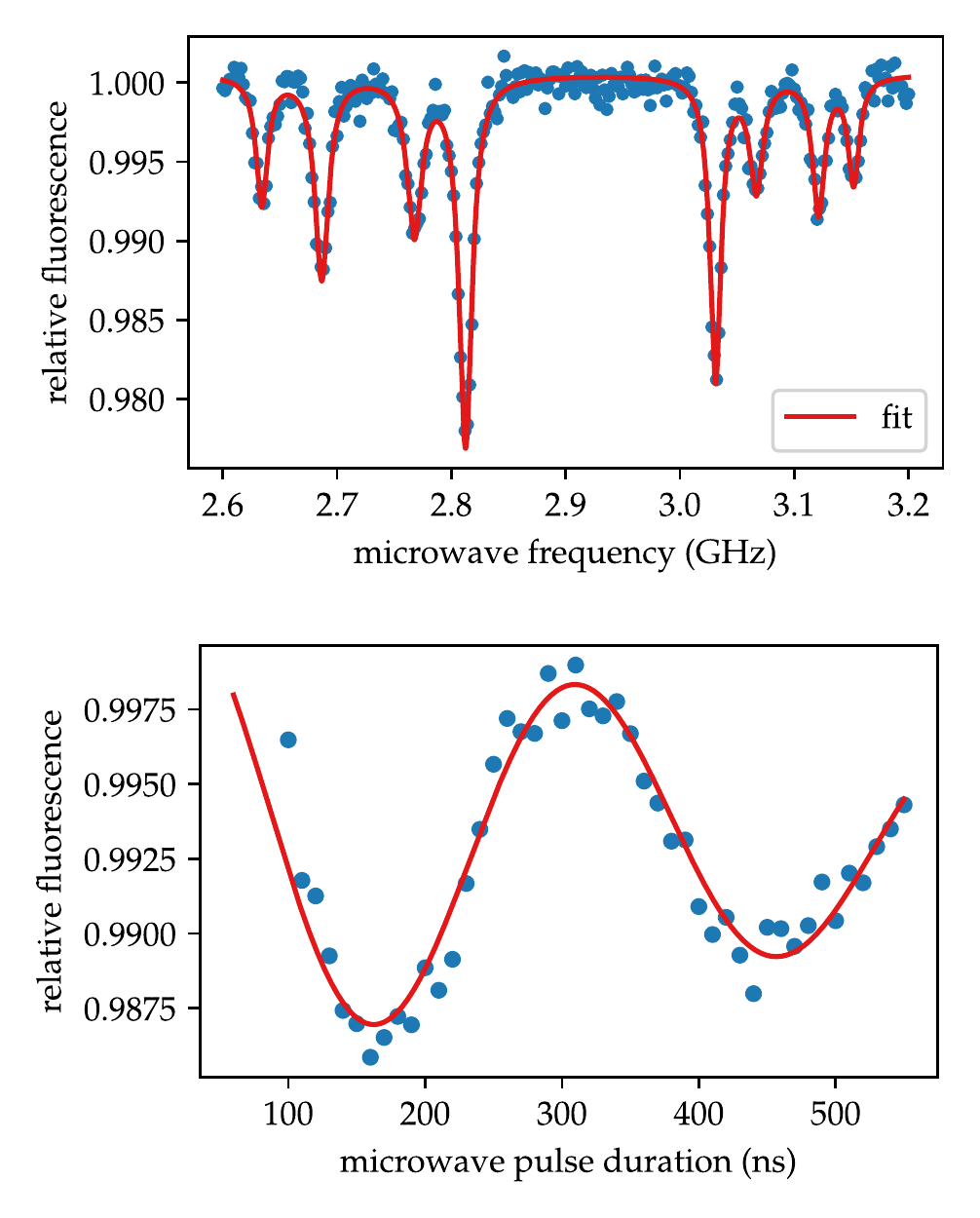}
         \put(2,96){(a)}
         \put(2,47){(b)}
        \end{overpic}
    \caption{\label{fig:ODMR}
    Typical ODMR spectrum and Rabi oscillations for the single nanodiamond investigated in this study in an external magnetic field, recorded with \SI{0.05}{\milli\watt} laser power.
    (a) ODMR spectrum recorded with \SI{6.3}{\watt} microwave power at the output of the microwave amplifier. Using a sum of eight Lorentzians, we determine the magnetic-field amplitude to \SI{11.89 \pm 0.01}{\milli\tesla}. The resonance at \SI{2.811}{\giga\hertz} was chosen for Rabi oscillations and relaxometry.
    (b) Rabi oscillations at \SI{2.811}{\giga\hertz} and a microwave power of \SI{16}{\watt}. Using a fit function of form $A e^{-t/T_2} \cos(\Omega t + \Phi) + c$, we determine the $\pi$-pulse duration to \SI{170}{\nano\second}. The motivation for this short $\pi$ pulse at high microwave power is to achieve a fast transition from the system's spin ground state to its spin excited state in a short duration compared to the system's coherence time. This way, we can achieve high contrast in our relaxometry measurements.
    }
    \end{figure}

	\section{Methods}
	To understand the NV centers' fluorescence evolution as a function of $\tau$ in terms of charge conversion, we map the fluorescence count ratio detected in both SPCMs to a ratio of $\mathrm{NV}^-$ and $\mathrm{NV}^0$ throughout the spin-relaxation measurement.
	For this, we combine the results of recorded NV spectra and spin-relaxation measurements.
	We choose a single nanodiamond and record fluorescence spectra at different laser powers using the setup in the configuration shown in Fig.~\ref{fig:exp_setup}~(a). 
	Both charge states, $\mathrm{NV}^-$ and $\mathrm{NV}^0$, contribute to the recorded spectra between \SI{500}{\nano\meter} and \SI{750}{\nano\meter} because of the charge states' overlapping phononic sidebands. 
	For further analysis, we decompose the obtained spectra into $\mathrm{NV}^-$ and $\mathrm{NV}^0$ basis functions as described by \cite{Alsid.2019} using the spectra we recorded at the highest and lowest laser power.
	Employing our extracted basis functions, we obtain the fluorescence ratio of both NV charge states for all other laser powers with the help of MATLAB's function nlinfit. 
	We access the NV-charge-state ratio from the fluorescence ratio after determining the necessary correction factor $\kappa_{520}$ \cite{Alsid.2019}. 
	A detailed description of $\kappa_{520}$'s derivation is given in Appendix~\ref{sec:520}.

	Next, we assign the concentration ratio to a count ratio in our SPCM detectors. 
	We alter the setup according to Fig.~\ref{fig:exp_setup}~(b).
	We illuminate the nanodiamond for \SI{1}{\second} with a given laser power and record the fluorescence counts in both SPCMs.
	Using the data for each laser power, we map the NV concentration ratio to a count ratio in both SPCMs. 
	At this point, we stress that we do not obtain the NV concentration ratio through fluorescence count ratios in SPCMs, but by analysis of the NV centers' fluorescence spectra. 
	This method provides the advantage that any influence of $\mathrm{NV}^0$ fluorescence in $\mathrm{SPCM}_2$ ($> \SI{665}{\nano\meter}$) can be neglected because only a count ratio is considered in our analysis and a mapping to previously-assigned concentration ratios performed. 
	
	\section{\label{sec:fluorescence_spectra} NV fluorescence spectra}
	
	\subsection{\label{sec:spectrometer} Setup}
    The incoming fluorescence light is dispersed at a grating (\SI{600}{grooves \per \milli\meter}), and an achromatic tube lens translates the angle dispersion into a spatial dispersion.
	Thus, the detection of light of different wavelengths at different positions of a camera's chip is facilitated, and spectra are obtained from \SI{500}{\nano\meter} to \SI{760}{\nano\meter}. 
	With this setup, we achieve a resolution of $\Delta \lambda \approx \SI{0.19}{\nano \meter \per pixel}$.
    Each spectrum consists of a mean of at least 20 spectra recorded at each laser power.
	We correct the spectra for the wavelength-dependent properties of optical elements in the beam path and subtract a background.

	\subsection{\label{sec:520}Determination of $\kappa_{520}$}
	\renewcommand{\thefigure}{C\arabic{figure}}
    \renewcommand{\theHfigure}{C\arabic{figure}}
    \setcounter{figure}{0}    
    This section describes how we retrieve the correction factor $\kappa_{520}$ from our measurement data. We derive $\kappa_{520}$ similarly to as described in \cite{Alsid.2019}.
	
\begin{figure}[b]
            \includegraphics[width=86mm]{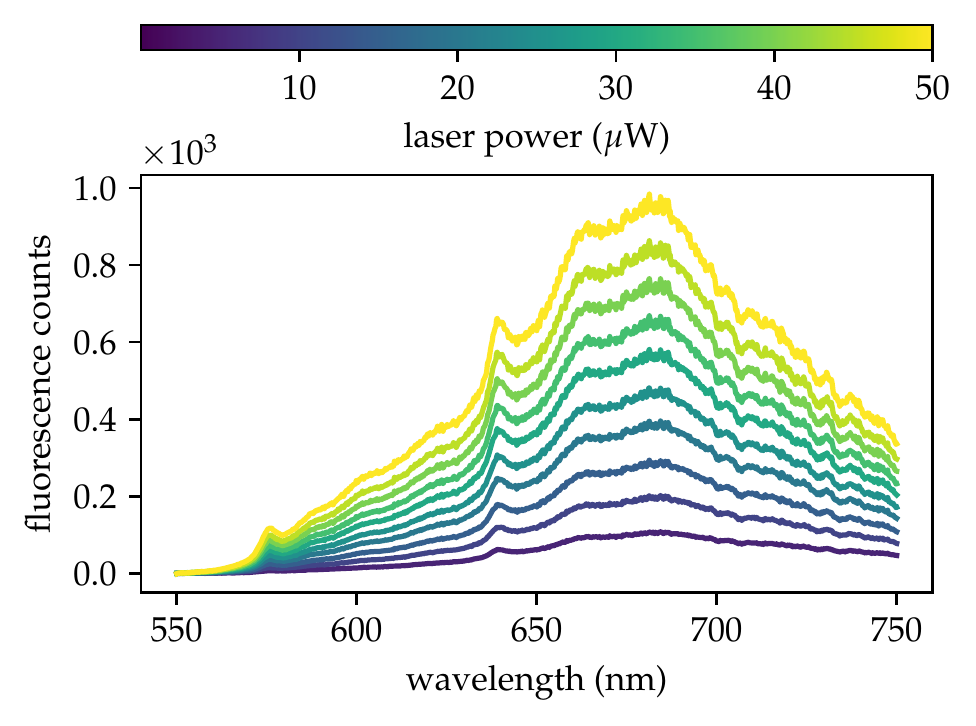}
            \caption{\label{fig:spectra_OD_1} 
            NV fluorescence spectra recorded at laser powers from \SI{5}{\micro\watt} to \SI{50}{\micro\watt}. The laser power is kept well below saturation intensity with a maximum laser power of $\sim$ \SI{50}{\micro\watt} ($\sim$~\SI{25}{\kilo\watt\per\centi\meter\squared}). The spectra were corrected for different camera exposure times used. An overall increase in the fluorescence counts is observed with increasing laser power.}
    \end{figure}

	We recorded fluorescence spectra of the single diamond crystal with laser powers well below saturation intensity with our setup shown in Fig.~\ref{fig:exp_setup} (a).
	To achieve these laser powers, an additional ND filter was used in our laser-beam path.
	We correct the spectra for different exposure times we set in our camera due to the different NV luminescence intensities at different laser powers.
	We show the spectra we obtain for different laser powers in Fig.~\ref{fig:spectra_OD_1}. 
	As can be seen, the overall fluorescence counts increase with increasing laser power. 
	We perform the spectra analysis as described in the main text to derive the coefficients $c_-$ and $c_0$.
	
	Below saturation intensity, the luminescence of $\mathrm{NV}^-$ and $\mathrm{NV}^0$ should scale linearly with the laser power \cite{Alsid.2019}. 
	However, due to charge conversion, we observe deviations from this linearity. 
	The coefficients $c_-$ and $c_0$ we obtain directly represent the amount of $\mathrm{NV}^-$ and $\mathrm{NV}^0$ fluorescence in the given spectra. 
	We scale these factors with the total integration value of the spectra in Fig.~\ref{fig:spectra_OD_1} for each laser power and obtain measured fluorescence counts for both NV charge states at each laser power.
	Further, we take the fluorescence counts for $\mathrm{NV}^-$ and $\mathrm{NV}^0$ of the lowest-intensity spectrum recorded and scale it with the laser power. 
	This way, we obtain calculated fluorescence counts for each NV charge state that strictly increase linearly with the laser power.
	
\begin{figure}[ht]
	    \includegraphics[width=86mm]{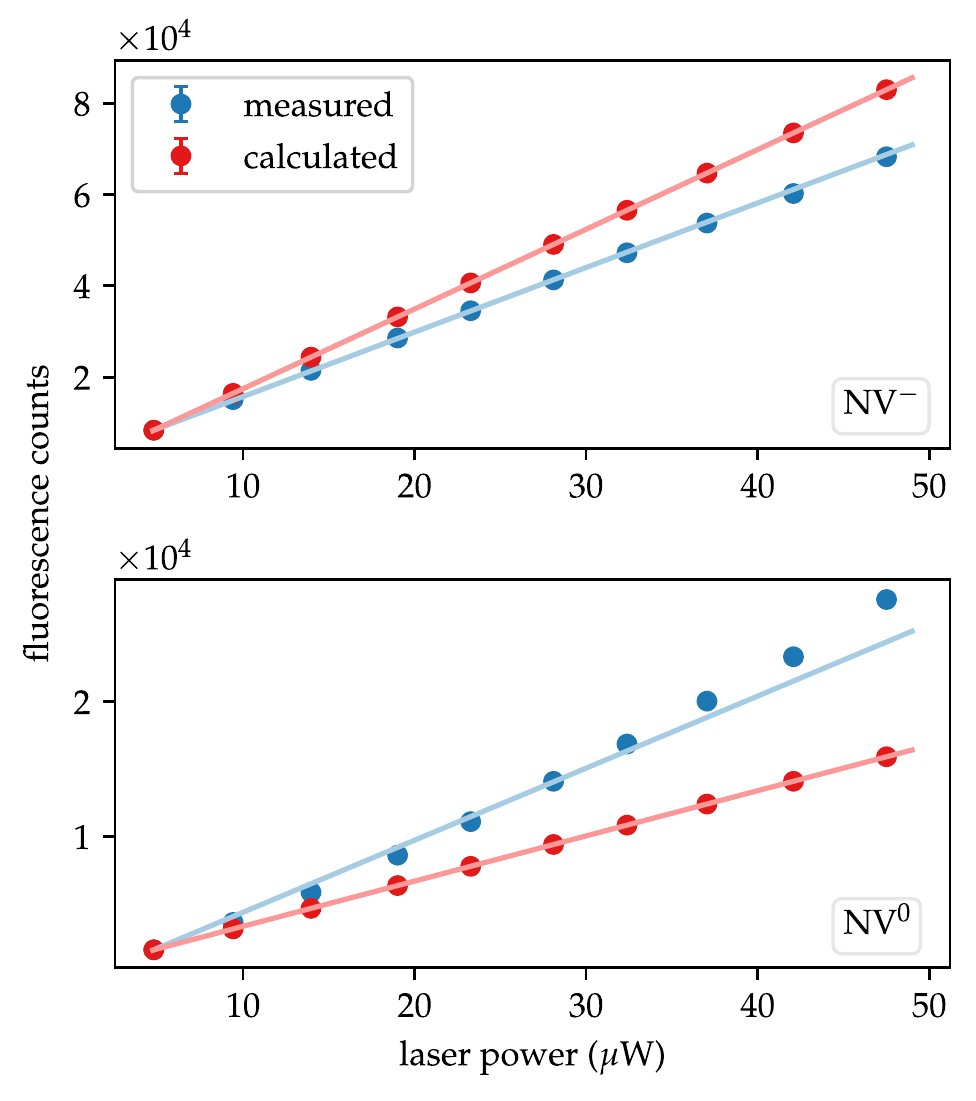}
        \caption{\label{fig:k_520} 
            Determination of $\kappa_{520}$. 
            Fluorescence counts as a function of the laser power for $\mathrm{NV}^-$ and $\mathrm{NV}^0$. The red curve displays the fluorescence counts obtained from scaling the counts at the lowest laser power with the laser power. 
            The blue curve depicts the fluorescence counts for $\mathrm{NV}^-$ and $\mathrm{NV}^0$ as a function of the laser power as we obtain it from the spectra.
            Error bars are derived from the statistical errors for $c_-$ and $c_0$ and are smaller than the data points shown in this graph.
            }
    \end{figure}

	These fluorescence counts for $\mathrm{NV}^-$ and $\mathrm{NV}^0$, measured and calculated, are shown in Fig.~\ref{fig:k_520} as a function of the laser power. 
	We note that the measured $\mathrm{NV}^-$ fluorescence is lower than the calculated linear integration value, while the $\mathrm{NV}^0$ fluorescence is higher. 
	We perform a weighted linear fit (inverse-variance weighting) for each data set and compare the slopes to one another for each NV charge state.
	We divide the two slope ratios by each other and obtain $\kappa_{520} = 1.97 \pm 0.07$, while we derive the error from the statistical error of the fits we performed.

    \subsection{\label{sec:ten_spectra}Charge conversion: Statistics}
\begin{figure}[ht]
	    \includegraphics[width=86mm]{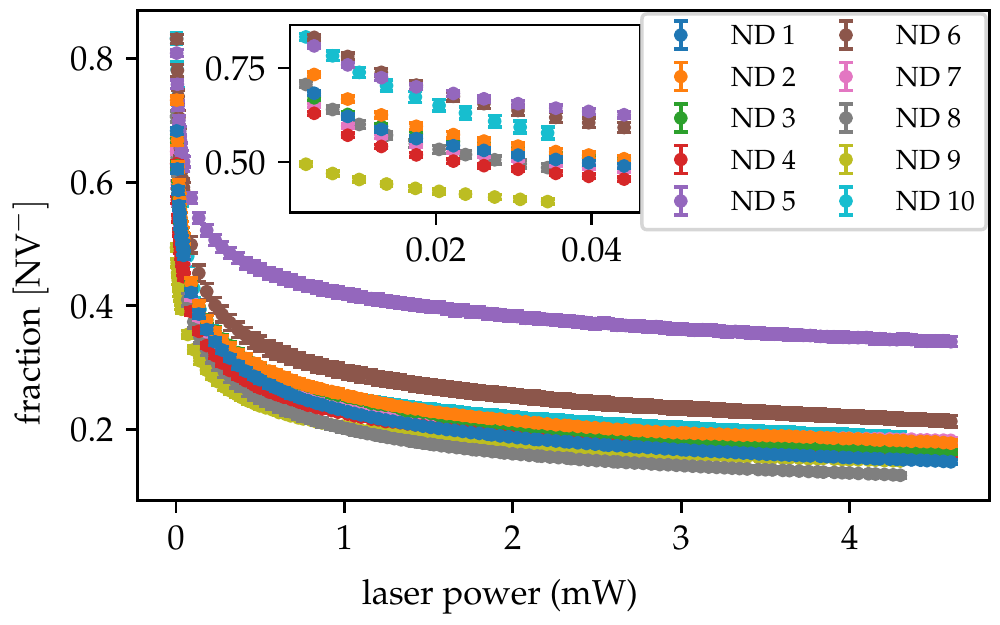}
        \caption{\label{fig:10_spectra} 
           Fraction of $[\mathrm{NV}^-]$ as a function of the laser power for ten additional nanodiamonds derived from the spectral analysis described in the main text. In all examined nanodiamonds, we see a decrease in the $[\mathrm{NV}^-]$ fraction as a function of the laser power.
            }
    \end{figure}

    To demonstrate statistical consistency in our measurement results, we performed the spectral analysis described in Section~\ref{sec:fl_spec} for ten other nanodiamonds to verify the observed laser-power-dependent charge conversion. We present the result in Fig.~\ref{fig:10_spectra}. The fraction $\mathrm{NV}^-$ decreases for increasing laser power in all examined nanodiamonds. From the analysis, as described above, we derive $\kappa_{520} = \SI{2.2 \pm 0.5}{}$ for all nanodiamonds, including the one discussed in the main text. The error denotes the standard deviation. We performed this experiment in the absence of a magnetic bias field.
    	
    \FloatBarrier
    
    \section{\label{sec:App_Relaxometry}Supporting relaxometry data} 
    \setcounter{figure}{0} 
    \renewcommand{\thefigure}{D\arabic{figure}}
    \renewcommand{\theHfigure}{D\arabic{figure}}
    
\begin{figure}[ht]
     \begin{overpic}[width=86mm]{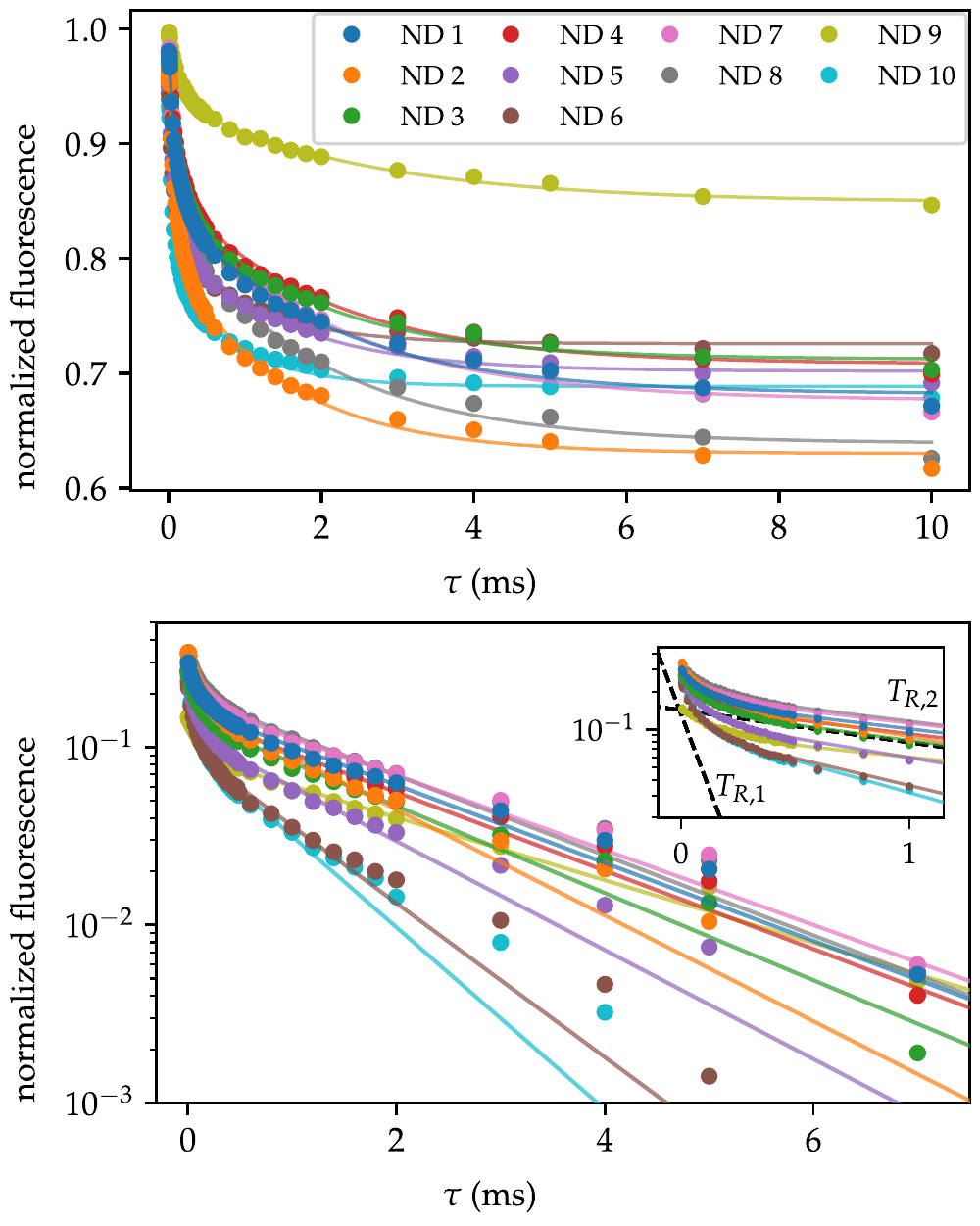}
         \put(0,98){(a)}
         \put(0,48){(b)}
        \end{overpic}
        \caption{\label{fig:Relaxometry_NV0_10} 
            Sequence $P_1$. 
            (a) $\mathrm{NV}^0$ fluorescence as a function of $\tau$, measured with 10 different nanodiamonds with a laser power of $\sim \SI{0.50}{\milli\watt}$. Solid lines show biexponential fits to the experimental data. 
            (b) Semi-logarithmic presentation of the data in (a). We subtracted the offset from the measurement data and the fit function. The inset shows the same data for low values of $\tau$. Dashed lines represent the mean values for all ten nanodiamonds we obtain for $T_{R,1}$ and $T_{R,2}$ and their amplitudes.
            }
    \end{figure}

    Fig.~\ref{fig:Relaxometry_NV0_10} provides data for the $\mathrm{NV}^0$ fluorescence as a function of $\tau$ as recorded with sequence $P_1$ (second half) at similar conditions as mentioned in the main text (no magnetic bias field was applied) for ten additional nanodiamonds. For the ten nanodiamonds, we find $T_{R,1} = \SI{91 \pm 26}{\micro \second}$ and $T_{R,2} = \SI{1.7 \pm 0.5}{\milli \second}$. The error denotes the standard deviation.
    
\begin{figure}[ht]
	    \includegraphics[width=86mm]{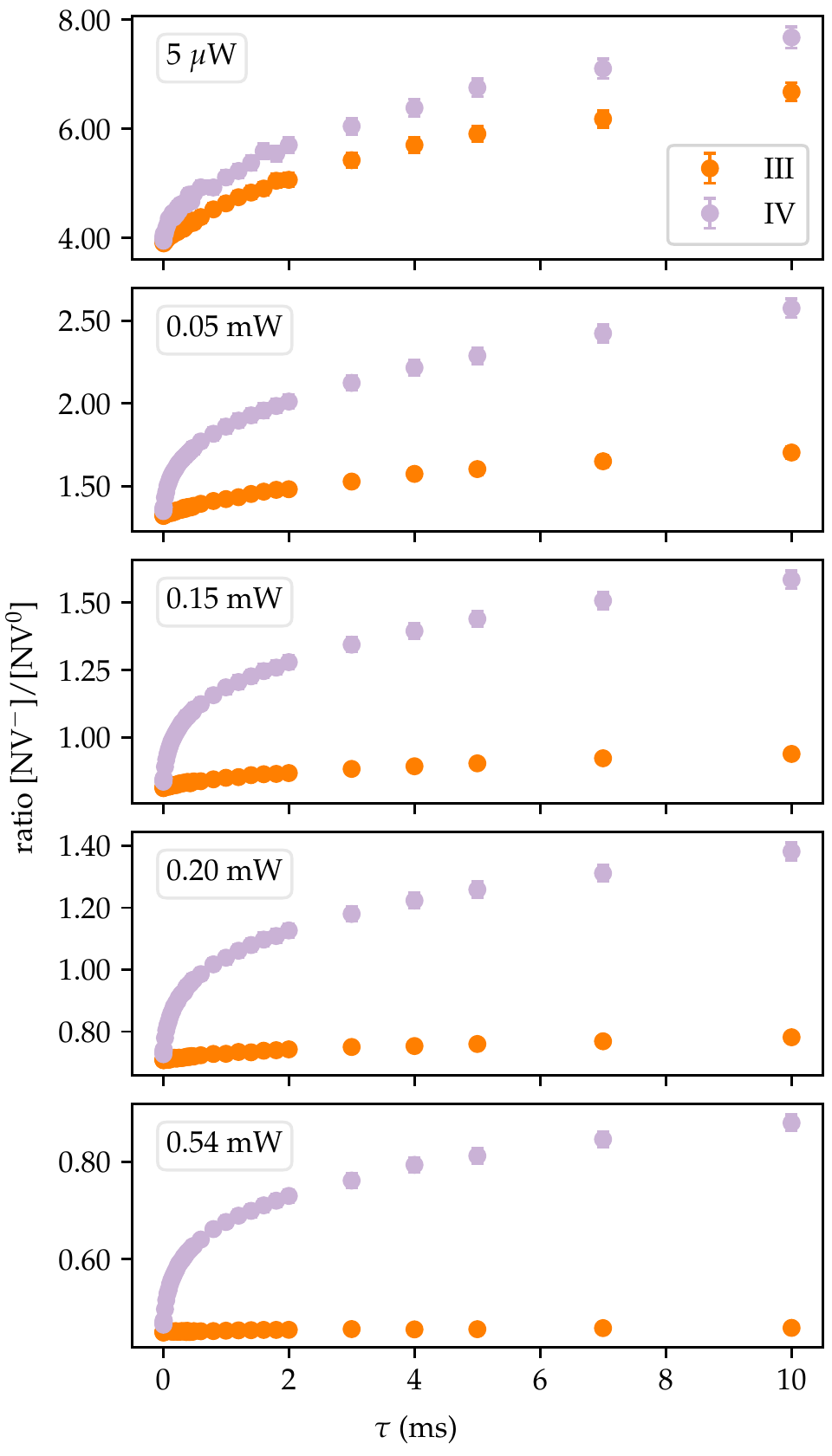}
        \caption{\label{fig:Relaxometry_ratios} 
            NV-charge-state ratios as a function of $\tau$, obtained from relaxometry measurements with sequence $P_1$ and spectra analysis. The ratios in IV and III are derived from the count-rate ratios of both SPCMs in the respective collection windows.
            The ratio $[\mathrm{NV}^-]/[\mathrm{NV}^0]$ increases in the readout collection window IV for all laser powers as a function of $\tau$. For lower laser powers, the ratio $[\mathrm{NV}^-]/[\mathrm{NV}^0]$ increases in the control collection window III, while for the highest laser power, it is constant. 
            While for the lowest power, the ratio $[\mathrm{NV}^-]/[\mathrm{NV}^0]$ is always larger than 1 over the variation of $\tau$, $[\mathrm{NV}^0]$ outweighs $[\mathrm{NV}^-]$ at \SI{0.54}{\milli\watt} laser power throughout the entire relaxation measurement.
            }
    \end{figure}

\begin{figure}[ht]
	    \includegraphics[width=86mm]{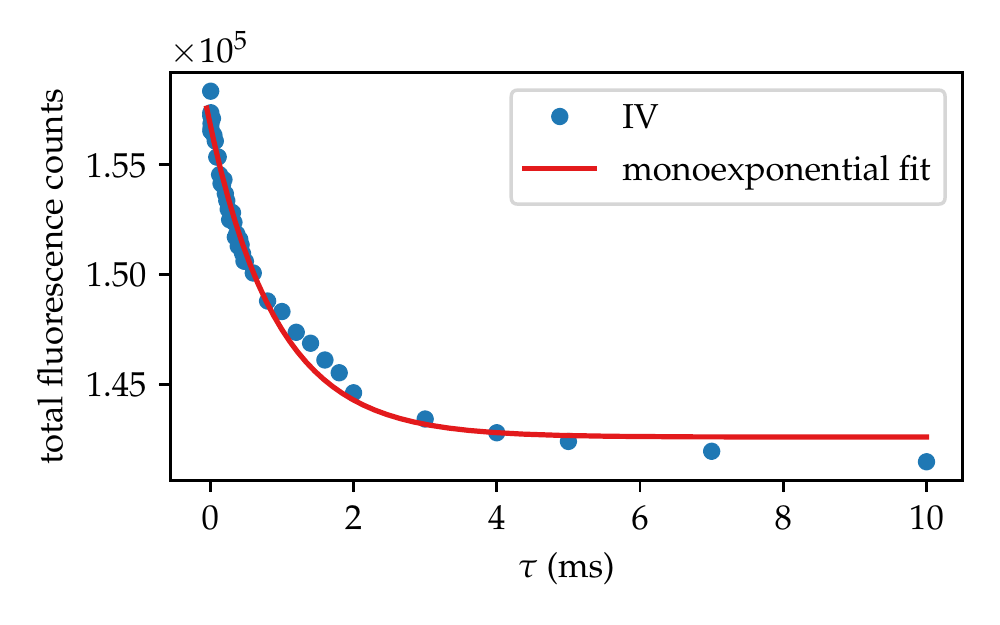}
        \caption{\label{fig:Relaxometry_ohne_norm} 
            $\mathrm{NV}^-$ fluorescence obtained from sequence $P_1$ at \SI{5}{\micro\watt} laser power in collection window IV. The data shown was not normalized by division by the fluorescence counts in detection window III. Fitting a monoexponential function to the data yields $T_1 = \SI{0.94 \pm 0.05}{\milli\second}$, which deviates drastically from the $T_1$ time obtained in the full sequence $P_1$ and in the case of normalization with III.
            }
    \end{figure}

\begin{figure}[ht]
	    \includegraphics[width=86mm]{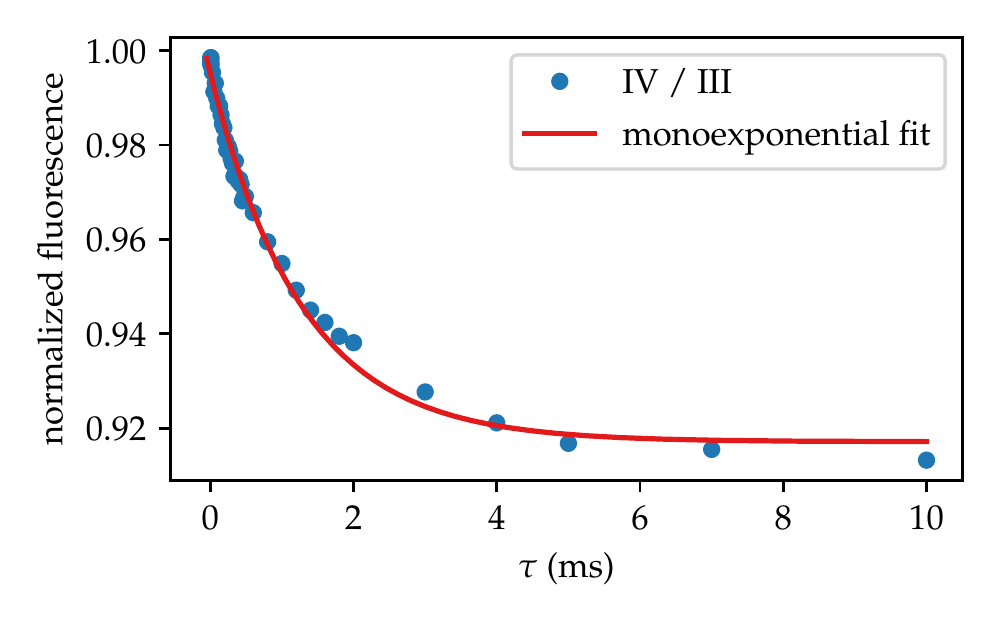}
        \caption{\label{fig:Relaxometry_64_mV_monoexp} 
            $\mathrm{NV}^-$ fluorescence obtained from sequence $P_1$ at \SI{0.05}{\milli\watt} laser power by division of the fluorescence counts in IV by the counts in III. Fitting a monoexponential function instead of the triexponential function yields $T_1 = \SI{1.28 \pm 0.06}{\milli\second}$, which does not match the value determined for $T_1$ in the full sequence $P_1$.
            }
    \end{figure}

\begin{figure}[ht]
	    \includegraphics[width=86mm]{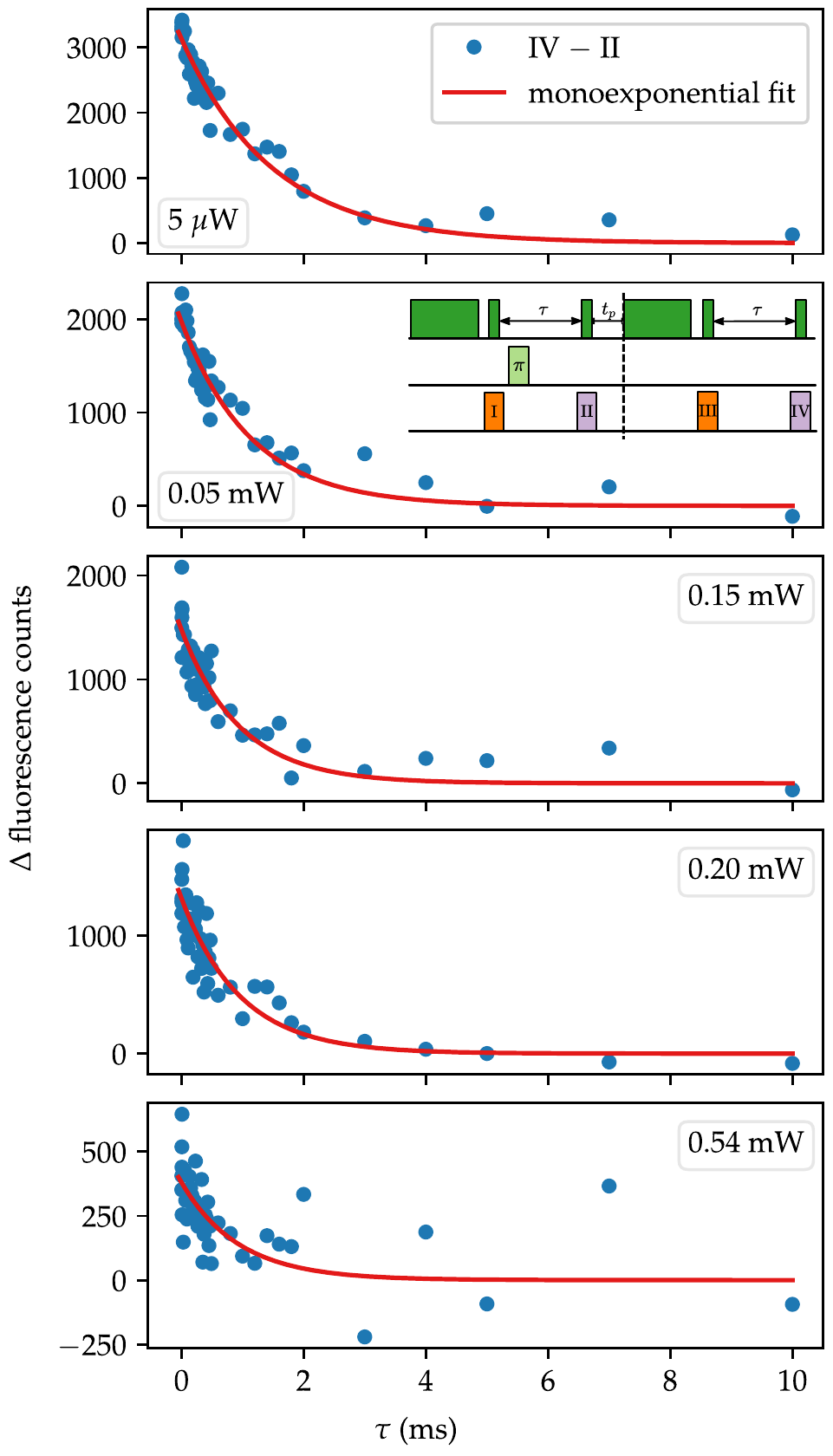}
        \caption{\label{fig:Relaxometry_MW} 
            Spin polarization of the $\mathrm{NV}^-$ ensemble as obtained from the full sequence $P_1$, subtracting the fluorescence counts in II from the counts in IV. Unlike Fig.~\ref{fig:relaxometry}~(a), we observe an exponential decay at all laser powers in this measurement data. However, with increasing laser power, we observe a decrease in the amplitude of the exponential function. Fitting a monoexponential function to the data at \SI{5}{\micro\watt} laser power, we obtain $T_1 = \SI{1.5 \pm 0.1}{\milli\second}$, consistent with the $T_1$ time we find in Fig.~\ref{fig:relaxometry}~(a) at the same laser power. 
            }
    \end{figure}

\begin{figure}[ht]
	    \includegraphics[width=86mm]{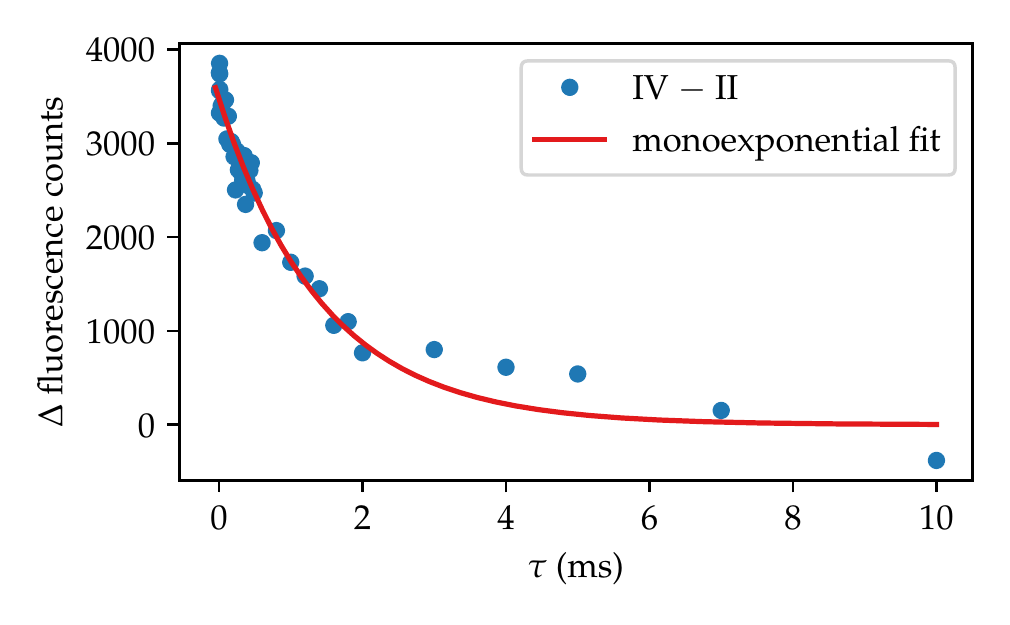}
        \caption{\label{fig:Relaxometry_GWR_OD_1} 
            $\mathrm{NV}^-$ spin polarization as obtained from the full sequence $P_2$ at \SI{5}{\micro\watt} laser power by subtracting the counts in II from the counts in IV. Fitting a monoexponential function to the data yields $T_1 = \SI{1.45 \pm 0.09}{\milli\second}$, which matches the previously determined values for $T_1$ in sequence $P_1$.
            }
    \end{figure}

\begin{figure}[ht]
	    \includegraphics[width=86mm]{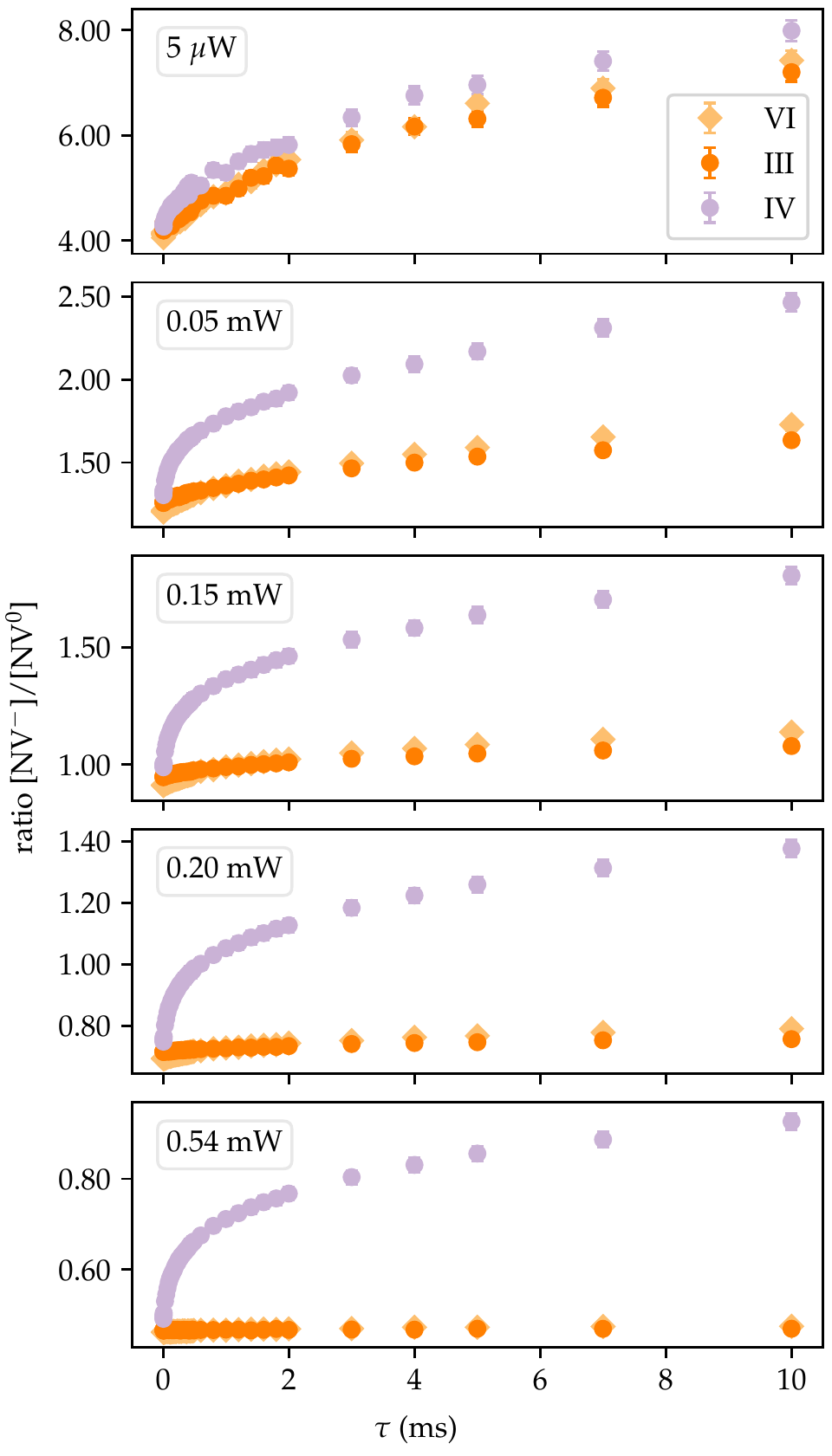}
        \caption{\label{fig:Relaxometry_ratios_GWR} 
            Sequence $P_2$. The ratio $[\mathrm{NV}^-]/[\mathrm{NV}^0]$ as a function of $\tau$ behaves similarly as in sequence $P_1$. However, the NV-charge-state ratios as a function of $\tau$ are different in III and VI, indicating charge-conversion processes during the measurement.
            }
    \end{figure}

\begin{figure}[ht]
	    \includegraphics[width=86mm]{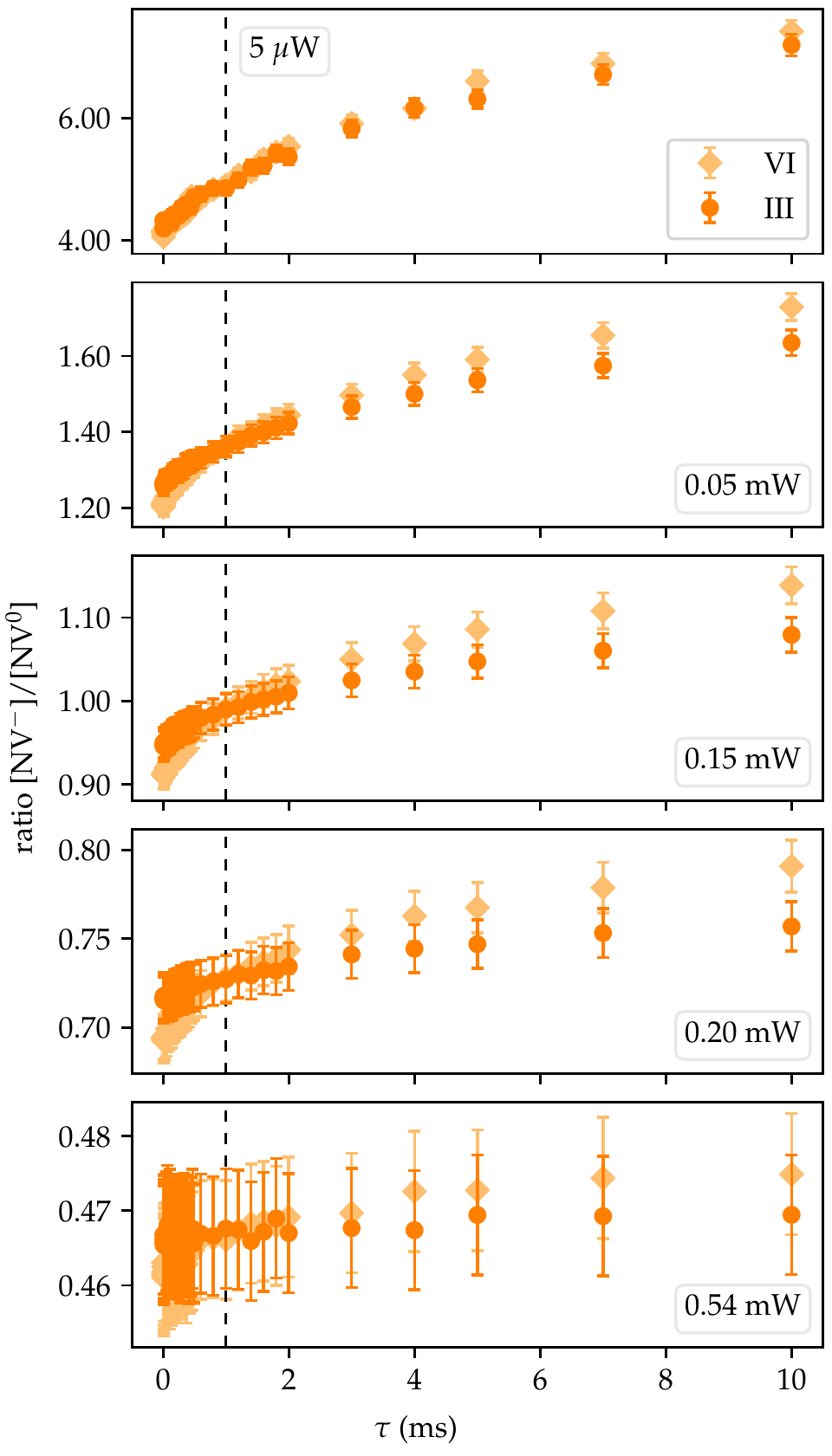}
        \caption{\label{fig:Relaxometry_ratios_GWR_2} 
            Sequence $P_2$. For better visibility, the ratios $[\mathrm{NV}^-]/[\mathrm{NV}^0]$ for III and VI as a function of $\tau$ are displayed from Fig.~\ref{fig:Relaxometry_ratios_GWR}. At laser powers up to \SI{0.20}{\milli\watt}, the ratio $[\mathrm{NV}^-]/[\mathrm{NV}^0]$ is smaller for VI than for III for $\tau \lesssim \SI{1}{\milli\second}$. For values $\tau \gtrsim \SI{1}{\milli\second}$, the opposite is the case. For visualization, $\tau = \SI{1}{\milli\second}$ is marked with a dashed line. At \SI{0.54}{\milli\watt} laser power, the ratios $[\mathrm{NV}^-]/[\mathrm{NV}^0]$ are equal in III and VI.
            }
    \end{figure}

    In Fig.~\ref{fig:Relaxometry_ratios} to Fig.~\ref{fig:Relaxometry_MW}, supporting data recorded with sequence $P_1$ is shown. 
    Additionally, we show supporting relaxometry data in Fig.~\ref{fig:Relaxometry_GWR_OD_1} to Fig.~\ref{fig:Relaxometry_ratios_GWR_2} recorded with sequence $P_2$.
    We obtained the data as described in the main text. 
    
    \FloatBarrier

\begin{thebibliography}{51}%
\makeatletter
\providecommand \@ifxundefined [1]{%
 \@ifx{#1\undefined}
}%
\providecommand \@ifnum [1]{%
 \ifnum #1\expandafter \@firstoftwo
 \else \expandafter \@secondoftwo
 \fi
}%
\providecommand \@ifx [1]{%
 \ifx #1\expandafter \@firstoftwo
 \else \expandafter \@secondoftwo
 \fi
}%
\providecommand \natexlab [1]{#1}%
\providecommand \enquote  [1]{``#1''}%
\providecommand \bibnamefont  [1]{#1}%
\providecommand \bibfnamefont [1]{#1}%
\providecommand \citenamefont [1]{#1}%
\providecommand \href@noop [0]{\@secondoftwo}%
\providecommand \href [0]{\begingroup \@sanitize@url \@href}%
\providecommand \@href[1]{\@@startlink{#1}\@@href}%
\providecommand \@@href[1]{\endgroup#1\@@endlink}%
\providecommand \@sanitize@url [0]{\catcode `\\12\catcode `\$12\catcode
  `\&12\catcode `\#12\catcode `\^12\catcode `\_12\catcode `\%12\relax}%
\providecommand \@@startlink[1]{}%
\providecommand \@@endlink[0]{}%
\providecommand \url  [0]{\begingroup\@sanitize@url \@url }%
\providecommand \@url [1]{\endgroup\@href {#1}{\urlprefix }}%
\providecommand \urlprefix  [0]{URL }%
\providecommand \Eprint [0]{\href }%
\providecommand \doibase [0]{https://doi.org/}%
\providecommand \selectlanguage [0]{\@gobble}%
\providecommand \bibinfo  [0]{\@secondoftwo}%
\providecommand \bibfield  [0]{\@secondoftwo}%
\providecommand \translation [1]{[#1]}%
\providecommand \BibitemOpen [0]{}%
\providecommand \bibitemStop [0]{}%
\providecommand \bibitemNoStop [0]{.\EOS\space}%
\providecommand \EOS [0]{\spacefactor3000\relax}%
\providecommand \BibitemShut  [1]{\csname bibitem#1\endcsname}%
\let\auto@bib@innerbib\@empty
\bibitem [{\citenamefont {Acosta}\ \emph {et~al.}(2009)\citenamefont {Acosta},
  \citenamefont {Bauch}, \citenamefont {Ledbetter}, \citenamefont {Santori},
  \citenamefont {Fu}, \citenamefont {Barclay}, \citenamefont {Beausoleil},
  \citenamefont {Linget}, \citenamefont {Roch}, \citenamefont {Treussart},
  \citenamefont {Chemerisov}, \citenamefont {Gawlik},\ and\ \citenamefont
  {Budker}}]{Acosta.2009}%
  \BibitemOpen
  \bibfield  {author} {\bibinfo {author} {\bibfnamefont {V.~M.}\ \bibnamefont
  {Acosta}}, \bibinfo {author} {\bibfnamefont {E.}~\bibnamefont {Bauch}},
  \bibinfo {author} {\bibfnamefont {M.~P.}\ \bibnamefont {Ledbetter}}, \bibinfo
  {author} {\bibfnamefont {C.}~\bibnamefont {Santori}}, \bibinfo {author}
  {\bibfnamefont {K.-M.~C.}\ \bibnamefont {Fu}}, \bibinfo {author}
  {\bibfnamefont {P.~E.}\ \bibnamefont {Barclay}}, \bibinfo {author}
  {\bibfnamefont {R.~G.}\ \bibnamefont {Beausoleil}}, \bibinfo {author}
  {\bibfnamefont {H.}~\bibnamefont {Linget}}, \bibinfo {author} {\bibfnamefont
  {J.~F.}\ \bibnamefont {Roch}}, \bibinfo {author} {\bibfnamefont
  {F.}~\bibnamefont {Treussart}}, \bibinfo {author} {\bibfnamefont
  {S.}~\bibnamefont {Chemerisov}}, \bibinfo {author} {\bibfnamefont
  {W.}~\bibnamefont {Gawlik}},\ and\ \bibinfo {author} {\bibfnamefont
  {D.}~\bibnamefont {Budker}},\ }\bibfield  {title} {\bibinfo {title}
  {{Diamonds with a high density of nitrogen-vacancy centers for magnetometry
  applications}},\ }\href {https://doi.org/10.1103/PhysRevB.80.115202}
  {\bibfield  {journal} {\bibinfo  {journal} {{Physical Review B}}\ }\textbf
  {\bibinfo {volume} {80}},\ \bibinfo {pages} {115202} (\bibinfo {year}
  {2009})}\BibitemShut {NoStop}%
\bibitem [{\citenamefont {Balasubramanian}\ \emph {et~al.}(2008)\citenamefont
  {Balasubramanian}, \citenamefont {Chan}, \citenamefont {Kolesov},
  \citenamefont {Al-Hmoud}, \citenamefont {Tisler}, \citenamefont {Shin},
  \citenamefont {Kim}, \citenamefont {Wojcik}, \citenamefont {Hemmer},
  \citenamefont {Krueger}, \citenamefont {Hanke}, \citenamefont
  {Leitenstorfer}, \citenamefont {Bratschitsch}, \citenamefont {Jelezko},\ and\
  \citenamefont {Wrachtrup}}]{Balasubramanian.2008}%
  \BibitemOpen
  \bibfield  {author} {\bibinfo {author} {\bibfnamefont {G.}~\bibnamefont
  {Balasubramanian}}, \bibinfo {author} {\bibfnamefont {I.~Y.}\ \bibnamefont
  {Chan}}, \bibinfo {author} {\bibfnamefont {R.}~\bibnamefont {Kolesov}},
  \bibinfo {author} {\bibfnamefont {M.}~\bibnamefont {Al-Hmoud}}, \bibinfo
  {author} {\bibfnamefont {J.}~\bibnamefont {Tisler}}, \bibinfo {author}
  {\bibfnamefont {C.}~\bibnamefont {Shin}}, \bibinfo {author} {\bibfnamefont
  {C.}~\bibnamefont {Kim}}, \bibinfo {author} {\bibfnamefont {A.}~\bibnamefont
  {Wojcik}}, \bibinfo {author} {\bibfnamefont {P.~R.}\ \bibnamefont {Hemmer}},
  \bibinfo {author} {\bibfnamefont {A.}~\bibnamefont {Krueger}}, \bibinfo
  {author} {\bibfnamefont {T.}~\bibnamefont {Hanke}}, \bibinfo {author}
  {\bibfnamefont {A.}~\bibnamefont {Leitenstorfer}}, \bibinfo {author}
  {\bibfnamefont {R.}~\bibnamefont {Bratschitsch}}, \bibinfo {author}
  {\bibfnamefont {F.}~\bibnamefont {Jelezko}},\ and\ \bibinfo {author}
  {\bibfnamefont {J.}~\bibnamefont {Wrachtrup}},\ }\bibfield  {title} {\bibinfo
  {title} {{Nanoscale imaging magnetometry with diamond spins under ambient
  conditions}},\ }\href {https://doi.org/10.1038/nature07278} {\bibfield
  {journal} {\bibinfo  {journal} {{Nature}}\ }\textbf {\bibinfo {volume}
  {455}},\ \bibinfo {pages} {648} (\bibinfo {year} {2008})}\BibitemShut
  {NoStop}%
\bibitem [{\citenamefont {Maze}\ \emph {et~al.}(2008)\citenamefont {Maze},
  \citenamefont {Stanwix}, \citenamefont {Hodges}, \citenamefont {Hong},
  \citenamefont {Taylor}, \citenamefont {Cappellaro}, \citenamefont {Jiang},
  \citenamefont {Dutt}, \citenamefont {Togan}, \citenamefont {Zibrov},
  \citenamefont {Yacoby}, \citenamefont {Walsworth},\ and\ \citenamefont
  {Lukin}}]{Maze.2008}%
  \BibitemOpen
  \bibfield  {author} {\bibinfo {author} {\bibfnamefont {J.~R.}\ \bibnamefont
  {Maze}}, \bibinfo {author} {\bibfnamefont {P.~L.}\ \bibnamefont {Stanwix}},
  \bibinfo {author} {\bibfnamefont {J.~S.}\ \bibnamefont {Hodges}}, \bibinfo
  {author} {\bibfnamefont {S.}~\bibnamefont {Hong}}, \bibinfo {author}
  {\bibfnamefont {J.~M.}\ \bibnamefont {Taylor}}, \bibinfo {author}
  {\bibfnamefont {P.}~\bibnamefont {Cappellaro}}, \bibinfo {author}
  {\bibfnamefont {L.}~\bibnamefont {Jiang}}, \bibinfo {author} {\bibfnamefont
  {M.~V.~G.}\ \bibnamefont {Dutt}}, \bibinfo {author} {\bibfnamefont
  {E.}~\bibnamefont {Togan}}, \bibinfo {author} {\bibfnamefont {A.~S.}\
  \bibnamefont {Zibrov}}, \bibinfo {author} {\bibfnamefont {A.}~\bibnamefont
  {Yacoby}}, \bibinfo {author} {\bibfnamefont {R.~L.}\ \bibnamefont
  {Walsworth}},\ and\ \bibinfo {author} {\bibfnamefont {M.~D.}\ \bibnamefont
  {Lukin}},\ }\bibfield  {title} {\bibinfo {title} {{Nanoscale magnetic sensing
  with an individual electronic spin in diamond}},\ }\href
  {https://doi.org/10.1038/nature07279} {\bibfield  {journal} {\bibinfo
  {journal} {{Nature}}\ }\textbf {\bibinfo {volume} {455}},\ \bibinfo {pages}
  {644} (\bibinfo {year} {2008})}\BibitemShut {NoStop}%
\bibitem [{\citenamefont {Degen}(2008)}]{Degen.2008}%
  \BibitemOpen
  \bibfield  {author} {\bibinfo {author} {\bibfnamefont {C.~L.}\ \bibnamefont
  {Degen}},\ }\bibfield  {title} {\bibinfo {title} {{Scanning magnetic field
  microscope with a diamond single-spin sensor}},\ }\href
  {https://doi.org/10.1063/1.2943282} {\bibfield  {journal} {\bibinfo
  {journal} {{Applied Physics Letters}}\ }\textbf {\bibinfo {volume} {92}},\
  \bibinfo {pages} {243111} (\bibinfo {year} {2008})}\BibitemShut {NoStop}%
\bibitem [{\citenamefont {Taylor}\ \emph {et~al.}(2008)\citenamefont {Taylor},
  \citenamefont {Cappellaro}, \citenamefont {Childress}, \citenamefont {Jiang},
  \citenamefont {Budker}, \citenamefont {Hemmer}, \citenamefont {Yacoby},
  \citenamefont {Walsworth},\ and\ \citenamefont {Lukin}}]{Taylor.2008}%
  \BibitemOpen
  \bibfield  {author} {\bibinfo {author} {\bibfnamefont {J.~M.}\ \bibnamefont
  {Taylor}}, \bibinfo {author} {\bibfnamefont {P.}~\bibnamefont {Cappellaro}},
  \bibinfo {author} {\bibfnamefont {L.}~\bibnamefont {Childress}}, \bibinfo
  {author} {\bibfnamefont {L.}~\bibnamefont {Jiang}}, \bibinfo {author}
  {\bibfnamefont {D.}~\bibnamefont {Budker}}, \bibinfo {author} {\bibfnamefont
  {P.~R.}\ \bibnamefont {Hemmer}}, \bibinfo {author} {\bibfnamefont
  {A.}~\bibnamefont {Yacoby}}, \bibinfo {author} {\bibfnamefont
  {R.}~\bibnamefont {Walsworth}},\ and\ \bibinfo {author} {\bibfnamefont
  {M.~D.}\ \bibnamefont {Lukin}},\ }\bibfield  {title} {\bibinfo {title}
  {{High-sensitivity diamond magnetometer with nanoscale resolution}},\ }\href
  {https://doi.org/10.1038/nphys1075} {\bibfield  {journal} {\bibinfo
  {journal} {{Nature Physics}}\ }\textbf {\bibinfo {volume} {4}},\ \bibinfo
  {pages} {810} (\bibinfo {year} {2008})}\BibitemShut {NoStop}%
\bibitem [{\citenamefont {Laraoui}\ \emph {et~al.}(2010)\citenamefont
  {Laraoui}, \citenamefont {Hodges},\ and\ \citenamefont
  {Meriles}}]{Laraoui.2010}%
  \BibitemOpen
  \bibfield  {author} {\bibinfo {author} {\bibfnamefont {A.}~\bibnamefont
  {Laraoui}}, \bibinfo {author} {\bibfnamefont {J.~S.}\ \bibnamefont
  {Hodges}},\ and\ \bibinfo {author} {\bibfnamefont {C.~A.}\ \bibnamefont
  {Meriles}},\ }\bibfield  {title} {\bibinfo {title} {{Magnetometry of random
  ac magnetic fields using a single nitrogen-vacancy center}},\ }\href
  {https://doi.org/10.1063/1.3497004} {\bibfield  {journal} {\bibinfo
  {journal} {{Applied Physics Letters}}\ }\textbf {\bibinfo {volume} {97}},\
  \bibinfo {pages} {143104} (\bibinfo {year} {2010})}\BibitemShut {NoStop}%
\bibitem [{\citenamefont {Schirhagl}\ \emph {et~al.}(2014)\citenamefont
  {Schirhagl}, \citenamefont {Chang}, \citenamefont {Loretz},\ and\
  \citenamefont {Degen}}]{Schirhagl.2014}%
  \BibitemOpen
  \bibfield  {author} {\bibinfo {author} {\bibfnamefont {R.}~\bibnamefont
  {Schirhagl}}, \bibinfo {author} {\bibfnamefont {K.}~\bibnamefont {Chang}},
  \bibinfo {author} {\bibfnamefont {M.}~\bibnamefont {Loretz}},\ and\ \bibinfo
  {author} {\bibfnamefont {C.~L.}\ \bibnamefont {Degen}},\ }\bibfield  {title}
  {\bibinfo {title} {{Nitrogen-Vacancy Centers in Diamond: Nanoscale Sensors
  for Physics and Biology}},\ }\href
  {https://doi.org/10.1146/annurev-physchem-040513-103659} {\bibfield
  {journal} {\bibinfo  {journal} {{Annual Review of Physical Chemistry}}\
  }\textbf {\bibinfo {volume} {65}},\ \bibinfo {pages} {83} (\bibinfo {year}
  {2014})}\BibitemShut {NoStop}%
\bibitem [{\citenamefont {Thiel}\ \emph {et~al.}(2019)\citenamefont {Thiel},
  \citenamefont {Wang}, \citenamefont {Tschudin}, \citenamefont {Rohner},
  \citenamefont {Guti{\'e}rrez-Lezama}, \citenamefont {Ubrig}, \citenamefont
  {Gibertini}, \citenamefont {Giannini}, \citenamefont {Morpurgo},\ and\
  \citenamefont {Maletinsky}}]{Thiel.2019}%
  \BibitemOpen
  \bibfield  {author} {\bibinfo {author} {\bibfnamefont {L.}~\bibnamefont
  {Thiel}}, \bibinfo {author} {\bibfnamefont {Z.}~\bibnamefont {Wang}},
  \bibinfo {author} {\bibfnamefont {M.~A.}\ \bibnamefont {Tschudin}}, \bibinfo
  {author} {\bibfnamefont {D.}~\bibnamefont {Rohner}}, \bibinfo {author}
  {\bibfnamefont {I.}~\bibnamefont {Guti{\'e}rrez-Lezama}}, \bibinfo {author}
  {\bibfnamefont {N.}~\bibnamefont {Ubrig}}, \bibinfo {author} {\bibfnamefont
  {M.}~\bibnamefont {Gibertini}}, \bibinfo {author} {\bibfnamefont
  {E.}~\bibnamefont {Giannini}}, \bibinfo {author} {\bibfnamefont {A.~F.}\
  \bibnamefont {Morpurgo}},\ and\ \bibinfo {author} {\bibfnamefont
  {P.}~\bibnamefont {Maletinsky}},\ }\bibfield  {title} {\bibinfo {title}
  {{Probing magnetism in 2D materials at the nanoscale with single-spin
  microscopy}},\ }\href {https://doi.org/10.1126/science.aav6926} {\bibfield
  {journal} {\bibinfo  {journal} {{Science (New York, N.Y.)}}\ }\textbf
  {\bibinfo {volume} {364}},\ \bibinfo {pages} {973} (\bibinfo {year}
  {2019})}\BibitemShut {NoStop}%
\bibitem [{\citenamefont {Dix}\ \emph {et~al.}(2022)\citenamefont {Dix},
  \citenamefont {Gutsche}, \citenamefont {Waller}, \citenamefont {von
  Freymann},\ and\ \citenamefont {Widera}}]{Dix.2022}%
  \BibitemOpen
  \bibfield  {author} {\bibinfo {author} {\bibfnamefont {S.}~\bibnamefont
  {Dix}}, \bibinfo {author} {\bibfnamefont {J.}~\bibnamefont {Gutsche}},
  \bibinfo {author} {\bibfnamefont {E.}~\bibnamefont {Waller}}, \bibinfo
  {author} {\bibfnamefont {G.}~\bibnamefont {von Freymann}},\ and\ \bibinfo
  {author} {\bibfnamefont {A.}~\bibnamefont {Widera}},\ }\bibfield  {title}
  {\bibinfo {title} {{Fiber-tip endoscope for optical and microwave control}},\
  }\href {https://doi.org/10.1063/5.0100330} {\bibfield  {journal} {\bibinfo
  {journal} {{The Review of Scientific Instruments}}\ }\textbf {\bibinfo
  {volume} {93}},\ \bibinfo {pages} {095104} (\bibinfo {year}
  {2022})}\BibitemShut {NoStop}%
\bibitem [{\citenamefont {Dolde}\ \emph {et~al.}(2011)\citenamefont {Dolde},
  \citenamefont {Fedder}, \citenamefont {Doherty}, \citenamefont {N{\"o}bauer},
  \citenamefont {Rempp}, \citenamefont {Balasubramanian}, \citenamefont {Wolf},
  \citenamefont {Reinhard}, \citenamefont {Hollenberg}, \citenamefont
  {Jelezko},\ and\ \citenamefont {Wrachtrup}}]{Dolde.2011}%
  \BibitemOpen
  \bibfield  {author} {\bibinfo {author} {\bibfnamefont {F.}~\bibnamefont
  {Dolde}}, \bibinfo {author} {\bibfnamefont {H.}~\bibnamefont {Fedder}},
  \bibinfo {author} {\bibfnamefont {M.~W.}\ \bibnamefont {Doherty}}, \bibinfo
  {author} {\bibfnamefont {T.}~\bibnamefont {N{\"o}bauer}}, \bibinfo {author}
  {\bibfnamefont {F.}~\bibnamefont {Rempp}}, \bibinfo {author} {\bibfnamefont
  {G.}~\bibnamefont {Balasubramanian}}, \bibinfo {author} {\bibfnamefont
  {T.}~\bibnamefont {Wolf}}, \bibinfo {author} {\bibfnamefont {F.}~\bibnamefont
  {Reinhard}}, \bibinfo {author} {\bibfnamefont {L.~C.~L.}\ \bibnamefont
  {Hollenberg}}, \bibinfo {author} {\bibfnamefont {F.}~\bibnamefont
  {Jelezko}},\ and\ \bibinfo {author} {\bibfnamefont {J.}~\bibnamefont
  {Wrachtrup}},\ }\bibfield  {title} {\bibinfo {title} {{Electric-field sensing
  using single diamond spins}},\ }\href {https://doi.org/10.1038/nphys1969}
  {\bibfield  {journal} {\bibinfo  {journal} {{Nature Physics}}\ }\textbf
  {\bibinfo {volume} {7}},\ \bibinfo {pages} {459} (\bibinfo {year}
  {2011})}\BibitemShut {NoStop}%
\bibitem [{\citenamefont {Rollo}\ \emph {et~al.}(2021)\citenamefont {Rollo},
  \citenamefont {Finco}, \citenamefont {Tanos}, \citenamefont {Fabre},
  \citenamefont {Devolder}, \citenamefont {Robert-Philip},\ and\ \citenamefont
  {Jacques}}]{Rollo.2021}%
  \BibitemOpen
  \bibfield  {author} {\bibinfo {author} {\bibfnamefont {M.}~\bibnamefont
  {Rollo}}, \bibinfo {author} {\bibfnamefont {A.}~\bibnamefont {Finco}},
  \bibinfo {author} {\bibfnamefont {R.}~\bibnamefont {Tanos}}, \bibinfo
  {author} {\bibfnamefont {F.}~\bibnamefont {Fabre}}, \bibinfo {author}
  {\bibfnamefont {T.}~\bibnamefont {Devolder}}, \bibinfo {author}
  {\bibfnamefont {I.}~\bibnamefont {Robert-Philip}},\ and\ \bibinfo {author}
  {\bibfnamefont {V.}~\bibnamefont {Jacques}},\ }\bibfield  {title} {\bibinfo
  {title} {{Quantitative study of the response of a single NV defect in diamond
  to magnetic noise}},\ }\href {https://doi.org/10.1103/PhysRevB.103.235418}
  {\bibfield  {journal} {\bibinfo  {journal} {{Physical Review B}}\ }\textbf
  {\bibinfo {volume} {103}},\ \bibinfo {pages} {235418} (\bibinfo {year}
  {2021})}\BibitemShut {NoStop}%
\bibitem [{\citenamefont {Sigaeva}\ \emph
  {et~al.}(2022{\natexlab{a}})\citenamefont {Sigaeva}, \citenamefont
  {Norouzi},\ and\ \citenamefont {Schirhagl}}]{Sigaeva.2022}%
  \BibitemOpen
  \bibfield  {author} {\bibinfo {author} {\bibfnamefont {A.}~\bibnamefont
  {Sigaeva}}, \bibinfo {author} {\bibfnamefont {N.}~\bibnamefont {Norouzi}},\
  and\ \bibinfo {author} {\bibfnamefont {R.}~\bibnamefont {Schirhagl}},\
  }\bibfield  {title} {\bibinfo {title} {{Intracellular Relaxometry,
  Challenges, and Future Directions}},\ }\href
  {https://doi.org/10.1021/acscentsci.2c00976} {\bibfield  {journal} {\bibinfo
  {journal} {{ACS Central Science}}\ }\textbf {\bibinfo {volume} {8}},\
  \bibinfo {pages} {1484} (\bibinfo {year} {2022}{\natexlab{a}})}\BibitemShut
  {NoStop}%
\bibitem [{\citenamefont {Cole}\ and\ \citenamefont
  {Hollenberg}(2009)}]{Cole.2009}%
  \BibitemOpen
  \bibfield  {author} {\bibinfo {author} {\bibfnamefont {J.~H.}\ \bibnamefont
  {Cole}}\ and\ \bibinfo {author} {\bibfnamefont {L.~C.~L.}\ \bibnamefont
  {Hollenberg}},\ }\bibfield  {title} {\bibinfo {title} {{Scanning quantum
  decoherence microscopy}},\ }\href
  {https://doi.org/10.1088/0957-4484/20/49/495401} {\bibfield  {journal}
  {\bibinfo  {journal} {{Nanotechnology}}\ }\textbf {\bibinfo {volume} {20}},\
  \bibinfo {pages} {495401} (\bibinfo {year} {2009})}\BibitemShut {NoStop}%
\bibitem [{\citenamefont {Hall}\ \emph {et~al.}(2009)\citenamefont {Hall},
  \citenamefont {Cole}, \citenamefont {Hill},\ and\ \citenamefont
  {Hollenberg}}]{Hall.2009}%
  \BibitemOpen
  \bibfield  {author} {\bibinfo {author} {\bibfnamefont {L.~T.}\ \bibnamefont
  {Hall}}, \bibinfo {author} {\bibfnamefont {J.~H.}\ \bibnamefont {Cole}},
  \bibinfo {author} {\bibfnamefont {C.~D.}\ \bibnamefont {Hill}},\ and\
  \bibinfo {author} {\bibfnamefont {L.~C.~L.}\ \bibnamefont {Hollenberg}},\
  }\bibfield  {title} {\bibinfo {title} {{Sensing of Fluctuating Nanoscale
  Magnetic Fields Using Nitrogen-Vacancy Centers in Diamond}},\ }\href
  {https://doi.org/10.1103/PhysRevLett.103.220802} {\bibfield  {journal}
  {\bibinfo  {journal} {{Physical Review Letters}}\ }\textbf {\bibinfo {volume}
  {103}},\ \bibinfo {pages} {220802} (\bibinfo {year} {2009})}\BibitemShut
  {NoStop}%
\bibitem [{\citenamefont {Steinert}\ \emph {et~al.}(2013)\citenamefont
  {Steinert}, \citenamefont {Ziem}, \citenamefont {Hall}, \citenamefont
  {Zappe}, \citenamefont {Schweikert}, \citenamefont {G{\"o}tz}, \citenamefont
  {Aird}, \citenamefont {Balasubramanian}, \citenamefont {Hollenberg},\ and\
  \citenamefont {Wrachtrup}}]{Steinert.2013}%
  \BibitemOpen
  \bibfield  {author} {\bibinfo {author} {\bibfnamefont {S.}~\bibnamefont
  {Steinert}}, \bibinfo {author} {\bibfnamefont {F.}~\bibnamefont {Ziem}},
  \bibinfo {author} {\bibfnamefont {L.~T.}\ \bibnamefont {Hall}}, \bibinfo
  {author} {\bibfnamefont {A.}~\bibnamefont {Zappe}}, \bibinfo {author}
  {\bibfnamefont {M.}~\bibnamefont {Schweikert}}, \bibinfo {author}
  {\bibfnamefont {N.}~\bibnamefont {G{\"o}tz}}, \bibinfo {author}
  {\bibfnamefont {A.}~\bibnamefont {Aird}}, \bibinfo {author} {\bibfnamefont
  {G.}~\bibnamefont {Balasubramanian}}, \bibinfo {author} {\bibfnamefont
  {L.}~\bibnamefont {Hollenberg}},\ and\ \bibinfo {author} {\bibfnamefont
  {J.}~\bibnamefont {Wrachtrup}},\ }\bibfield  {title} {\bibinfo {title}
  {{Magnetic spin imaging under ambient conditions with sub-cellular
  resolution}},\ }\href {https://doi.org/10.1038/ncomms2588} {\bibfield
  {journal} {\bibinfo  {journal} {{Nature Communications}}\ }\textbf {\bibinfo
  {volume} {4}},\ \bibinfo {pages} {1607} (\bibinfo {year} {2013})}\BibitemShut
  {NoStop}%
\bibitem [{\citenamefont {Schmid-Lorch}\ \emph {et~al.}(2015)\citenamefont
  {Schmid-Lorch}, \citenamefont {H{\"a}berle}, \citenamefont {Reinhard},
  \citenamefont {Zappe}, \citenamefont {Slota}, \citenamefont {Bogani},
  \citenamefont {Finkler},\ and\ \citenamefont {Wrachtrup}}]{SchmidLorch.2015}%
  \BibitemOpen
  \bibfield  {author} {\bibinfo {author} {\bibfnamefont {D.}~\bibnamefont
  {Schmid-Lorch}}, \bibinfo {author} {\bibfnamefont {T.}~\bibnamefont
  {H{\"a}berle}}, \bibinfo {author} {\bibfnamefont {F.}~\bibnamefont
  {Reinhard}}, \bibinfo {author} {\bibfnamefont {A.}~\bibnamefont {Zappe}},
  \bibinfo {author} {\bibfnamefont {M.}~\bibnamefont {Slota}}, \bibinfo
  {author} {\bibfnamefont {L.}~\bibnamefont {Bogani}}, \bibinfo {author}
  {\bibfnamefont {A.}~\bibnamefont {Finkler}},\ and\ \bibinfo {author}
  {\bibfnamefont {J.}~\bibnamefont {Wrachtrup}},\ }\bibfield  {title} {\bibinfo
  {title} {{Relaxometry and Dephasing Imaging of Superparamagnetic Magnetite
  Nanoparticles Using a Single Qubit}},\ }\href
  {https://doi.org/10.1021/acs.nanolett.5b00679} {\bibfield  {journal}
  {\bibinfo  {journal} {{Nano Letters}}\ }\textbf {\bibinfo {volume} {15}},\
  \bibinfo {pages} {4942} (\bibinfo {year} {2015})}\BibitemShut {NoStop}%
\bibitem [{\citenamefont {Tetienne}\ \emph {et~al.}(2013)\citenamefont
  {Tetienne}, \citenamefont {Hingant}, \citenamefont {Rondin}, \citenamefont
  {Cavaill{\`e}s}, \citenamefont {Mayer}, \citenamefont {Dantelle},
  \citenamefont {Gacoin}, \citenamefont {Wrachtrup}, \citenamefont {Roch},\
  and\ \citenamefont {Jacques}}]{Tetienne.2013}%
  \BibitemOpen
  \bibfield  {author} {\bibinfo {author} {\bibfnamefont {J.-P.}\ \bibnamefont
  {Tetienne}}, \bibinfo {author} {\bibfnamefont {T.}~\bibnamefont {Hingant}},
  \bibinfo {author} {\bibfnamefont {L.}~\bibnamefont {Rondin}}, \bibinfo
  {author} {\bibfnamefont {A.}~\bibnamefont {Cavaill{\`e}s}}, \bibinfo {author}
  {\bibfnamefont {L.}~\bibnamefont {Mayer}}, \bibinfo {author} {\bibfnamefont
  {G.}~\bibnamefont {Dantelle}}, \bibinfo {author} {\bibfnamefont
  {T.}~\bibnamefont {Gacoin}}, \bibinfo {author} {\bibfnamefont
  {J.}~\bibnamefont {Wrachtrup}}, \bibinfo {author} {\bibfnamefont {J.-F.}\
  \bibnamefont {Roch}},\ and\ \bibinfo {author} {\bibfnamefont
  {V.}~\bibnamefont {Jacques}},\ }\bibfield  {title} {\bibinfo {title} {{Spin
  relaxometry of single nitrogen-vacancy defects in diamond nanocrystals for
  magnetic noise sensing}},\ }\href
  {https://doi.org/10.1103/PhysRevB.87.235436} {\bibfield  {journal} {\bibinfo
  {journal} {{Physical Review B}}\ }\textbf {\bibinfo {volume} {87}},\ \bibinfo
  {pages} {235436} (\bibinfo {year} {2013})}\BibitemShut {NoStop}%
\bibitem [{\citenamefont {Sushkov}\ \emph {et~al.}(2014)\citenamefont
  {Sushkov}, \citenamefont {Chisholm}, \citenamefont {Lovchinsky},
  \citenamefont {Kubo}, \citenamefont {Lo}, \citenamefont {Bennett},
  \citenamefont {Hunger}, \citenamefont {Akimov}, \citenamefont {Walsworth},
  \citenamefont {Park},\ and\ \citenamefont {Lukin}}]{Sushkov.2014}%
  \BibitemOpen
  \bibfield  {author} {\bibinfo {author} {\bibfnamefont {A.~O.}\ \bibnamefont
  {Sushkov}}, \bibinfo {author} {\bibfnamefont {N.}~\bibnamefont {Chisholm}},
  \bibinfo {author} {\bibfnamefont {I.}~\bibnamefont {Lovchinsky}}, \bibinfo
  {author} {\bibfnamefont {M.}~\bibnamefont {Kubo}}, \bibinfo {author}
  {\bibfnamefont {P.~K.}\ \bibnamefont {Lo}}, \bibinfo {author} {\bibfnamefont
  {S.~D.}\ \bibnamefont {Bennett}}, \bibinfo {author} {\bibfnamefont
  {D.}~\bibnamefont {Hunger}}, \bibinfo {author} {\bibfnamefont
  {A.}~\bibnamefont {Akimov}}, \bibinfo {author} {\bibfnamefont {R.~L.}\
  \bibnamefont {Walsworth}}, \bibinfo {author} {\bibfnamefont {H.}~\bibnamefont
  {Park}},\ and\ \bibinfo {author} {\bibfnamefont {M.~D.}\ \bibnamefont
  {Lukin}},\ }\bibfield  {title} {\bibinfo {title} {{All-Optical Sensing of a
  Single-Molecule Electron Spin}},\ }\href {https://doi.org/10.1021/nl502988n}
  {\bibfield  {journal} {\bibinfo  {journal} {{Nano Letters}}\ }\textbf
  {\bibinfo {volume} {14}},\ \bibinfo {pages} {6443} (\bibinfo {year}
  {2014})}\BibitemShut {NoStop}%
\bibitem [{\citenamefont {Pelliccione}\ \emph {et~al.}(2014)\citenamefont
  {Pelliccione}, \citenamefont {Myers}, \citenamefont {Pascal}, \citenamefont
  {Das},\ and\ \citenamefont {{Bleszynski Jayich}}}]{Pelliccione.2014}%
  \BibitemOpen
  \bibfield  {author} {\bibinfo {author} {\bibfnamefont {M.}~\bibnamefont
  {Pelliccione}}, \bibinfo {author} {\bibfnamefont {B.~A.}\ \bibnamefont
  {Myers}}, \bibinfo {author} {\bibfnamefont {L.~M.~A.}\ \bibnamefont
  {Pascal}}, \bibinfo {author} {\bibfnamefont {A.}~\bibnamefont {Das}},\ and\
  \bibinfo {author} {\bibfnamefont {A.~C.}\ \bibnamefont {{Bleszynski
  Jayich}}},\ }\bibfield  {title} {\bibinfo {title} {{Two-Dimensional Nanoscale
  Imaging of Gadolinium Spins via Scanning Probe Relaxometry with a Single Spin
  in Diamond}},\ }\href {https://doi.org/10.1103/PhysRevApplied.2.054014}
  {\bibfield  {journal} {\bibinfo  {journal} {{Physical Review Applied}}\
  }\textbf {\bibinfo {volume} {2}},\ \bibinfo {pages} {054014} (\bibinfo {year}
  {2014})}\BibitemShut {NoStop}%
\bibitem [{\citenamefont {Gorrini}\ \emph {et~al.}(2019)\citenamefont
  {Gorrini}, \citenamefont {Giri}, \citenamefont {Avalos}, \citenamefont
  {Tambalo}, \citenamefont {Mannucci}, \citenamefont {Basso}, \citenamefont
  {Bazzanella}, \citenamefont {Dorigoni}, \citenamefont {Cazzanelli},
  \citenamefont {Marzola}, \citenamefont {Miotello},\ and\ \citenamefont
  {Bifone}}]{Gorrini.2019}%
  \BibitemOpen
  \bibfield  {author} {\bibinfo {author} {\bibfnamefont {F.}~\bibnamefont
  {Gorrini}}, \bibinfo {author} {\bibfnamefont {R.}~\bibnamefont {Giri}},
  \bibinfo {author} {\bibfnamefont {C.~E.}\ \bibnamefont {Avalos}}, \bibinfo
  {author} {\bibfnamefont {S.}~\bibnamefont {Tambalo}}, \bibinfo {author}
  {\bibfnamefont {S.}~\bibnamefont {Mannucci}}, \bibinfo {author}
  {\bibfnamefont {L.}~\bibnamefont {Basso}}, \bibinfo {author} {\bibfnamefont
  {N.}~\bibnamefont {Bazzanella}}, \bibinfo {author} {\bibfnamefont
  {C.}~\bibnamefont {Dorigoni}}, \bibinfo {author} {\bibfnamefont
  {M.}~\bibnamefont {Cazzanelli}}, \bibinfo {author} {\bibfnamefont
  {P.}~\bibnamefont {Marzola}}, \bibinfo {author} {\bibfnamefont
  {A.}~\bibnamefont {Miotello}},\ and\ \bibinfo {author} {\bibfnamefont
  {A.}~\bibnamefont {Bifone}},\ }\bibfield  {title} {\bibinfo {title} {{Fast
  and Sensitive Detection of Paramagnetic Species Using Coupled Charge and Spin
  Dynamics in Strongly Fluorescent Nanodiamonds}},\ }\href
  {https://doi.org/10.1021/acsami.9b05779} {\bibfield  {journal} {\bibinfo
  {journal} {{ACS Applied Materials {\&} Interfaces}}\ }\textbf {\bibinfo
  {volume} {11}},\ \bibinfo {pages} {24412} (\bibinfo {year}
  {2019})}\BibitemShut {NoStop}%
\bibitem [{\citenamefont {Barton}\ \emph {et~al.}(2020)\citenamefont {Barton},
  \citenamefont {Gulka}, \citenamefont {Tarabek}, \citenamefont {Mindarava},
  \citenamefont {Wang}, \citenamefont {Schimer}, \citenamefont {Raabova},
  \citenamefont {Bednar}, \citenamefont {Plenio}, \citenamefont {Jelezko},
  \citenamefont {Nesladek},\ and\ \citenamefont {Cigler}}]{Barton.2020}%
  \BibitemOpen
  \bibfield  {author} {\bibinfo {author} {\bibfnamefont {J.}~\bibnamefont
  {Barton}}, \bibinfo {author} {\bibfnamefont {M.}~\bibnamefont {Gulka}},
  \bibinfo {author} {\bibfnamefont {J.}~\bibnamefont {Tarabek}}, \bibinfo
  {author} {\bibfnamefont {Y.}~\bibnamefont {Mindarava}}, \bibinfo {author}
  {\bibfnamefont {Z.}~\bibnamefont {Wang}}, \bibinfo {author} {\bibfnamefont
  {J.}~\bibnamefont {Schimer}}, \bibinfo {author} {\bibfnamefont
  {H.}~\bibnamefont {Raabova}}, \bibinfo {author} {\bibfnamefont
  {J.}~\bibnamefont {Bednar}}, \bibinfo {author} {\bibfnamefont {M.~B.}\
  \bibnamefont {Plenio}}, \bibinfo {author} {\bibfnamefont {F.}~\bibnamefont
  {Jelezko}}, \bibinfo {author} {\bibfnamefont {M.}~\bibnamefont {Nesladek}},\
  and\ \bibinfo {author} {\bibfnamefont {P.}~\bibnamefont {Cigler}},\
  }\bibfield  {title} {\bibinfo {title} {{Nanoscale Dynamic Readout of a
  Chemical Redox Process Using Radicals Coupled with Nitrogen-Vacancy Centers
  in Nanodiamonds}},\ }\href {https://doi.org/10.1021/acsnano.0c04010}
  {\bibfield  {journal} {\bibinfo  {journal} {{ACS Nano}}\ }\textbf {\bibinfo
  {volume} {14}},\ \bibinfo {pages} {12938} (\bibinfo {year}
  {2020})}\BibitemShut {NoStop}%
\bibitem [{\citenamefont {{Perona Mart{\'i}nez}}\ \emph
  {et~al.}(2020)\citenamefont {{Perona Mart{\'i}nez}}, \citenamefont
  {Nusantara}, \citenamefont {Chipaux}, \citenamefont {Padamati},\ and\
  \citenamefont {Schirhagl}}]{PeronaMartinez.2020}%
  \BibitemOpen
  \bibfield  {author} {\bibinfo {author} {\bibfnamefont {F.}~\bibnamefont
  {{Perona Mart{\'i}nez}}}, \bibinfo {author} {\bibfnamefont {A.~C.}\
  \bibnamefont {Nusantara}}, \bibinfo {author} {\bibfnamefont {M.}~\bibnamefont
  {Chipaux}}, \bibinfo {author} {\bibfnamefont {S.~K.}\ \bibnamefont
  {Padamati}},\ and\ \bibinfo {author} {\bibfnamefont {R.}~\bibnamefont
  {Schirhagl}},\ }\bibfield  {title} {\bibinfo {title} {{Nanodiamond
  Relaxometry-Based Detection of Free-Radical Species When Produced in Chemical
  Reactions in Biologically Relevant Conditions}},\ }\href
  {https://doi.org/10.1021/acssensors.0c01037} {\bibfield  {journal} {\bibinfo
  {journal} {{ACS Sensors}}\ }\textbf {\bibinfo {volume} {5}},\ \bibinfo
  {pages} {3862} (\bibinfo {year} {2020})}\BibitemShut {NoStop}%
\bibitem [{\citenamefont {Sch{\"a}fer-Nolte}\ \emph {et~al.}(2014)\citenamefont
  {Sch{\"a}fer-Nolte}, \citenamefont {Schlipf}, \citenamefont {Ternes},
  \citenamefont {Reinhard}, \citenamefont {Kern},\ and\ \citenamefont
  {Wrachtrup}}]{SchaferNolte.2014}%
  \BibitemOpen
  \bibfield  {author} {\bibinfo {author} {\bibfnamefont {E.}~\bibnamefont
  {Sch{\"a}fer-Nolte}}, \bibinfo {author} {\bibfnamefont {L.}~\bibnamefont
  {Schlipf}}, \bibinfo {author} {\bibfnamefont {M.}~\bibnamefont {Ternes}},
  \bibinfo {author} {\bibfnamefont {F.}~\bibnamefont {Reinhard}}, \bibinfo
  {author} {\bibfnamefont {K.}~\bibnamefont {Kern}},\ and\ \bibinfo {author}
  {\bibfnamefont {J.}~\bibnamefont {Wrachtrup}},\ }\bibfield  {title} {\bibinfo
  {title} {{Tracking Temperature-Dependent Relaxation Times of Ferritin
  Nanomagnets with a Wideband Quantum Spectrometer}},\ }\href
  {https://doi.org/10.1103/PhysRevLett.113.217204} {\bibfield  {journal}
  {\bibinfo  {journal} {{Physical Review Letters}}\ }\textbf {\bibinfo {volume}
  {113}},\ \bibinfo {pages} {217204} (\bibinfo {year} {2014})}\BibitemShut
  {NoStop}%
\bibitem [{\citenamefont {Nie}\ \emph {et~al.}(2021)\citenamefont {Nie},
  \citenamefont {Nusantara}, \citenamefont {Damle}, \citenamefont {Sharmin},
  \citenamefont {Evans}, \citenamefont {Hemelaar}, \citenamefont {{van der
  Laan}}, \citenamefont {Li}, \citenamefont {{Perona Martinez}}, \citenamefont
  {Vedelaar}, \citenamefont {Chipaux},\ and\ \citenamefont
  {Schirhagl}}]{Nie.2021}%
  \BibitemOpen
  \bibfield  {author} {\bibinfo {author} {\bibfnamefont {L.}~\bibnamefont
  {Nie}}, \bibinfo {author} {\bibfnamefont {A.~C.}\ \bibnamefont {Nusantara}},
  \bibinfo {author} {\bibfnamefont {V.~G.}\ \bibnamefont {Damle}}, \bibinfo
  {author} {\bibfnamefont {R.}~\bibnamefont {Sharmin}}, \bibinfo {author}
  {\bibfnamefont {E.~P.~P.}\ \bibnamefont {Evans}}, \bibinfo {author}
  {\bibfnamefont {S.~R.}\ \bibnamefont {Hemelaar}}, \bibinfo {author}
  {\bibfnamefont {K.~J.}\ \bibnamefont {{van der Laan}}}, \bibinfo {author}
  {\bibfnamefont {R.}~\bibnamefont {Li}}, \bibinfo {author} {\bibfnamefont
  {F.~P.}\ \bibnamefont {{Perona Martinez}}}, \bibinfo {author} {\bibfnamefont
  {T.}~\bibnamefont {Vedelaar}}, \bibinfo {author} {\bibfnamefont
  {M.}~\bibnamefont {Chipaux}},\ and\ \bibinfo {author} {\bibfnamefont
  {R.}~\bibnamefont {Schirhagl}},\ }\bibfield  {title} {\bibinfo {title}
  {{Quantum monitoring of cellular metabolic activities in single
  mitochondria}},\ }\href {https://doi.org/10.1126/sciadv.abf0573} {\bibfield
  {journal} {\bibinfo  {journal} {{Science Advances}}\ }\textbf {\bibinfo
  {volume} {7}},\ \bibinfo {pages} {eabf0573} (\bibinfo {year}
  {2021})}\BibitemShut {NoStop}%
\bibitem [{\citenamefont {Sharmin}\ \emph {et~al.}(2021)\citenamefont
  {Sharmin}, \citenamefont {Hamoh}, \citenamefont {Sigaeva}, \citenamefont
  {Mzyk}, \citenamefont {Damle}, \citenamefont {Morita}, \citenamefont
  {Vedelaar},\ and\ \citenamefont {Schirhagl}}]{Sharmin.2021}%
  \BibitemOpen
  \bibfield  {author} {\bibinfo {author} {\bibfnamefont {R.}~\bibnamefont
  {Sharmin}}, \bibinfo {author} {\bibfnamefont {T.}~\bibnamefont {Hamoh}},
  \bibinfo {author} {\bibfnamefont {A.}~\bibnamefont {Sigaeva}}, \bibinfo
  {author} {\bibfnamefont {A.}~\bibnamefont {Mzyk}}, \bibinfo {author}
  {\bibfnamefont {V.~G.}\ \bibnamefont {Damle}}, \bibinfo {author}
  {\bibfnamefont {A.}~\bibnamefont {Morita}}, \bibinfo {author} {\bibfnamefont
  {T.}~\bibnamefont {Vedelaar}},\ and\ \bibinfo {author} {\bibfnamefont
  {R.}~\bibnamefont {Schirhagl}},\ }\bibfield  {title} {\bibinfo {title}
  {{Fluorescent Nanodiamonds for Detecting Free-Radical Generation in Real Time
  during Shear Stress in Human Umbilical Vein Endothelial Cells}},\ }\href
  {https://doi.org/10.1021/acssensors.1c01582} {\bibfield  {journal} {\bibinfo
  {journal} {{ACS Sensors}}\ }\textbf {\bibinfo {volume} {6}},\ \bibinfo
  {pages} {4349} (\bibinfo {year} {2021})}\BibitemShut {NoStop}%
\bibitem [{\citenamefont {Sigaeva}\ \emph
  {et~al.}(2022{\natexlab{b}})\citenamefont {Sigaeva}, \citenamefont {Shirzad},
  \citenamefont {Martinez}, \citenamefont {Nusantara}, \citenamefont {Mougios},
  \citenamefont {Chipaux},\ and\ \citenamefont {Schirhagl}}]{Sigaeva.2022b}%
  \BibitemOpen
  \bibfield  {author} {\bibinfo {author} {\bibfnamefont {A.}~\bibnamefont
  {Sigaeva}}, \bibinfo {author} {\bibfnamefont {H.}~\bibnamefont {Shirzad}},
  \bibinfo {author} {\bibfnamefont {F.~P.}\ \bibnamefont {Martinez}}, \bibinfo
  {author} {\bibfnamefont {A.~C.}\ \bibnamefont {Nusantara}}, \bibinfo {author}
  {\bibfnamefont {N.}~\bibnamefont {Mougios}}, \bibinfo {author} {\bibfnamefont
  {M.}~\bibnamefont {Chipaux}},\ and\ \bibinfo {author} {\bibfnamefont
  {R.}~\bibnamefont {Schirhagl}},\ }\bibfield  {title} {\bibinfo {title}
  {{Diamond-Based Nanoscale Quantum Relaxometry for Sensing Free Radical
  Production in Cells}},\ }\href {https://doi.org/10.1002/smll.202105750}
  {\bibfield  {journal} {\bibinfo  {journal} {{Small (Weinheim an der
  Bergstrasse, Germany)}}\ }\textbf {\bibinfo {volume} {18}},\ \bibinfo {pages}
  {e2105750} (\bibinfo {year} {2022}{\natexlab{b}})}\BibitemShut {NoStop}%
\bibitem [{\citenamefont {Norouzi}\ \emph {et~al.}(2022)\citenamefont
  {Norouzi}, \citenamefont {Nusantara}, \citenamefont {Ong}, \citenamefont
  {Hamoh}, \citenamefont {Nie}, \citenamefont {Morita}, \citenamefont {Zhang},
  \citenamefont {Mzyk},\ and\ \citenamefont {Schirhagl}}]{Norouzi.2022}%
  \BibitemOpen
  \bibfield  {author} {\bibinfo {author} {\bibfnamefont {N.}~\bibnamefont
  {Norouzi}}, \bibinfo {author} {\bibfnamefont {A.~C.}\ \bibnamefont
  {Nusantara}}, \bibinfo {author} {\bibfnamefont {Y.}~\bibnamefont {Ong}},
  \bibinfo {author} {\bibfnamefont {T.}~\bibnamefont {Hamoh}}, \bibinfo
  {author} {\bibfnamefont {L.}~\bibnamefont {Nie}}, \bibinfo {author}
  {\bibfnamefont {A.}~\bibnamefont {Morita}}, \bibinfo {author} {\bibfnamefont
  {Y.}~\bibnamefont {Zhang}}, \bibinfo {author} {\bibfnamefont
  {A.}~\bibnamefont {Mzyk}},\ and\ \bibinfo {author} {\bibfnamefont
  {R.}~\bibnamefont {Schirhagl}},\ }\bibfield  {title} {\bibinfo {title}
  {{Relaxometry for detecting free radical generation during Bacteria's
  response to antibiotics}},\ }\href
  {https://doi.org/10.1016/j.carbon.2022.08.025} {\bibfield  {journal}
  {\bibinfo  {journal} {{Carbon}}\ }\textbf {\bibinfo {volume} {199}},\
  \bibinfo {pages} {444} (\bibinfo {year} {2022})}\BibitemShut {NoStop}%
\bibitem [{\citenamefont {Choi}\ \emph {et~al.}(2017)\citenamefont {Choi},
  \citenamefont {Choi}, \citenamefont {Kucsko}, \citenamefont {Maurer},
  \citenamefont {Shields}, \citenamefont {Sumiya}, \citenamefont {Onoda},
  \citenamefont {Isoya}, \citenamefont {Demler}, \citenamefont {Jelezko},
  \citenamefont {Yao},\ and\ \citenamefont {Lukin}}]{Choi.2017}%
  \BibitemOpen
  \bibfield  {author} {\bibinfo {author} {\bibfnamefont {J.}~\bibnamefont
  {Choi}}, \bibinfo {author} {\bibfnamefont {S.}~\bibnamefont {Choi}}, \bibinfo
  {author} {\bibfnamefont {G.}~\bibnamefont {Kucsko}}, \bibinfo {author}
  {\bibfnamefont {P.~C.}\ \bibnamefont {Maurer}}, \bibinfo {author}
  {\bibfnamefont {B.~J.}\ \bibnamefont {Shields}}, \bibinfo {author}
  {\bibfnamefont {H.}~\bibnamefont {Sumiya}}, \bibinfo {author} {\bibfnamefont
  {S.}~\bibnamefont {Onoda}}, \bibinfo {author} {\bibfnamefont
  {J.}~\bibnamefont {Isoya}}, \bibinfo {author} {\bibfnamefont
  {E.}~\bibnamefont {Demler}}, \bibinfo {author} {\bibfnamefont
  {F.}~\bibnamefont {Jelezko}}, \bibinfo {author} {\bibfnamefont {N.~Y.}\
  \bibnamefont {Yao}},\ and\ \bibinfo {author} {\bibfnamefont {M.~D.}\
  \bibnamefont {Lukin}},\ }\bibfield  {title} {\bibinfo {title}
  {{Depolarization Dynamics in a Strongly Interacting Solid-State Spin
  Ensemble}},\ }\href {https://doi.org/10.1103/PhysRevLett.118.093601}
  {\bibfield  {journal} {\bibinfo  {journal} {{Physical Review Letters}}\
  }\textbf {\bibinfo {volume} {118}},\ \bibinfo {pages} {093601} (\bibinfo
  {year} {2017})}\BibitemShut {NoStop}%
\bibitem [{\citenamefont {Giri}\ \emph {et~al.}(2018)\citenamefont {Giri},
  \citenamefont {Gorrini}, \citenamefont {Dorigoni}, \citenamefont {Avalos},
  \citenamefont {Cazzanelli}, \citenamefont {Tambalo},\ and\ \citenamefont
  {Bifone}}]{Giri.2018}%
  \BibitemOpen
  \bibfield  {author} {\bibinfo {author} {\bibfnamefont {R.}~\bibnamefont
  {Giri}}, \bibinfo {author} {\bibfnamefont {F.}~\bibnamefont {Gorrini}},
  \bibinfo {author} {\bibfnamefont {C.}~\bibnamefont {Dorigoni}}, \bibinfo
  {author} {\bibfnamefont {C.~E.}\ \bibnamefont {Avalos}}, \bibinfo {author}
  {\bibfnamefont {M.}~\bibnamefont {Cazzanelli}}, \bibinfo {author}
  {\bibfnamefont {S.}~\bibnamefont {Tambalo}},\ and\ \bibinfo {author}
  {\bibfnamefont {A.}~\bibnamefont {Bifone}},\ }\bibfield  {title} {\bibinfo
  {title} {{Coupled charge and spin dynamics in high-density ensembles of
  nitrogen-vacancy centers in diamond}},\ }\href
  {https://doi.org/10.1103/PhysRevB.98.045401} {\bibfield  {journal} {\bibinfo
  {journal} {{Physical Review B}}\ }\textbf {\bibinfo {volume} {98}},\ \bibinfo
  {pages} {045401} (\bibinfo {year} {2018})}\BibitemShut {NoStop}%
\bibitem [{\citenamefont {Giri}\ \emph {et~al.}(2019)\citenamefont {Giri},
  \citenamefont {Dorigoni}, \citenamefont {Tambalo}, \citenamefont {Gorrini},\
  and\ \citenamefont {Bifone}}]{Giri.2019}%
  \BibitemOpen
  \bibfield  {author} {\bibinfo {author} {\bibfnamefont {R.}~\bibnamefont
  {Giri}}, \bibinfo {author} {\bibfnamefont {C.}~\bibnamefont {Dorigoni}},
  \bibinfo {author} {\bibfnamefont {S.}~\bibnamefont {Tambalo}}, \bibinfo
  {author} {\bibfnamefont {F.}~\bibnamefont {Gorrini}},\ and\ \bibinfo {author}
  {\bibfnamefont {A.}~\bibnamefont {Bifone}},\ }\bibfield  {title} {\bibinfo
  {title} {{Selective measurement of charge dynamics in an ensemble of
  nitrogen-vacancy centers in nanodiamond and bulk diamond}},\ }\href
  {https://doi.org/10.1103/PhysRevB.99.155426} {\bibfield  {journal} {\bibinfo
  {journal} {{Physical Review B}}\ }\textbf {\bibinfo {volume} {99}},\ \bibinfo
  {pages} {155426} (\bibinfo {year} {2019})}\BibitemShut {NoStop}%
\bibitem [{\citenamefont {Gorrini}\ \emph {et~al.}(2021)\citenamefont
  {Gorrini}, \citenamefont {Dorigoni}, \citenamefont {Olivares-Postigo},
  \citenamefont {Giri}, \citenamefont {Apr{\`a}}, \citenamefont {Picollo},\
  and\ \citenamefont {Bifone}}]{Gorrini.2021}%
  \BibitemOpen
  \bibfield  {author} {\bibinfo {author} {\bibfnamefont {F.}~\bibnamefont
  {Gorrini}}, \bibinfo {author} {\bibfnamefont {C.}~\bibnamefont {Dorigoni}},
  \bibinfo {author} {\bibfnamefont {D.}~\bibnamefont {Olivares-Postigo}},
  \bibinfo {author} {\bibfnamefont {R.}~\bibnamefont {Giri}}, \bibinfo {author}
  {\bibfnamefont {P.}~\bibnamefont {Apr{\`a}}}, \bibinfo {author}
  {\bibfnamefont {F.}~\bibnamefont {Picollo}},\ and\ \bibinfo {author}
  {\bibfnamefont {A.}~\bibnamefont {Bifone}},\ }\bibfield  {title} {\bibinfo
  {title} {{Long-Lived Ensembles of Shallow NV- Centers in Flat and
  Nanostructured Diamonds by Photoconversion}},\ }\href
  {https://doi.org/10.1021/acsami.1c09825} {\bibfield  {journal} {\bibinfo
  {journal} {{ACS Applied Materials {\&} Interfaces}}\ }\textbf {\bibinfo
  {volume} {13}},\ \bibinfo {pages} {43221} (\bibinfo {year}
  {2021})}\BibitemShut {NoStop}%
\bibitem [{\citenamefont {{Raman Nair}}\ \emph {et~al.}(2020)\citenamefont
  {{Raman Nair}}, \citenamefont {Rogers}, \citenamefont {Vidal}, \citenamefont
  {Roberts}, \citenamefont {Abe}, \citenamefont {Ohshima}, \citenamefont
  {Yatsui}, \citenamefont {Greentree}, \citenamefont {Jeske},\ and\
  \citenamefont {Volz}}]{RamanNair.2020}%
  \BibitemOpen
  \bibfield  {author} {\bibinfo {author} {\bibfnamefont {S.}~\bibnamefont
  {{Raman Nair}}}, \bibinfo {author} {\bibfnamefont {L.~J.}\ \bibnamefont
  {Rogers}}, \bibinfo {author} {\bibfnamefont {X.}~\bibnamefont {Vidal}},
  \bibinfo {author} {\bibfnamefont {R.~P.}\ \bibnamefont {Roberts}}, \bibinfo
  {author} {\bibfnamefont {H.}~\bibnamefont {Abe}}, \bibinfo {author}
  {\bibfnamefont {T.}~\bibnamefont {Ohshima}}, \bibinfo {author} {\bibfnamefont
  {T.}~\bibnamefont {Yatsui}}, \bibinfo {author} {\bibfnamefont {A.~D.}\
  \bibnamefont {Greentree}}, \bibinfo {author} {\bibfnamefont {J.}~\bibnamefont
  {Jeske}},\ and\ \bibinfo {author} {\bibfnamefont {T.}~\bibnamefont {Volz}},\
  }\bibfield  {title} {\bibinfo {title} {{Amplification by stimulated emission
  of nitrogen-vacancy centres in a diamond-loaded fibre cavity}},\ }\href
  {https://doi.org/10.1515/nanoph-2020-0305} {\bibfield  {journal} {\bibinfo
  {journal} {{Nanophotonics}}\ }\textbf {\bibinfo {volume} {9}},\ \bibinfo
  {pages} {4505} (\bibinfo {year} {2020})}\BibitemShut {NoStop}%
\bibitem [{\citenamefont {Doherty}\ \emph {et~al.}(2011)\citenamefont
  {Doherty}, \citenamefont {Manson}, \citenamefont {Delaney},\ and\
  \citenamefont {Hollenberg}}]{Doherty.2011}%
  \BibitemOpen
  \bibfield  {author} {\bibinfo {author} {\bibfnamefont {M.~W.}\ \bibnamefont
  {Doherty}}, \bibinfo {author} {\bibfnamefont {N.~B.}\ \bibnamefont {Manson}},
  \bibinfo {author} {\bibfnamefont {P.}~\bibnamefont {Delaney}},\ and\ \bibinfo
  {author} {\bibfnamefont {L.~C.~L.}\ \bibnamefont {Hollenberg}},\ }\bibfield
  {title} {\bibinfo {title} {{The negatively charged nitrogen-vacancy centre in
  diamond: the electronic solution}},\ }\href
  {https://doi.org/10.1088/1367-2630/13/2/025019} {\bibfield  {journal}
  {\bibinfo  {journal} {{New Journal of Physics}}\ }\textbf {\bibinfo {volume}
  {13}},\ \bibinfo {pages} {025019} (\bibinfo {year} {2011})}\BibitemShut
  {NoStop}%
\bibitem [{\citenamefont {Felton}\ \emph {et~al.}(2008)\citenamefont {Felton},
  \citenamefont {Edmonds}, \citenamefont {Newton}, \citenamefont {Martineau},
  \citenamefont {Fisher},\ and\ \citenamefont {Twitchen}}]{Felton.2008}%
  \BibitemOpen
  \bibfield  {author} {\bibinfo {author} {\bibfnamefont {S.}~\bibnamefont
  {Felton}}, \bibinfo {author} {\bibfnamefont {A.~M.}\ \bibnamefont {Edmonds}},
  \bibinfo {author} {\bibfnamefont {M.~E.}\ \bibnamefont {Newton}}, \bibinfo
  {author} {\bibfnamefont {P.~M.}\ \bibnamefont {Martineau}}, \bibinfo {author}
  {\bibfnamefont {D.}~\bibnamefont {Fisher}},\ and\ \bibinfo {author}
  {\bibfnamefont {D.~J.}\ \bibnamefont {Twitchen}},\ }\bibfield  {title}
  {\bibinfo {title} {{Electron paramagnetic resonance studies of the neutral
  nitrogen vacancy in diamond}},\ }\href
  {https://doi.org/10.1103/PhysRevB.77.081201} {\bibfield  {journal} {\bibinfo
  {journal} {{Physical Review B}}\ }\textbf {\bibinfo {volume} {77}},\ \bibinfo
  {pages} {081201(R)} (\bibinfo {year} {2008})}\BibitemShut {NoStop}%
\bibitem [{\citenamefont {Levine}\ \emph {et~al.}(2019)\citenamefont {Levine},
  \citenamefont {Turner}, \citenamefont {Kehayias}, \citenamefont {Hart},
  \citenamefont {Langellier}, \citenamefont {Trubko}, \citenamefont {Glenn},
  \citenamefont {Fu},\ and\ \citenamefont {Walsworth}}]{Levine.2019}%
  \BibitemOpen
  \bibfield  {author} {\bibinfo {author} {\bibfnamefont {E.~V.}\ \bibnamefont
  {Levine}}, \bibinfo {author} {\bibfnamefont {M.~J.}\ \bibnamefont {Turner}},
  \bibinfo {author} {\bibfnamefont {P.}~\bibnamefont {Kehayias}}, \bibinfo
  {author} {\bibfnamefont {C.~A.}\ \bibnamefont {Hart}}, \bibinfo {author}
  {\bibfnamefont {N.}~\bibnamefont {Langellier}}, \bibinfo {author}
  {\bibfnamefont {R.}~\bibnamefont {Trubko}}, \bibinfo {author} {\bibfnamefont
  {D.~R.}\ \bibnamefont {Glenn}}, \bibinfo {author} {\bibfnamefont {R.~R.}\
  \bibnamefont {Fu}},\ and\ \bibinfo {author} {\bibfnamefont {R.~L.}\
  \bibnamefont {Walsworth}},\ }\bibfield  {title} {\bibinfo {title}
  {{Principles and techniques of the quantum diamond microscope}},\ }\href
  {https://doi.org/10.1515/nanoph-2019-0209} {\bibfield  {journal} {\bibinfo
  {journal} {{Nanophotonics}}\ }\textbf {\bibinfo {volume} {8}},\ \bibinfo
  {pages} {1945} (\bibinfo {year} {2019})}\BibitemShut {NoStop}%
\bibitem [{\citenamefont {Chen}\ \emph {et~al.}(2015)\citenamefont {Chen},
  \citenamefont {Zhou}, \citenamefont {Zou}, \citenamefont {Li}, \citenamefont
  {Dong}, \citenamefont {Sun},\ and\ \citenamefont {Guo}}]{Chen.2015}%
  \BibitemOpen
  \bibfield  {author} {\bibinfo {author} {\bibfnamefont {X.-D.}\ \bibnamefont
  {Chen}}, \bibinfo {author} {\bibfnamefont {L.-M.}\ \bibnamefont {Zhou}},
  \bibinfo {author} {\bibfnamefont {C.-L.}\ \bibnamefont {Zou}}, \bibinfo
  {author} {\bibfnamefont {C.-C.}\ \bibnamefont {Li}}, \bibinfo {author}
  {\bibfnamefont {Y.}~\bibnamefont {Dong}}, \bibinfo {author} {\bibfnamefont
  {F.-W.}\ \bibnamefont {Sun}},\ and\ \bibinfo {author} {\bibfnamefont {G.-C.}\
  \bibnamefont {Guo}},\ }\bibfield  {title} {\bibinfo {title} {{Spin
  depolarization effect induced by charge state conversion of nitrogen vacancy
  center in diamond}},\ }\href {https://doi.org/10.1103/PhysRevB.92.104301}
  {\bibfield  {journal} {\bibinfo  {journal} {{Physical Review B}}\ }\textbf
  {\bibinfo {volume} {92}},\ \bibinfo {pages} {104301} (\bibinfo {year}
  {2015})}\BibitemShut {NoStop}%
\bibitem [{\citenamefont {Meirzada}\ \emph {et~al.}(2018)\citenamefont
  {Meirzada}, \citenamefont {Hovav}, \citenamefont {Wolf},\ and\ \citenamefont
  {Bar-Gill}}]{Meirzada.2018}%
  \BibitemOpen
  \bibfield  {author} {\bibinfo {author} {\bibfnamefont {I.}~\bibnamefont
  {Meirzada}}, \bibinfo {author} {\bibfnamefont {Y.}~\bibnamefont {Hovav}},
  \bibinfo {author} {\bibfnamefont {S.~A.}\ \bibnamefont {Wolf}},\ and\
  \bibinfo {author} {\bibfnamefont {N.}~\bibnamefont {Bar-Gill}},\ }\bibfield
  {title} {\bibinfo {title} {{Negative charge enhancement of near-surface
  nitrogen vacancy centers by multicolor excitation}},\ }\href
  {https://doi.org/10.1103/PhysRevB.98.245411} {\bibfield  {journal} {\bibinfo
  {journal} {{Physical Review B}}\ }\textbf {\bibinfo {volume} {98}},\ \bibinfo
  {pages} {245411} (\bibinfo {year} {2018})}\BibitemShut {NoStop}%
\bibitem [{\citenamefont {Naydenov}\ \emph {et~al.}(2011)\citenamefont
  {Naydenov}, \citenamefont {Dolde}, \citenamefont {Hall}, \citenamefont
  {Shin}, \citenamefont {Fedder}, \citenamefont {Hollenberg}, \citenamefont
  {Jelezko},\ and\ \citenamefont {Wrachtrup}}]{Naydenov.2011}%
  \BibitemOpen
  \bibfield  {author} {\bibinfo {author} {\bibfnamefont {B.}~\bibnamefont
  {Naydenov}}, \bibinfo {author} {\bibfnamefont {F.}~\bibnamefont {Dolde}},
  \bibinfo {author} {\bibfnamefont {L.~T.}\ \bibnamefont {Hall}}, \bibinfo
  {author} {\bibfnamefont {C.}~\bibnamefont {Shin}}, \bibinfo {author}
  {\bibfnamefont {H.}~\bibnamefont {Fedder}}, \bibinfo {author} {\bibfnamefont
  {L.~C.~L.}\ \bibnamefont {Hollenberg}}, \bibinfo {author} {\bibfnamefont
  {F.}~\bibnamefont {Jelezko}},\ and\ \bibinfo {author} {\bibfnamefont
  {J.}~\bibnamefont {Wrachtrup}},\ }\bibfield  {title} {\bibinfo {title}
  {{Dynamical decoupling of a single-electron spin at room temperature}},\
  }\href {https://doi.org/10.1103/PhysRevB.83.081201} {\bibfield  {journal}
  {\bibinfo  {journal} {{Physical Review B}}\ }\textbf {\bibinfo {volume}
  {83}},\ \bibinfo {pages} {081201(R)} (\bibinfo {year} {2011})}\BibitemShut
  {NoStop}%
\bibitem [{\citenamefont {Jarmola}\ \emph {et~al.}(2012)\citenamefont
  {Jarmola}, \citenamefont {Acosta}, \citenamefont {Jensen}, \citenamefont
  {Chemerisov},\ and\ \citenamefont {Budker}}]{Jarmola.2012}%
  \BibitemOpen
  \bibfield  {author} {\bibinfo {author} {\bibfnamefont {A.}~\bibnamefont
  {Jarmola}}, \bibinfo {author} {\bibfnamefont {V.~M.}\ \bibnamefont {Acosta}},
  \bibinfo {author} {\bibfnamefont {K.}~\bibnamefont {Jensen}}, \bibinfo
  {author} {\bibfnamefont {S.}~\bibnamefont {Chemerisov}},\ and\ \bibinfo
  {author} {\bibfnamefont {D.}~\bibnamefont {Budker}},\ }\bibfield  {title}
  {\bibinfo {title} {{Temperature- and Magnetic-Field-Dependent Longitudinal
  Spin Relaxation in Nitrogen-Vacancy Ensembles in Diamond}},\ }\href
  {https://doi.org/10.1103/PhysRevLett.108.197601} {\bibfield  {journal}
  {\bibinfo  {journal} {{Physical Review Letters}}\ }\textbf {\bibinfo {volume}
  {108}},\ \bibinfo {pages} {197601} (\bibinfo {year} {2012})}\BibitemShut
  {NoStop}%
\bibitem [{\citenamefont {Romach}\ \emph {et~al.}(2015)\citenamefont {Romach},
  \citenamefont {M{\"u}ller}, \citenamefont {Unden}, \citenamefont {Rogers},
  \citenamefont {Isoda}, \citenamefont {Itoh}, \citenamefont {Markham},
  \citenamefont {Stacey}, \citenamefont {Meijer}, \citenamefont {Pezzagna},
  \citenamefont {Naydenov}, \citenamefont {McGuinness}, \citenamefont
  {Bar-Gill},\ and\ \citenamefont {Jelezko}}]{Romach.2015}%
  \BibitemOpen
  \bibfield  {author} {\bibinfo {author} {\bibfnamefont {Y.}~\bibnamefont
  {Romach}}, \bibinfo {author} {\bibfnamefont {C.}~\bibnamefont {M{\"u}ller}},
  \bibinfo {author} {\bibfnamefont {T.}~\bibnamefont {Unden}}, \bibinfo
  {author} {\bibfnamefont {L.~J.}\ \bibnamefont {Rogers}}, \bibinfo {author}
  {\bibfnamefont {T.}~\bibnamefont {Isoda}}, \bibinfo {author} {\bibfnamefont
  {K.~M.}\ \bibnamefont {Itoh}}, \bibinfo {author} {\bibfnamefont
  {M.}~\bibnamefont {Markham}}, \bibinfo {author} {\bibfnamefont
  {A.}~\bibnamefont {Stacey}}, \bibinfo {author} {\bibfnamefont
  {J.}~\bibnamefont {Meijer}}, \bibinfo {author} {\bibfnamefont
  {S.}~\bibnamefont {Pezzagna}}, \bibinfo {author} {\bibfnamefont
  {B.}~\bibnamefont {Naydenov}}, \bibinfo {author} {\bibfnamefont {L.~P.}\
  \bibnamefont {McGuinness}}, \bibinfo {author} {\bibfnamefont
  {N.}~\bibnamefont {Bar-Gill}},\ and\ \bibinfo {author} {\bibfnamefont
  {F.}~\bibnamefont {Jelezko}},\ }\bibfield  {title} {\bibinfo {title}
  {{Spectroscopy of Surface-Induced Noise Using Shallow Spins in Diamond}},\
  }\href {https://doi.org/10.1103/PhysRevLett.114.017601} {\bibfield  {journal}
  {\bibinfo  {journal} {{Physical Review Letters}}\ }\textbf {\bibinfo {volume}
  {114}},\ \bibinfo {pages} {017601} (\bibinfo {year} {2015})}\BibitemShut
  {NoStop}%
\bibitem [{\citenamefont {Mr{\'o}zek}\ \emph {et~al.}(2015)\citenamefont
  {Mr{\'o}zek}, \citenamefont {Rudnicki}, \citenamefont {Kehayias},
  \citenamefont {Jarmola}, \citenamefont {Budker},\ and\ \citenamefont
  {Gawlik}}]{Mrozek.2015}%
  \BibitemOpen
  \bibfield  {author} {\bibinfo {author} {\bibfnamefont {M.}~\bibnamefont
  {Mr{\'o}zek}}, \bibinfo {author} {\bibfnamefont {D.}~\bibnamefont
  {Rudnicki}}, \bibinfo {author} {\bibfnamefont {P.}~\bibnamefont {Kehayias}},
  \bibinfo {author} {\bibfnamefont {A.}~\bibnamefont {Jarmola}}, \bibinfo
  {author} {\bibfnamefont {D.}~\bibnamefont {Budker}},\ and\ \bibinfo {author}
  {\bibfnamefont {W.}~\bibnamefont {Gawlik}},\ }\bibfield  {title} {\bibinfo
  {title} {{Longitudinal spin relaxation in nitrogen-vacancy ensembles in
  diamond}},\ }\bibfield  {journal} {\bibinfo  {journal} {{EPJ Quantum
  Technology}}\ }\textbf {\bibinfo {volume} {2}},\ \href
  {https://doi.org/10.1140/epjqt/s40507-015-0035-z}
  {10.1140/epjqt/s40507-015-0035-z} (\bibinfo {year} {2015})\BibitemShut
  {NoStop}%
\bibitem [{\citenamefont {Manson}\ \emph {et~al.}(2018)\citenamefont {Manson},
  \citenamefont {Hedges}, \citenamefont {Barson}, \citenamefont {Ahlefeldt},
  \citenamefont {Doherty}, \citenamefont {Abe}, \citenamefont {Ohshima},\ and\
  \citenamefont {Sellars}}]{Manson.2018}%
  \BibitemOpen
  \bibfield  {author} {\bibinfo {author} {\bibfnamefont {N.~B.}\ \bibnamefont
  {Manson}}, \bibinfo {author} {\bibfnamefont {M.}~\bibnamefont {Hedges}},
  \bibinfo {author} {\bibfnamefont {M.~S.~J.}\ \bibnamefont {Barson}}, \bibinfo
  {author} {\bibfnamefont {R.}~\bibnamefont {Ahlefeldt}}, \bibinfo {author}
  {\bibfnamefont {M.~W.}\ \bibnamefont {Doherty}}, \bibinfo {author}
  {\bibfnamefont {H.}~\bibnamefont {Abe}}, \bibinfo {author} {\bibfnamefont
  {T.}~\bibnamefont {Ohshima}},\ and\ \bibinfo {author} {\bibfnamefont {M.~J.}\
  \bibnamefont {Sellars}},\ }\bibfield  {title} {\bibinfo {title} {{NV $-$ --N
  + pair centre in 1b diamond}},\ }\href
  {https://doi.org/10.1088/1367-2630/aaec58} {\bibfield  {journal} {\bibinfo
  {journal} {{New Journal of Physics}}\ }\textbf {\bibinfo {volume} {20}},\
  \bibinfo {pages} {113037} (\bibinfo {year} {2018})}\BibitemShut {NoStop}%
\bibitem [{\citenamefont {Juan}\ \emph {et~al.}(2017)\citenamefont {Juan},
  \citenamefont {Bradac}, \citenamefont {Besga}, \citenamefont {Johnsson},
  \citenamefont {Brennen}, \citenamefont {Molina-Terriza},\ and\ \citenamefont
  {Volz}}]{Juan.2017}%
  \BibitemOpen
  \bibfield  {author} {\bibinfo {author} {\bibfnamefont {M.~L.}\ \bibnamefont
  {Juan}}, \bibinfo {author} {\bibfnamefont {C.}~\bibnamefont {Bradac}},
  \bibinfo {author} {\bibfnamefont {B.}~\bibnamefont {Besga}}, \bibinfo
  {author} {\bibfnamefont {M.}~\bibnamefont {Johnsson}}, \bibinfo {author}
  {\bibfnamefont {G.}~\bibnamefont {Brennen}}, \bibinfo {author} {\bibfnamefont
  {G.}~\bibnamefont {Molina-Terriza}},\ and\ \bibinfo {author} {\bibfnamefont
  {T.}~\bibnamefont {Volz}},\ }\bibfield  {title} {\bibinfo {title}
  {{Cooperatively enhanced dipole forces from artificial atoms in trapped
  nanodiamonds}},\ }\href {https://doi.org/10.1038/nphys3940} {\bibfield
  {journal} {\bibinfo  {journal} {{Nature Physics}}\ }\textbf {\bibinfo
  {volume} {13}},\ \bibinfo {pages} {241} (\bibinfo {year} {2017})}\BibitemShut
  {NoStop}%
\bibitem [{\citenamefont {Alsid}\ \emph {et~al.}(2019)\citenamefont {Alsid},
  \citenamefont {Barry}, \citenamefont {Pham}, \citenamefont {Schloss},
  \citenamefont {O'Keeffe}, \citenamefont {Cappellaro},\ and\ \citenamefont
  {Braje}}]{Alsid.2019}%
  \BibitemOpen
  \bibfield  {author} {\bibinfo {author} {\bibfnamefont {S.~T.}\ \bibnamefont
  {Alsid}}, \bibinfo {author} {\bibfnamefont {J.~F.}\ \bibnamefont {Barry}},
  \bibinfo {author} {\bibfnamefont {L.~M.}\ \bibnamefont {Pham}}, \bibinfo
  {author} {\bibfnamefont {J.~M.}\ \bibnamefont {Schloss}}, \bibinfo {author}
  {\bibfnamefont {M.~F.}\ \bibnamefont {O'Keeffe}}, \bibinfo {author}
  {\bibfnamefont {P.}~\bibnamefont {Cappellaro}},\ and\ \bibinfo {author}
  {\bibfnamefont {D.~A.}\ \bibnamefont {Braje}},\ }\bibfield  {title} {\bibinfo
  {title} {{Photoluminescence Decomposition Analysis: A Technique to
  Characterize N - V Creation in Diamond}},\ }\href
  {https://doi.org/10.1103/PhysRevApplied.12.044003} {\bibfield  {journal}
  {\bibinfo  {journal} {{Physical Review Applied}}\ }\textbf {\bibinfo {volume}
  {12}},\ \bibinfo {pages} {044003} (\bibinfo {year} {2019})}\BibitemShut
  {NoStop}%
\bibitem [{\citenamefont {Wolf}\ \emph {et~al.}(2015)\citenamefont {Wolf},
  \citenamefont {Neumann}, \citenamefont {Nakamura}, \citenamefont {Sumiya},
  \citenamefont {Ohshima}, \citenamefont {Isoya},\ and\ \citenamefont
  {Wrachtrup}}]{Wolf.2015}%
  \BibitemOpen
  \bibfield  {author} {\bibinfo {author} {\bibfnamefont {T.}~\bibnamefont
  {Wolf}}, \bibinfo {author} {\bibfnamefont {P.}~\bibnamefont {Neumann}},
  \bibinfo {author} {\bibfnamefont {K.}~\bibnamefont {Nakamura}}, \bibinfo
  {author} {\bibfnamefont {H.}~\bibnamefont {Sumiya}}, \bibinfo {author}
  {\bibfnamefont {T.}~\bibnamefont {Ohshima}}, \bibinfo {author} {\bibfnamefont
  {J.}~\bibnamefont {Isoya}},\ and\ \bibinfo {author} {\bibfnamefont
  {J.}~\bibnamefont {Wrachtrup}},\ }\bibfield  {title} {\bibinfo {title}
  {{Subpicotesla Diamond Magnetometry}},\ }\href
  {https://doi.org/10.1103/PhysRevX.5.041001} {\bibfield  {journal} {\bibinfo
  {journal} {{Physical Review X}}\ }\textbf {\bibinfo {volume} {5}},\ \bibinfo
  {pages} {041001} (\bibinfo {year} {2015})}\BibitemShut {NoStop}%
\bibitem [{\citenamefont {Robledo}\ \emph {et~al.}(2011)\citenamefont
  {Robledo}, \citenamefont {Bernien}, \citenamefont {{van der Sar}},\ and\
  \citenamefont {Hanson}}]{Robledo.2011}%
  \BibitemOpen
  \bibfield  {author} {\bibinfo {author} {\bibfnamefont {L.}~\bibnamefont
  {Robledo}}, \bibinfo {author} {\bibfnamefont {H.}~\bibnamefont {Bernien}},
  \bibinfo {author} {\bibfnamefont {T.}~\bibnamefont {{van der Sar}}},\ and\
  \bibinfo {author} {\bibfnamefont {R.}~\bibnamefont {Hanson}},\ }\bibfield
  {title} {\bibinfo {title} {{Spin dynamics in the optical cycle of single
  nitrogen-vacancy centres in diamond}},\ }\href
  {https://doi.org/10.1088/1367-2630/13/2/025013} {\bibfield  {journal}
  {\bibinfo  {journal} {{New Journal of Physics}}\ }\textbf {\bibinfo {volume}
  {13}},\ \bibinfo {pages} {025013} (\bibinfo {year} {2011})}\BibitemShut
  {NoStop}%
\bibitem [{\citenamefont {de~Guillebon}\ \emph {et~al.}(2020)\citenamefont
  {de~Guillebon}, \citenamefont {Vindolet}, \citenamefont {Roch}, \citenamefont
  {Jacques},\ and\ \citenamefont {Rondin}}]{Guillebon.2020}%
  \BibitemOpen
  \bibfield  {author} {\bibinfo {author} {\bibfnamefont {T.}~\bibnamefont
  {de~Guillebon}}, \bibinfo {author} {\bibfnamefont {B.}~\bibnamefont
  {Vindolet}}, \bibinfo {author} {\bibfnamefont {J.-F.}\ \bibnamefont {Roch}},
  \bibinfo {author} {\bibfnamefont {V.}~\bibnamefont {Jacques}},\ and\ \bibinfo
  {author} {\bibfnamefont {L.}~\bibnamefont {Rondin}},\ }\bibfield  {title}
  {\bibinfo {title} {{Temperature dependence of the longitudinal spin
  relaxation time T1 of single nitrogen-vacancy centers in nanodiamonds}},\
  }\href {https://doi.org/10.1103/PhysRevB.102.165427} {\bibfield  {journal}
  {\bibinfo  {journal} {{Physical Review B}}\ }\textbf {\bibinfo {volume}
  {102}},\ \bibinfo {pages} {165427} (\bibinfo {year} {2020})}\BibitemShut
  {NoStop}%
\bibitem [{\citenamefont {Rondin}\ \emph {et~al.}(2010)\citenamefont {Rondin},
  \citenamefont {Dantelle}, \citenamefont {Slablab}, \citenamefont {Grosshans},
  \citenamefont {Treussart}, \citenamefont {Bergonzo}, \citenamefont
  {Perruchas}, \citenamefont {Gacoin}, \citenamefont {Chaigneau}, \citenamefont
  {Chang}, \citenamefont {Jacques},\ and\ \citenamefont {Roch}}]{Rondin.2010}%
  \BibitemOpen
  \bibfield  {author} {\bibinfo {author} {\bibfnamefont {L.}~\bibnamefont
  {Rondin}}, \bibinfo {author} {\bibfnamefont {G.}~\bibnamefont {Dantelle}},
  \bibinfo {author} {\bibfnamefont {A.}~\bibnamefont {Slablab}}, \bibinfo
  {author} {\bibfnamefont {F.}~\bibnamefont {Grosshans}}, \bibinfo {author}
  {\bibfnamefont {F.}~\bibnamefont {Treussart}}, \bibinfo {author}
  {\bibfnamefont {P.}~\bibnamefont {Bergonzo}}, \bibinfo {author}
  {\bibfnamefont {S.}~\bibnamefont {Perruchas}}, \bibinfo {author}
  {\bibfnamefont {T.}~\bibnamefont {Gacoin}}, \bibinfo {author} {\bibfnamefont
  {M.}~\bibnamefont {Chaigneau}}, \bibinfo {author} {\bibfnamefont {H.-C.}\
  \bibnamefont {Chang}}, \bibinfo {author} {\bibfnamefont {V.}~\bibnamefont
  {Jacques}},\ and\ \bibinfo {author} {\bibfnamefont {J.-F.}\ \bibnamefont
  {Roch}},\ }\bibfield  {title} {\bibinfo {title} {{Surface-induced charge
  state conversion of nitrogen-vacancy defects in nanodiamonds}},\ }\href
  {https://doi.org/10.1103/PhysRevB.82.115449} {\bibfield  {journal} {\bibinfo
  {journal} {{Physical Review B}}\ }\textbf {\bibinfo {volume} {82}},\ \bibinfo
  {pages} {115449} (\bibinfo {year} {2010})}\BibitemShut {NoStop}%
\bibitem [{\citenamefont {Wilson}\ \emph {et~al.}(2019)\citenamefont {Wilson},
  \citenamefont {Parker}, \citenamefont {Orth}, \citenamefont {Nunn},
  \citenamefont {Torelli}, \citenamefont {Shenderova}, \citenamefont {Gibson},\
  and\ \citenamefont {Reineck}}]{Wilson.2019}%
  \BibitemOpen
  \bibfield  {author} {\bibinfo {author} {\bibfnamefont {E.~R.}\ \bibnamefont
  {Wilson}}, \bibinfo {author} {\bibfnamefont {L.~M.}\ \bibnamefont {Parker}},
  \bibinfo {author} {\bibfnamefont {A.}~\bibnamefont {Orth}}, \bibinfo {author}
  {\bibfnamefont {N.}~\bibnamefont {Nunn}}, \bibinfo {author} {\bibfnamefont
  {M.}~\bibnamefont {Torelli}}, \bibinfo {author} {\bibfnamefont
  {O.}~\bibnamefont {Shenderova}}, \bibinfo {author} {\bibfnamefont {B.~C.}\
  \bibnamefont {Gibson}},\ and\ \bibinfo {author} {\bibfnamefont
  {P.}~\bibnamefont {Reineck}},\ }\bibfield  {title} {\bibinfo {title} {{The
  effect of particle size on nanodiamond fluorescence and colloidal properties
  in biological media}},\ }\href {https://doi.org/10.1088/1361-6528/ab283d}
  {\bibfield  {journal} {\bibinfo  {journal} {{Nanotechnology}}\ }\textbf
  {\bibinfo {volume} {30}},\ \bibinfo {pages} {385704} (\bibinfo {year}
  {2019})}\BibitemShut {NoStop}%
\bibitem [{\citenamefont {Dhomkar}\ \emph {et~al.}(2018)\citenamefont
  {Dhomkar}, \citenamefont {Jayakumar}, \citenamefont {Zangara},\ and\
  \citenamefont {Meriles}}]{Dhomkar.2018}%
  \BibitemOpen
  \bibfield  {author} {\bibinfo {author} {\bibfnamefont {S.}~\bibnamefont
  {Dhomkar}}, \bibinfo {author} {\bibfnamefont {H.}~\bibnamefont {Jayakumar}},
  \bibinfo {author} {\bibfnamefont {P.~R.}\ \bibnamefont {Zangara}},\ and\
  \bibinfo {author} {\bibfnamefont {C.~A.}\ \bibnamefont {Meriles}},\
  }\bibfield  {title} {\bibinfo {title} {{Charge Dynamics in near-Surface,
  Variable-Density Ensembles of Nitrogen-Vacancy Centers in Diamond}},\ }\href
  {https://doi.org/10.1021/acs.nanolett.8b01739} {\bibfield  {journal}
  {\bibinfo  {journal} {{Nano Letters}}\ }\textbf {\bibinfo {volume} {18}},\
  \bibinfo {pages} {4046} (\bibinfo {year} {2018})}\BibitemShut {NoStop}%
\bibitem [{Zen(2023)}]{Zenodo}%
  \BibitemOpen
  \href {https://doi.org/10.5281/zenodo.7599850} {\bibinfo {title} {Zenodo data
  repository}},\ \bibinfo {howpublished} {doi: 10.5281/zenodo.7599850}
  (\bibinfo {year} {2023})\BibitemShut {NoStop}%
\end{thebibliography}
    
%

\end{document}